\documentclass[11pt,twoside]{article}

\usepackage[utf8]{inputenc}
\usepackage{epsfig,amsfonts,color}
\usepackage{chngcntr}
\usepackage{mathrsfs} 
\usepackage{amsmath, bbm, mathabx, amsthm}
\usepackage[font=small]{caption}
\usepackage[normalem]{ulem}
\usepackage{amssymb, palatino, geometry,url}
\usepackage{booktabs}
\usepackage{tabularx}
\usepackage{array}
\usepackage[colorlinks=true,linkcolor=blue,citecolor=blue,urlcolor=blue]{hyperref}
\usepackage{subcaption}
\usepackage{multirow}

\usepackage{makecell}
\usepackage{graphicx} 
\newcolumntype{Y}{>{\centering\arraybackslash}X}

\numberwithin{equation}{section}

\usepackage{lineno}
\usepackage{float}
\usepackage[section]{placeins}
\usepackage[table]{xcolor}

\title{On the Automatic Discovery of Optimal Hybrid First Principles – Machine Learning Model Architecture}

\author{Angan Mukherjee$^{1,2}$, Nishant V. Giridhar$^1$, and Debangsu Bhattacharyya$^{1}$\thanks{Corresponding Author. Tel: +1-3042939335, Fax:+1-3042934139, Email: Debangsu.Bhattacharyya@mail.wvu.edu}\\
\\
    {\small $^1$Department of Chemical \& Biomedical Engineering, West Virginia University, Morgantown, WV 26506, USA}\\
    {\small $^2$Department of Chemical \& Biological Engineering, University of Wisconsin-Madison, Madison, WI 53706, USA}
    }
\date{}

\begin{document}

\maketitle

\begin{abstract}

This paper presents ADOPT-FPML (Automatic Discovery of OPTimal hybrid First Principles - Machine Learning models), a novel systematic approach to the automatic discovery of optimal hybrid first principles – machine learning (FPML) model architectures. We introduce a generic approach for superstructure of hybrid models represented by binary decision variables that encode data flow between measured, intermediate, and predicted variables. Sizes of the input and output variables for the machine learning (ML) model(s) are implicitly governed by the superstructure decision variables. The proposed framework is agnostic to the specific ML model used. The structure of the hybrid FPML model is optimized given the training data by using an objective function based on an information-theoretic criterion, that accounts for the trade-off between predictive accuracy, computational expense, and model sparsity. The developed ADOPT-FPML approach has been evaluated through three case studies involving two nonlinear dynamic process systems, while also illustrating how the proposed superstructure can be implemented  on interconnected process systems in a modular manner, where different process sections might be represented by different optimal hybrid FPML configurations depending on the underlying physics, available measurements, and prediction accuracy. It is observed that the proposed algorithm for superstructure optimization can automatically discover the hybrid structures that are non-intuitive and can be time-consuming or difficult to come up with heuristically. This work  provides a systematic first step towards the automated discovery of optimal hybrid FPML models, that can be used for real-time monitoring, soft sensing,  control, and optimal decision making of complex process systems.    
\\
\\
\\


\end{abstract}
\newpage

\section{Introduction} \label{sec:introduction}

First-principles (FP) (or mechanistic) models can provide predictive capabilities in wide operating regimes while satisfying the physics or chemistry of the process. They do not require a large amount of data. Thus, FP models are highly desired where there is a lack of data or large extrapolation is desired beyond what the available data represent or  where data collection is infeasible given the current state of measurement technology. However, developing accurate FP models, especially for complex nonlinear transient process systems, may be computationally expensive, difficult to scale up, time/resource consuming, and intractable for online adaptation. Furthermore, development of FP models requires process/domain knowledge and often must be simplified by suitable assumptions to reliably solve \cite{saini2026oxide, das2026bcd} that may compromise their predictive accuracy. In contrast, machine learning (ML) or black-box models are relatively easier to develop, simulate, and adapt online, even for complex and ill-defined processes \cite{mukherjee2025nnvsbml,mukherjee2023hybserpar}. One of the key limitations of data-driven models is the anticipated lack of predictive or extrapolative capabilities especially when there is an information gap in the data used for model development. One way of synergistic utilization of the strengths of FP and ML models is through the development of hybrid first principles – machine learning (FPML) models \cite{mukherjee2025pcml,shah2022dnnhybrid}. However, existing state-of-the-art approaches for hybrid modeling are typically derived from domain knowledge and / or heuristics \cite{mukherjee2025hybridfpai}. There is a {\textit{critical}} need for developing a systematic method for synthesis of FPML models \cite{dicaprio2026sindybrid}. This work develops ADOPT-FPML (Automatic Discovery of OPTimal hybrid First Principles - Machine Learning models), a systematic framework that uses disjunctive superstructure optimization \cite{perez2024gdp, linan2023superstructure} to automatically identify the optimal structure of hybrid FPML models. 

Over the last few decades, hybrid FPML modeling frameworks have been extensively explored across a broad range of applications \cite{mukherjee2025hybridfpai,chen2021hybco2,saini2026oxide} in process systems engineering (PSE). One common implementation paradigm typically involves lumped parameter-based hybridization, where the ML model estimates key static parameters that are subsequently embedded within the FP model \cite{psichogios1992hybfpai}. Other alternative hybridization strategies include modeling the residuals of incomplete / partial FP models using data-driven models, with demonstrated applications in chemical reactors \cite{rajulapati2022hybreview}, power and energy systems \cite{machalek2022hybenergy}, and biological \cite{lee2020intracellular} processes. More recently, researchers have also proposed physics-constrained or physics-informed ML approaches that enforce system conservation principles such as mass \cite{mukherjee2024mcnn}, energy \cite{mukherjee2024mecnn}, thermodynamics \cite{carranza2022asnn,mukherjee2025metcnn}, and momentum \cite{raissi2019pinn} balances directly into the training algorithm, alongside digital twin platforms for energy systems \cite{yu2022digitaltwin}. Collectively, such approaches offer enhanced interpretability, robustness, and improved extrapolation capabilities as compared to pure ML models \cite{bradley2022perspective}. Despite these advancements, significant opportunities still exist to more effectively harness the complementary strengths of FP and ML models. Furthermore, it is also desired that the synthesis of hybrid FPML architectures explicitly accounts for model complexity, computational expense, and predictive accuracy.

While modeling many practical real-world processes, it may even be necessary to employ multiple ML models simultaneously within a unified hybrid framework \cite{mukherjee2025hybridfpai, kudva2026hybrid}. The hybridization of FP and ML models can be represented by different configurations, e.g., parallel, series, integrated, or more complex composite structures, each designed to exchange information between FP and ML models in distinct ways. Parallel FPML architectures are typically used when the unmodeled physical phenomena are separable from the FP representation, such that both FP and ML models operate independently on shared inputs without mutual interaction. In such cases, the data-driven model often serves as direct corrective functions to account for bias in the FP predictions \cite{chen2020hyb}. For non-separable discrepancies, where either the outputs of the FP model serve as additional inputs to the ML model or vice-versa (but not both), series-type architectures are typically preferred and have shown relatively superior extrapolation capabilities than parallel configurations \cite{von2014serbetterpar}. Lastly, when bidirectional information exchange is necessary, i.e., FP model outputs impact the ML model while ML model outputs are used to simultaneously update the FP model, an integrated architecture is particularly favorable, as they enable continuous or frequent refinement of internal states, parameters, or other desired variables \cite{mukherjee2025hybridfpai,saini2026oxide}. In most existing hybrid FPML implementations, however, a single global structure is selected for the full model, such that all outputs are predicted through the same interaction architecture between the FP and ML model components. This can be restrictive for multivariate systems, where different predicted outputs might require different forms of coupling between the FP and ML model components. 

Despite the recent advancements, the development of hybrid FPML models, in practice, remains largely problem-specific and experience-driven, where users need to make decisions about which states require correction, which variables should serve as ML inputs, and how the mechanistic and data-driven components should interact \cite{bradley2022perspective}. Consequently, the resulting architectures may not always represent the best structure for a given system. Furthermore, existing approaches generally do not consider different structural interactions for different predicted outputs within one unified framework. More generally, for interconnected systems, different sections of the overall process might be represented by different hybrid FPML structures depending on the local physics, measurement availability, and prediction objectives. Enabling such output-specific and subsystem-specific hybridization requires a more general superstructure representation than the fixed global architectures commonly used in the literature. On the other hand, structure optimization and automated model discovery have seen significant developments in areas such as symbolic regression and sparse model discovery. Approaches such as SINDy-based frameworks, Bayesian discovery methods, and hybrid symbolic data-driven learning have demonstrated the benefits of automated search over complex structural spaces to identify parsimonious yet accurate nonlinear models \cite{dicaprio2026sindybrid, adeyemo2024bidsam, adeyemo2026sparse}. These approaches explicitly account for tradeoffs between predictive accuracy and structural sparsity and have shown promise in discovering models directly from data given the candidate basis functions. However, these strategies fundamentally differ from hybrid FPML synthesis problems, because FPML architectures involve information exchange between physics-based and learning-based modules rather than only discovering algebraic and/or dynamical relationships. Therefore, the systematic approaches being developed for data-driven  model-building are not applicable to the automatic discovery of FPML model architectures.

A key challenge to synthesizing the optimal FPML model architecture is in deciding what information would be exchanged between FP and ML models. This would depend on several decisions including (i) which output(s) of the FP model require correction(s), (ii) which external inputs and which intermediate or measured state(s) should serve as ML inputs, (iii) how many hybrid stages are needed, and (iv) which configuration (parallel, series, integrated, or combination of series and parallel, etc.) would yield the best performance. As system dimensionality increases, the number of potential decisions to the items noted before grows combinatorially, thus resulting in an excessively large discrete search space. For example, with respect to item(ii), the selection of all available (state) variables as ML inputs is generally undesirable, as it would induce redundancy, and can increase model overfitting, while excessively restrictive selections may degrade predictive accuracy and robustness \cite{azevedo2024inpoutsel}. Thus, input–output selection in FPML structures becomes a challenging integer programming (IP) problem that must simultaneously balance model accuracy, extrapolation capability, identifiability, and computational tractability \cite{mukherjee2024thesis}. 

A related line of work in neural architecture search (NAS) \cite{wistuba2019survey} addresses large discrete architecture search spaces either through continuous relaxations \cite{wang2024pinn} of the discrete search space or through problem decomposition and bilevel optimization \cite{jiang2020efficient}. In the latter, architecture selection is posed as a combinatorial optimization problem, while the likely intractability of simultaneous architecture and parameter optimization problem is addressed by a bi-level approach, where the discrete problem for the optimal architecture is solved at the higher level followed by the nonlinear programming problem for  training. In practice, the resulting combinatorial search is often conducted using metaheuristic strategies such as evolutionary or population-based methods. These approaches do not guarantee global optimality and would incur substantial computational cost, as a large number of candidate architectures must be trained and evaluated to assess relative performance. Moreover, NAS methods focus exclusively on neural network architecture design and do not address the fundamental challenge of what information would be exchanged between FP and ML models for discovering optimal FPML model configurations.

In the ML literature, the input selection problem has been mainly addressed   through feature selection, in which feature relevance can be learned separately using an unsupervised approach or learned jointly with model parameters. When learned jointly, typically learnable input masks are introduced that are optimized along with network weights using gradient-based training, typically via continuous relaxations of discrete selection \cite{dona2021differentiable}. Such methods enable automated input selection without explicit combinatorial enumeration and scale favorably to high-dimensional settings \cite{cancela2022e2e}. However, these approaches are developed primarily for purely data-driven models and do not account for the structured information exchange, physical interpretability, and mechanistic consistency requirements inherent to hybrid first-principles and machine learning systems. Unlike conventional ML model selection problems that are purely data-driven, it is desired that the optimal FPML configuration also preserves meaningful flow of information between FP and ML model components and ensures compatibility with mechanistic constraints. This creates additional layers of coupling between continuous parameters (e.g., FP and ML model parameters) and discrete decisions (e.g., optimal FPML architecture, input-output variable selection), that can be represented in the form of a mixed-integer nonlinear programming (MINLP) problem. To the authors' knowledge, currently there exists no generalized, automated, and systematic framework that rigorously explores the full discrete space of hybrid FPML model architectures, while taking into account the trade-offs between model sparsity and predictive accuracy.

Motivated by these challenges, we propose ADOPT-FPML,  where the hybrid FPML synthesis is posed as a disjunctive superstructure optimization problem, in which feasible FPML architectures are represented through binary routing decisions that encode information exchange between the FP and ML model components. These binary arrays serve as mapping matrices that determine which state variables would be used as inputs or outputs for the ML model. The resulting superstructure also provides the flexibility to construct hybrid architectures in which predicted outputs may have different interaction structure between FP and ML components. In addition, the same superstructure can be deployed as a modular modeling block for interconnected process systems, where different process subsystems may be modeled based on their respective optimal hybrid FPML configurations. Such partitions may be defined based on process connectivity, available measurements, or the scope of the mechanistic model associated with each subsystem. As a result, the structure of the ML model, including the size of the input and output layers and thus the number of trainable parameters related to the architecture, is implicitly governed by the superstructure decision variables. The proposed framework remains agnostic to the specific ML model used. The hybrid FPML model architecture is automatically discovered  given observational data. The resulting disjunctive MINLP formulation is solved using a stage-wise branch-and-prune strategy, with the corrected Akaike Information Criterion (AICc) as the objective function to penalize model complexity while rewarding predictive capability \cite{mukherjee2025nnvsbml, adeyemo2024bidsam, ferguson2019aicbic}. ADOPT-FPML enables the automated discovery of non-trivial hybrid FPML architectures that can be cumbersome or non-intuitive to synthesize through heuristics. This work presents a systematic approach to the automated discovery of hybrid FPML architectures and provides a unified methodology by combining strategies for optimal selection of input-output variables and model configuration, thus facilitating advanced process systems applications.

In summary, the {\textit {contributions}} of this paper are as follows:

\begin{itemize}

\item We propose ADOPT-FPML, a  systematic disjunctive superstructure optimization framework for the automated discovery of optimal hybrid FPML model architectures.   

\item The proposed algorithm simultaneously incorporates strategies for optimal input and output variable selection, as well as optimal configuration of the FP and ML components through disjunctive routing decisions, while enabling the construction of composite hybrid architectures where outputs are better represented by different interaction structures. 

\item We use an information-theoretic model selection criterion (i.e., AICc) as the objective function for the optimization (MINLP) problem, thus enabling  trade-off between predictive capability and model sparsity, thus addressing potential risk of overfitting in high-dimensional transient systems.

\item We evaluate the proposed approach through three case studies involving two nonlinear process systems: two subsystem envelopes of an interconnected acetic anhydride production process, encompassing an electrically heated reactor and downstream heat-exchange and flash-separation operations \cite{Jeffreys1964Book,Fogler2017Book}, and an industrial superheater system associated with a natural gas combined cycle power plant. We further demonstrate how the same superstructure can be deployed across interconnected process systems through modular decomposition into subsystem-level FPML building blocks. 

\end{itemize}

The rest of the paper is organized as follows: Section \ref{sec:structures} introduces the individual hybrid FPML substructures and the resulting superstructure. Section \ref{sec:optimization} discusses the optimization problem and its solution approach. Section \ref{sec:case_studies} describes the case studies. Results for the case studies are presented in Section \ref{sec:results}. Finally, Section \ref{sec:conclusions} summarizes the concluding remarks and discusses future work.

\section{Hybrid FPML Model Architecture} \label{sec:structures}

This section discusses the overall problem statement and introduces the proposed hybrid FPML superstructure that forms the basis of ADOPT-FPML. We first mathematically define the problem statement for optimal synthesis of the hybrid FPML model architecture, followed by the generalized superstructure representation and give examples of the feasible hybrid substructures that can be generated from it. The section also clarifies how canonical hybrid structures arise as special cases of the superstructure and how more general composite structures can be constructed for multivariate and interconnected  systems.

\subsection{Problem Statement} \label{sec:problem_statement}

Consider a nonlinear dynamic process defined over a time horizon $t \in [0,T]$, with measured input variables ${\bf u}(t) \in \mathbb{R}^{n_u}$ and measured output variables ${\bf y}(t) \in \mathbb{R}^{n_y}$. The objective is to develop a  model for desired process outputs denoted as:
\begin{equation}
{\bf y}(t) \in \mathbb{R}^{n_y}
\end{equation}

\noindent Let the available FP model be represented as:

\begin{equation}
\mathcal{M}_{FP}: \mathbb{R}^{n_u} \rightarrow \mathbb{R}^{n_{FP}}
\end{equation}
where $n_{FP}$ denotes the total number of variables predicted by the FP model. These quantities may include desired measured outputs as well as additional unmeasured internal states, intermediate variables, spatially distributed profiles, or other mechanistic quantities generated by the FP model. Therefore, $n_{FP}$ is not restricted by the number of desired outputs $n_y$ and may, in general, be larger than $n_y$. Among the $n_y$ desired outputs in ${\bf y}(t)$, let $\mathcal{F}\subseteq{1,2,\dots,n_y}$ denote the subset for which corresponding predictions are available from the FP model, with cardinality $|\mathcal{F}|=F$. Thus, $F$ denotes the number of \textit{desired outputs} predicted by the FP model.

If the FP model yields sufficient accuracy for all desired outputs, no hybridization would be necessary. However, suppose that among the $F$ FP-predicted desired outputs, only $H$ variables satisfy the specified admissible tolerance $\varepsilon_{p,j}$, i.e.,
\begin{equation}
\left\| y_j^{FP}(t) - y_j^{data}(t) \right\| \leq \varepsilon_{p,j},
\quad \forall j \in \mathcal{H}^*
\end{equation}
where $\mathcal{H}^*$ denotes the set of satisfactorily predicted variables with cardinality $|\mathcal{H}^*| = H$.

Consequently, it is desired that the FP model be hybridized with data-driven model components for the remaining $F - H$ variables to satisfy the specified admissible tolerance for all desired outputs. Furthermore, if $F < n_y$, the remaining $n_y - F$ desired outputs that are not predicted by the FP model must be estimated entirely by a ML model. Let the ML model for the ML model-only case be represented as:
\begin{equation}
\mathcal{M}_{ML}: \mathbb{R}^{n_{u}} \rightarrow \mathbb{R}^{n_y-F}
\end{equation}
where the output dimension $n_y-F$ corresponds to the subset of desired outputs that are not estimated in the FP model. Therefore, the hybrid FPML architecture must be constructed such that it predicts exactly $F - H$ outputs through appropriate integration of FP and ML components, i.e.,
\begin{equation}
\mathcal{M}_{FPML}: \mathbb{R}^{n_{u}} \rightarrow \mathbb{R}^{F-H}
\end{equation}

\noindent Specifically, the fundamental challenge lies in determining: (a) which subset of FP outputs requires correction or augmentation by using the hybrid model? (b) which variables should serve as inputs to the ML component? (c) how information should flow between FP and ML modules? (d) which hybrid configuration (parallel, series, or integrated, or other hybrid such as series combined with parallel) would yield the optimal trade-off between predictive accuracy and model complexity?

As the system dimensionality increases, the number of feasible FP–ML architectures grows combinatorially. Therefore, hybrid FPML synthesis becomes a structured discrete decision-making problem coupled with continuous problem for parameter estimation, thus leading to an MINLP problem. Subsequent subsections introduce a hybrid FPML superstructure and associated substructure representations that enable systematic optimization over this combinatorial space.

\subsection{FPML Superstructure and Substructures} \label{sec:super-and-substructure}

\subsubsection{FPML Superstructure Formulation} \label{sec:superstructure}

\begin{figure}[htbp]
    \centering
    \includegraphics[width=0.95\textwidth]{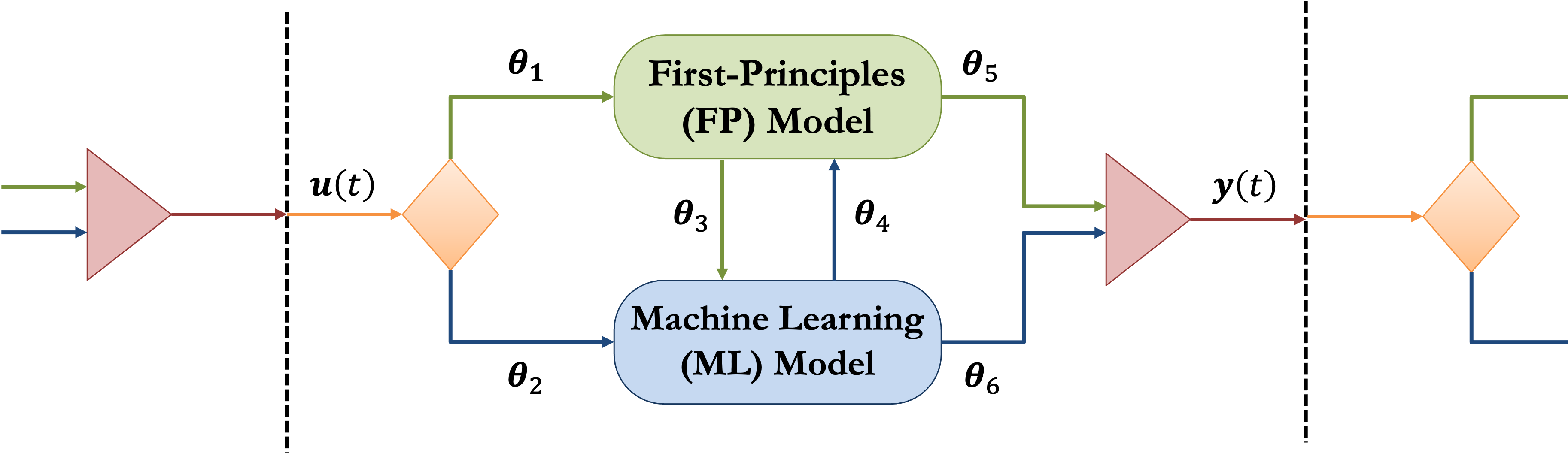}
    \caption{Representation of the hybrid FPML superstructure underlying the ADOPT-FPML framework.}
    \label{fig:superstructure}
\end{figure}

Figure \ref{fig:superstructure} illustrates the hybrid FPML superstructure used in ADOPT-FPML. Let ${\bf u}(t) \in \mathbb{R}^{n_u}$ denote the vector of measured input variables and ${\bf y}(t) \in \mathbb{R}^{n_y}$ denote the vector of measured output variables. The superstructure includes binary routing operators, denoted by $\theta_i \quad \forall i \in \{1,2,\dots,6\}$, of appropriate dimensions. Each $\theta_i$ is a binary routing matrix or vector that encodes admissible information flow to, from, and between the FP and ML modules in the hybrid FPML superstructure. Specifically, throughout this paper, $\theta_i = \mathbf{0}$ indicates that all entries of the corresponding routing operator are zero, i.e., the associated information pathway is inactive. Likewise, $\theta_i \neq \mathbf{0}$ indicates that at least one entry of the respective routing operator is nonzero, i.e., the corresponding pathway is active for at least one variable / parameter.    

The proposed superstructure, as illustrated in Figure \ref{fig:superstructure}, encodes all feasible hybrid FPML architectures through systematic combinations of the 6 binary mapping matrices $\theta_i ~\forall i \in \{1,2,\dots,6\}$. These routing operators are classified into three groups based on their role in governing information flow within the hybrid FPML superstructure:

\begin{itemize}
    \item \textit{Input routing operators} ($\theta_1$, $\theta_2$): determine which measured inputs ${\bf u}(t)$ are used in the FP and ML models, respectively. It should be noted that cardinality of $\theta_1$ and that of $\theta_2$ can be different. 
    \item \textit{Interaction routing operators} ($\theta_3$, $\theta_4$): determine the exchange of information between the FP and ML components, where $\theta_3$ routes FP model outputs to the ML model and $\theta_4$ routes ML model outputs to the FP model. Admissible interpretation of these interaction pathways is discussed in Section \ref{sec:consistency_constraints}.
    \item \textit{Output routing operators} ($\theta_5$, $\theta_6$): determine which outputs from the FP and ML model components, respectively, contribute to the final predicted output ${\bf y}(t)$.
\end{itemize}

The final predicted output ${\bf y}(t)$ is therefore computed based on $\theta_5$ and $\theta_6$. Obviously, setting any routing operator $(\theta_i)$ to zero entirely eliminates the corresponding information pathway from the superstructure.

\subsubsection{Classical / Canonical FPML Substructures} \label{sec:classicalsubstructures}

\begin{figure}[htbp]
    \centering
    \includegraphics[width=0.98\textwidth]{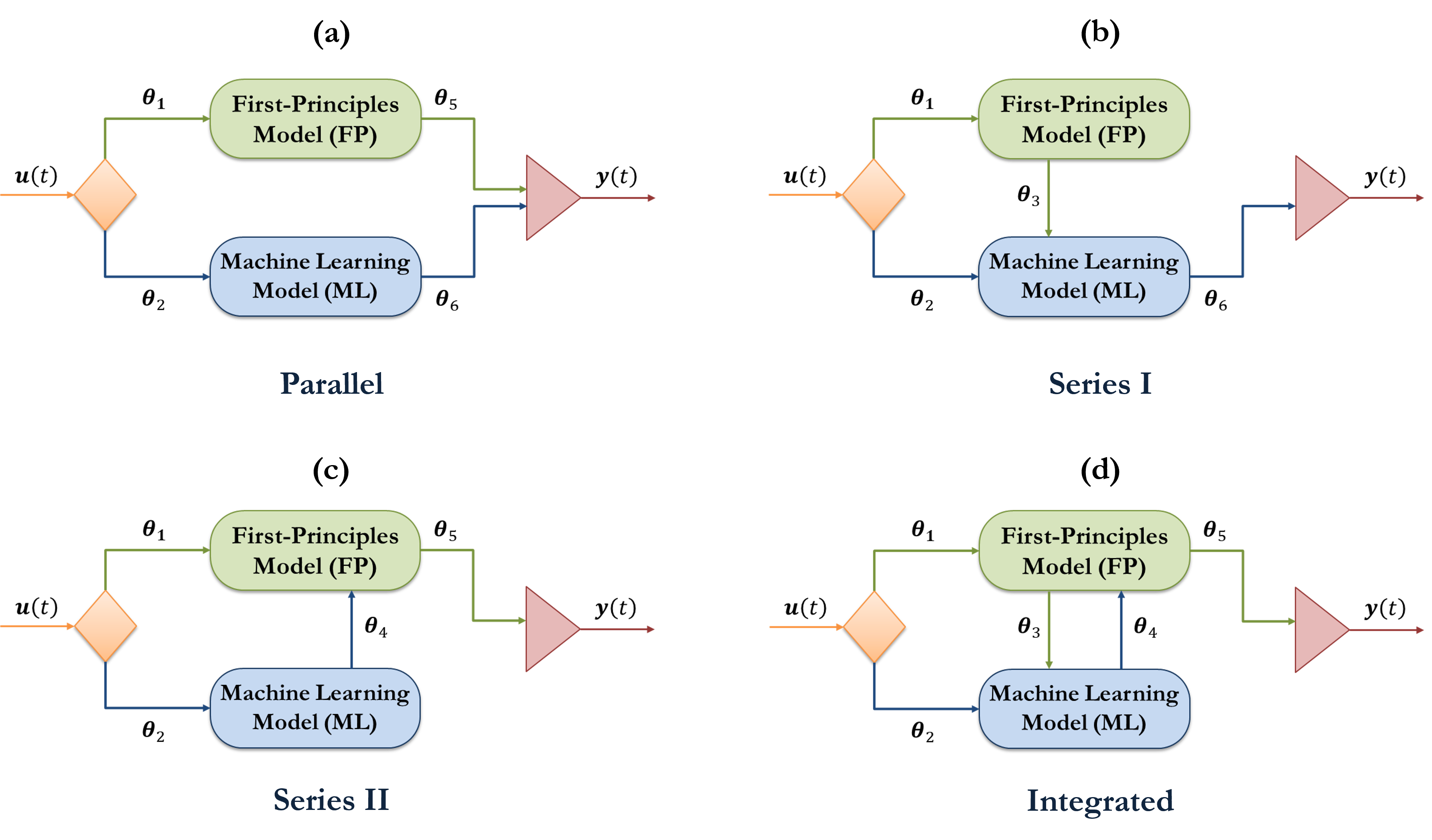}
    \caption{Classical / canonical hybrid FPML substructures recovered as special cases of the proposed superstructure: (a) parallel configuration, (b) series configuration I (FP $\rightarrow$ ML), (c) series configuration II (ML $\rightarrow$ FP), and (d) integrated configuration.}
    \label{fig:classical_substructures}
\end{figure}

The interaction routing operators $\theta_3$ and $\theta_4$ fundamentally govern the nature of coupling between the FP and ML components. By systematically activating or deactivating these routing operators, the following classical / canonical hybrid FPML configurations reported in the literature can be recovered as special cases of the proposed superstructure, as shown in Figure \ref{fig:classical_substructures}:

\begin{itemize}
    \item \textit{Parallel configuration}: $\theta_3 = \mathbf{0}$ and $\theta_4 = \mathbf{0}$. The FP and ML models operate independently with no mutual interaction. The ML model may receive process inputs directly from ${\bf u}(t)$ determined by $\theta_2$. ML model outputs are then combined with FP model predictions determined by $\theta_5$ and $\theta_6$, yielding ${\bf y}(t)$. In this configuration, the ML model typically serves as a corrective function to account for systematic biases in the FP model predictions or unmodeled phenomena in the FP model \cite{chen2020hyb, mukherjee2025hybridfpai, rajulapati2022hybreview}.
    
    \item \textit{Series I configuration (FP $\rightarrow$ ML)}: $\theta_3 \neq \mathbf{0}$ and $\theta_4 = \mathbf{0}$. Outputs of the FP model serve as additional inputs to the ML model, enabling the data-driven component to learn corrections as a function of mechanistic predictions \cite{yan2022gasifier}. In the canonical arrangement considered here, the final desired outputs are obtained from the ML model, i.e., $\theta_5 = \mathbf{0}$. A special variant arises when $\theta_2 = \mathbf{0}$, corresponding to the classical FP-to-ML series architecture in which the ML model receives inputs exclusively from the FP model \cite{mukherjee2025hybridfpai}.
    
    \item \textit{Series II configuration (ML $\rightarrow$ FP)}: $\theta_3 = \mathbf{0}$ and $\theta_4 \neq \mathbf{0}$. Outputs of the ML model are routed to the FP model, allowing data-driven refinement of internal mechanistic quantities such as FP model parameters or terms \cite{psichogios1992hybfpai, bangi2020hybhydraulic, shah2022dnnhybrid}. In the canonical arrangement considered here, the final desired outputs are obtained from the FP model, therefore, $\theta_6 = \mathbf{0}$.
    
    \item \textit{Integrated configuration}: $\theta_3 \neq \mathbf{0}$ and $\theta_4 \neq \mathbf{0}$. Bidirectional information exchange is allowed between the FP and ML components, allowing fully coupled refinement of states, parameters, or other mechanistic quantities \cite{saini2026oxide, mukherjee2025hybridfpai}.
\end{itemize}

\noindent While previous studies have considered one or more of these predefined canonical configurations for specific modeling tasks, the proposed superstructure framework generalizes model development by treating these canonical forms as subsets of a broader structural search space. The mathematical formulation of the admissible interaction patterns in terms of $\theta_3$ and $\theta_4$, along with the associated output-routing assignments for $\theta_5$ and $\theta_6$, is developed in Section \ref{sec:consistency_constraints}.

\subsubsection{Composite FPML Substructures} \label{sec:compositesubstructure}

A key distinguishing feature of the proposed superstructure formulation is the consideration of binary mapping matrices $\theta_5$ and $\theta_6$ for individual output variables. These matrices enable independent representation of each output variable by the FP model only, the ML model only, or the hybrid FPML model.

In a general interconnected process system, each subsystem envelope may be associated with its own FP model, defined, for example, by process connectivity, instrumentation, or the scope of the available mechanistic model. Within a given envelope, multiple superstructure blocks may be employed sequentially to represent different subsets of target outputs, with the same FP model used across these blocks. Additionally, different subsystems need not share the same hybrid configuration, thereby allowing the overall process model to be assembled from multiple optimal FPML building blocks. Accordingly, the proposed superstructure should be interpreted as a  modular building block for hybridization between FP and ML components, rather than as a monolithic representation of the entire process system. In such configurations, the intermediate output variables generated by one modular building block may serve as additional candidate inputs for the next one, if needed.

In this context, a large-scale interconnected process \cite{mukherjee2025metcnn} may be partitioned into multiple subsystem envelopes, each associated with its own FP model, local measurements, and prediction objectives. While the proposed superstructure could, in principle, be applied to the interconnected process as a whole, this would result in a combinatorial search space that might scale poorly. Moreover, different subsystem  may exhibit distinct sources of modeling error and input-output structures. By decomposing the interconnected process into different subsystems and applying the superstructure independently to each subsystem, optimal hybrid FPML model architecture for each system can be obtained. 

However, the selection of subsystems is not trivial in the context of interconnected process systems. The present work focuses on the automatic discovery of the optimal hybrid FPML model architecture for a given system or a subsystem, including the possibility of sequential optimization of multiple superstructure blocks through the proposed branch-and-prune algorithm. For interconnected systems, the current analysis therefore proceeds sequentially based on the availability of measurements and the existence of an appropriate FP model for each envelope, which makes the resulting hybrid FPML model architectures  relatively easier to train, interpret, and deploy in a modular manner. The automatic partition or decomposition of a fully interconnected process into optimal subsystem envelopes, followed by simultaneous discovery of the corresponding hybrid FPML configurations, would require further investigation.

\begin{figure}[htbp]
    \centering
    \includegraphics[width=0.98\textwidth]{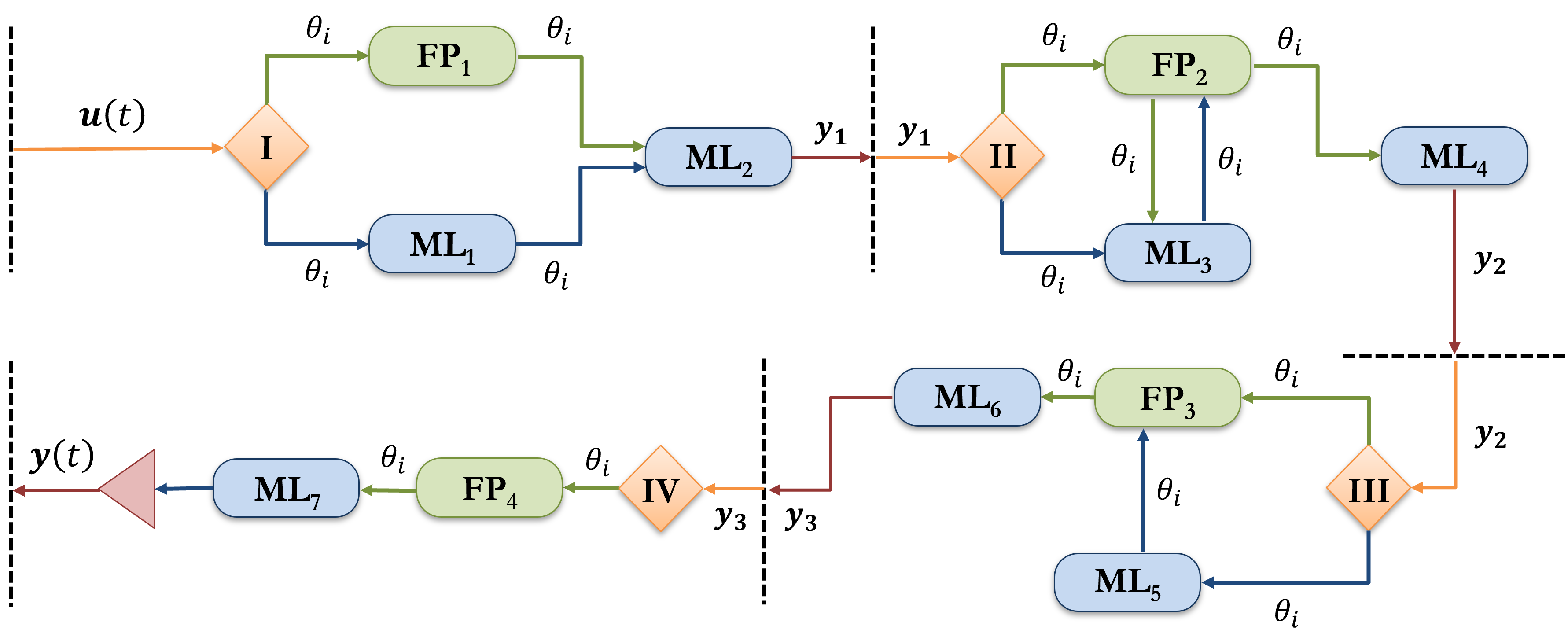}
    \caption{Representative composite hybrid FPML structure formed by combining repeated superstructure blocks.}
    \label{fig:composite_structures}
\end{figure}

Figure \ref{fig:composite_structures} illustrates a representative composite hybrid FPML structure obtained by combining repeated superstructure blocks across different sections of an interconnected process. The overall composite architecture thus generated does not necessarily reduce to any single canonical parallel, series, or integrated configuration. Instead, the composite hybrid structure naturally evolves from multiple local FPML architectures, with each superstructure block represented by its own hybrid configuration. 

The hybrid configurations introduced in this section define the discrete structural design space over which the optimal FPML architecture is to be identified. The next section develops the MINLP-based superstructure optimization problem and the associated branch-and-prune solution strategy.

\section{Optimization Approach} \label{sec:optimization}

For clarity, the optimization framework is first developed for a \textit{single} process system or subsystem envelope, represented by one FP model and its associated prediction objectives. Within a given envelope, the optimal hybrid FPML architecture may require a sequence of superstructure blocks, indexed by $b$, with each block corresponding to a stage of the proposed branch-and-prune algorithm. The FP model remains the same across these blocks, while the target output set, candidate ML input library, and selected FPML configuration may differ from one block to the next. At each block, the corresponding optimization problem determines the appropriate FP-ML interaction structure and the subsets of target outputs and ML inputs. The same optimization framework can then be applied independently to additional subsystem envelopes, each associated with its own FP model, when modeling an interconnected process system, as discussed in Section \ref{sec:compositesubstructure}.  

The objective of ADOPT-FPML is to automatically identify the optimal hybrid FPML structure given the data  by simultaneously selecting: (i) the subset of outputs for which hybridization or representation by purely data-driven model would be desired, (ii) the subset of candidate inputs for the ML model \cite{adeyemo2024bidsam, mukherjee2024thesis}, and (iii) optimal selection of the routing variables that specify the integration between the FP and ML model components \cite{mukherjee2025hybridfpai}. Since these decisions include discrete variables representing the routing variables that lead to variations in the configuration of the FPML model and since  the model (ML) parameters that are continuous need to be optimally estimated given the routing variables and the data, it leads to a disjunctive MINLP problem.

Let $\mathcal{D}=\{ {\bf u}(t_j), {\bf y}^{data}(t_j)\}_{j=1}^{n_s}$ denote the available transient data consisting of $n_s$ samples of inputs-outputs. For a given feasible configuration of the FPML superstructure, the corresponding optimization problem is solved subject to the FPML model equations, routing constraints, and the associated parameter-estimation (NLP) subproblem. The information-theoretic objective function that is used \cite{ferguson2019aicbic, mukherjee2023hybserpar} is given by:

\begin{equation}
\min_{{\boldsymbol \psi}} \;\; \mathrm{AIC}_c = -2 \ln \!\left(\mathcal{L}(\hat{\boldsymbol{\psi}})\right)
+ 2k + \frac{2k(k+1)}{n_s-k-1}
\label{eq:overall_minlp_generic}
\end{equation}
where $\mathcal{L}(\hat{\boldsymbol{\psi}})$ denotes the likelihood evaluated at the estimated model parameters $\boldsymbol{\psi}$ and $k$ is the total number of adjustable parameters in the resulting hybrid FPML model.

It is worth noting that the parameter count $k$ depends implicitly on the routing decisions because the selected structure determines: (i) whether FP outputs are directly used, corrected, or bypassed, (ii) the number of ML outputs to be predicted, and (iii) the number of candidate inputs to the ML model. Hence, structural complexity and predictive accuracy are considered simultaneously through the AICc objective function. To extend this approach beyond a single FPML superstructure block, let $\mathcal{J}^{(1)}$ denote the initial set of outputs that could not be satisfactorily represented by the FP model alone and let $\mathcal{U}^{(1)}$ denote the set of corresponding initial candidate inputs. For the first superstructure block, $\mathcal{U}^{(1)}$ may be taken as the original measured input set or an augmented feature library constructed from the raw measured inputs. If additional superstructure blocks are used, then, for each subsequent block index $b=1,2,\ldots$, the reduced superstructure optimization problem is solved over the current pair $(\mathcal{J}^{(b)},\mathcal{U}^{(b)})$. Let $\mathcal{H}^{(b)} \subseteq \mathcal{J}^{(b)}$ denote the subset of outputs satisfactorily represented by the optimal hybridization strategy in block $b$, according to the admissible threshold criterion. The set of outputs that are still not satisfactorily represented is then updated as:
\begin{equation}
\mathcal{J}^{(b+1)} = \mathcal{J}^{(b)} \setminus \mathcal{H}^{(b)}
\label{eq:J_update}
\end{equation}
Additionally, the candidate input library is updated as:
\begin{equation}
\mathcal{U}^{(b+1)} = \mathcal{U}^{(b)} \cup \mathcal{S}^{(b)}
\label{eq:U_update}
\end{equation}
where $\mathcal{S}^{(b)}$ denotes the screened set of synthetic variables generated by block $b$ that can be utilized as additional inputs to the subsequent blocks. Thus, the composite FPML architecture for the system or envelope under consideration is synthesized as a sequence of superstructure blocks, each selected optimally for the outputs that are still not satisfactorily represented by the preceding block(s).

In principle, one could solve Eq. \eqref{eq:overall_minlp_generic} by exhaustive enumeration of all feasible routing configurations, output subsets, and input subsets. However, the number of candidate structures grows combinatorially with the number of outputs, candidate inputs, and allowable FP-ML interactions. Such comprehensive enumeration quickly becomes computationally prohibitive because each candidate structure requires one or more nonlinear parameter estimation problems to be solved. Therefore, the proposed framework decomposes the overall MINLP into a sequence of structured subproblems and applies a {\textit{branch-and-prune}} strategy to eliminate inferior structural decisions early, thereby significantly reducing the number of model evaluations required to identify the optimal hybrid FPML model.

\subsection{Structural Consistency and Admissible Interaction Space} \label{sec:consistency_constraints}

For a given superstructure block, the routing operators must satisfy logical consistency conditions so that the resulting hybrid FPML architecture remains physically and structurally meaningful. Let $i$ index the candidate input routing associated with $\theta_1$ and $\theta_2$, with their respective index sets not necessarily having the same cardinality. Similarly, let $\ell$ index the candidate information routing from the FP model to the ML model associated with $\theta_3$, and let $p$ index the candidate information routing from the ML model to the FP model associated with $\theta_4$. Finally, let $j\in\{1,2,\dots,n_y\}$ index the desired output variables, with $\theta_{5,j}$ and $\theta_{6,j}$ denoting the corresponding binary output-routing decisions associated with the FP and ML components, respectively. Each desired output must have at least one source, yielding the following constraint:

\begin{equation}
1 \leq \theta_{5,j} + \theta_{6,j} \leq 2, \qquad \forall j \in \{1,2,\dots,n_y\}
\label{eq:output_consistency}
\end{equation}

\noindent Eq. \eqref{eq:output_consistency} allows three possibilities for representing each desired output: FP-only, ML-only, or hybrid FPML.

In addition to the output-routing constraint in Eq. \eqref{eq:output_consistency}, the generic FPML superstructure must satisfy structural consistency conditions governing the activation of the FP and ML model components. Specifically, an active model component must have at least one active (i.e., non-zero) input routing and at least one active output routing, while a component with no active input routing must have no active output routing. For the active ML model, this requirement can be expressed as:

\begin{equation}
\left(\sum_i \theta_{2,i}+\sum_{\ell}\theta_{3,\ell}\geq 1\right)
\Longleftrightarrow
\left(\sum_p \theta_{4,p}+\sum_j \theta_{6,j}\geq 1\right)
\label{eq:ml_activity_consistency}
\end{equation}

\noindent Similarly, the corresponding condition for the FP model (considered to be always active in a given superstructure block) is:

\begin{equation}
\left(\sum_i \theta_{1,i}+\sum_p \theta_{4,p}\geq 1\right)
\Longleftrightarrow
\left(\sum_{\ell}\theta_{3,\ell}+\sum_j \theta_{5,j}\geq 1\right)
\label{eq:fp_activity_consistency}
\end{equation}
\noindent Eq. \eqref{eq:ml_activity_consistency} and \eqref{eq:fp_activity_consistency} express the generic input-output activity requirements for the ML and FP components, respectively, without prescribing a particular FPML interaction configuration. The structural consistency conditions define the feasible interaction set for a single superstructure block.

As established in the problem statement in Section \ref{sec:problem_statement}, the $n_y$ desired outputs can be partitioned into three categories: (i) $H$ outputs that are satisfactorily predicted by the standalone FP model, (ii) $F-H$ outputs for which FP predictions are available but require hybridization, and (iii) $n_y-F$ outputs that are not predicted by the FP model and must therefore be represented by an ML model. Consequently, outputs in the first category can be retained directly from the FP model, while those in the third category can be represented through purely ML-based prediction. The superstructure optimization for hybridization can thus focus on the remaining $F-H$ outputs, reducing the number of relevant routing combinations.

Within this reduced structural search space, the interaction routing operators $\theta_3$ and $\theta_4$ govern the exchange of information between the FP and ML components. Since these routing operators are binary, each interaction pathway can either be inactive with all corresponding entries equal to zero, or active with at least one nonzero entry. Consequently, the interaction space can be partitioned into the following 4 \textit{mutually exclusive} disjunctive configurations for each superstructure block:

\begin{align}
\text{Config 1:}\quad
& \sum_{\ell}\theta_{3,\ell}=0,\qquad
\sum_p\theta_{4,p}=0
\label{eq:config1}
\\
\text{Config 2:}\quad
& \sum_{\ell}\theta_{3,\ell}\geq 1,\qquad
\sum_p\theta_{4,p}=0
\label{eq:config2}
\\
\text{Config 3:}\quad
& \sum_{\ell}\theta_{3,\ell}=0,\qquad
\sum_p\theta_{4,p}\geq 1
\label{eq:config3}
\\
\text{Config 4:}\quad
& \sum_{\ell}\theta_{3,\ell}\geq 1,\qquad
\sum_p\theta_{4,p}\geq 1
\label{eq:config4}
\end{align}

\noindent These 4 configurations collectively represent all possible activation combinations of the two interaction pathways. In particular, Config 4 requires at least one active information pathway from the FP model to the ML model through $\theta_3$ and at least one active pathway from the ML model to the FP model through $\theta_4$. Consequently, Config 4 remains distinct from Config 3, for which $\theta_3=\mathbf{0}$. These disjunctive configurations recover the canonical structures introduced in Section \ref{sec:classicalsubstructures}. Specifically, Config 1 corresponds to the parallel interaction pattern, Config 2 to Series I $(\mathrm{FP}\rightarrow\mathrm{ML})$, Config 3 to Series II $(\mathrm{ML}\rightarrow\mathrm{FP})$, and Config 4 to the integrated interaction pattern. Thus, the four-way disjunction provides a reduced representation of the generic FPML superstructure by classifying the admissible interaction patterns according to the activation states of $\theta_3$ and $\theta_4$.

Moreover, within each disjunctive configuration, the corresponding input-routing decisions depend on the input-selection algorithm (see Section \ref{sec:input_selection}). Accordingly, the inputs supplied to the ML model are determined by the activated entries of $\theta_2$ and $\theta_3$. Following the problem definition in Section \ref{sec:problem_statement} and the corresponding interaction mechanisms associated with Configs 1-4, the output-routing assignments $\theta_5$ and $\theta_6$ are uniquely determined by the selected configuration. Specifically, for each target output $j$ assigned to a given superstructure block:

\begin{equation}
(\theta_{5,j},\theta_{6,j}) =
\begin{cases}
(1,1), & \text{Config 1}\\
(0,1), & \text{Config 2}\\
(1,0), & \text{Configs 3 and 4}
\end{cases}
\label{eq:config_output_routing}
\end{equation}

\noindent In Config 1, the final output combines FP and ML contributions through the parallel correction mechanism. In Config 2, the ML model generates the final output using information received from the FP model. In Configs 3 and 4, the ML model supplies information to the FP model, which subsequently generates the final output. Accordingly, the output-routing assignments in Eq. \eqref{eq:config_output_routing} follow directly from the problem definition and the corresponding configuration logic. These assignments satisfy Eq. \eqref{eq:output_consistency} and do not constitute additional independent disjunctive branching decisions.

Although the superstructure formulation is general, the admissible interpretation of the interaction routing operators is expected to be restricted for computational tractability and structural clarity, particularly for large problems. Specifically, in the present work, $\theta_3$ is restricted to routing intermediate variables computed by the FP model  \cite{mukherjee2025hybridfpai} to represent additional inputs for the data-driven model. Other forms of information transfer are already represented through the remaining routing operators and therefore need not be embedded in $\theta_3$. Likewise, $\theta_4$ is restricted to routing only parameters or correction to the parameters used in the FP model. In principle, $\theta_4$ could be defined more broadly to include corrections to states, internal or output variables, or other internal quantities. However, such corrections may be represented elsewhere in the proposed substructure framework. For example, additive output corrections are captured through the parallel configuration, while state variable corrections by using the ML model are represented through the series II configuration. Consequently, to avoid redundancy in the admissible interaction space and for computational tractability, $\theta_4$ is considered here to correspond strictly to ML-assisted correction of FP model parameters. It should be noted, however, that this restriction is specific to the implementations considered in the present work and may be relaxed in future extensions of the framework.

Even under this restriction, the candidate set of parameter corrections associated with $\theta_4$ may still be large. While the FPML model including all routing decision variables and parameters of the FP and ML models can be optimized together, such optimization problems can be computationally intractable even for moderate sized problems, particularly due to large computational expense of the FP model (assuming a FP model of at least reasonable fidelity  is being used), potential ill-conditioning due to large admissible parameter space and potential poor-conditioning of the FP model, and potential use of different optimization algorithms and/or software for the FP model versus the ML model. One way of addressing these issues can be to decompose the parameter estimation problem (the NLP problem) separately for the FP model and ML model. In that case, the training data for $\theta_4$ must be generated for training the ML model. One option is to solve an auxiliary optimization or estimation problem by using the FP model for each hypothesized parameter correction that would lead to minimization of an error function with respect to the measured data. These estimated corrections can then be used as training targets. Such an approach has been proposed by some of the authors of this paper earlier \cite{mukherjee2025hybridfpai}. A naive treatment of all such possibilities would therefore lead to excessive computational expense and poor scalability. To mitigate this, the candidate parameter set associated with $\theta_4$ may be screened prior to superstructure optimization using problem-specific relevance measures, such as global sensitivity analysis \cite{saltelli2019sensitivity, iooss2015globalsensitivity, das2026vibes}, local sensitivity metrics \cite{chiron2023localsens, tarantola2024sensitivityreview}, or variance-based screening methods \cite{zhang2014variancesensitivity}. In this way, only the most influential FP parameters, i.e., those contributing most significantly to output variation or model mismatch, are retained as admissible candidates for ML correction. 

Furthermore, when multiple superstructure blocks are used, the admissible interaction space must also account for information propagated forward from previously optimized FPML blocks. Let $\mathcal{C}^{(b)}$ denote the raw set of candidate synthetic variables generated by the superstructure block $b$, comprising selected resolved outputs and/or admissible intermediate variables. However, these variables are not appended automatically to the input library of the subsequent block. Instead, a screened subset $\mathcal{S}^{(b)} \subseteq \mathcal{C}^{(b)}$ is retained according to admissibility, informativeness (or richness), and redundancy criteria, to enhance scalability of the overall approach.

Specifically, let $a_s^{(b)} \in \{0,1\}$ denote a binary admissibility indicator for candidate synthetic variable $s \in \mathcal{C}^{(b)}$, where $a_s^{(b)}=1$ enables the variable to be structurally admissible for propagation to the subsequent superstructure block, only if it is not already present in the current input library, and is predicted with acceptable reliability by the retained block. For each admissible candidate, let $r_s^{(b)}$ denote a relevance score with respect to the current set of inadequately modeled outputs $\mathcal{J}^{(b+1)}$, and let $d_s^{(b)}$ denote a redundancy measure with respect to the current input library $\mathcal{U}^{(b)}$. Then the screened propagated input set may be defined as:
\begin{equation}
\mathcal{S}^{(b)}=\left\{s \in \mathcal{C}^{(b)}\;\middle|\;a_s^{(b)}=1,\;
r_s^{(b)} \geq \tau_r,\;
d_s^{(b)} \leq \delta_r
\right\}
\label{eq:S_screened}
\end{equation}
where $\tau_r$ and $\delta_r$ denote prescribed relevance and redundancy thresholds, respectively. It should be noted that the choice of these screening metrics, such as admissibility, informativeness, and redundancy, is neither unique nor exhaustive; additional and/or alternative metrics may be incorporated as appropriate to define the screening algorithm for a given system. To prevent unbounded growth of the candidate input library across successive blocks, the retained synthetic set may further be restricted by a cardinality cap:
\begin{equation}
|\mathcal{S}^{(b)}| \leq N_I
\label{eq:NI_cap}
\end{equation}
where $N_I$ denotes the maximum number of propagated synthetic inputs admitted at any block. The next subsection builds on this feasible set and shows how the resulting MINLP is decomposed into structure-selection, output-selection, and input-selection subproblems that can be solved more efficiently using the proposed branch-and-prune strategy.

\subsection{Decomposition of the MINLP Problem} \label{sec:minlp_decomposition}

The overall MINLP can be decomposed into three coupled classes of discrete decisions:

\begin{itemize}
    \item \textit{Structure selection}: selection of the feasible interaction parameters between the FP and ML components, as represented by the admissible routing operators governing canonical or more general hybrid configurations.
    \item \textit{Output selection}: identification of the outputs or states that require direct FP prediction, hybrid correction, or purely data-driven estimation.
    \item \textit{Input selection}: identification of the subset of candidate inputs supplied to the ML model for the selected structure and output set.
\end{itemize}

This decomposition is motivated by the hierarchical nature of the discrete decisions. The structure-selection problem determines the feasible set of information pathways within the superstructure, the output-selection problem determines which quantities must be corrected or predicted by the selected hybrid configuration, and the input-selection problem determines the dimension and complexity of the ML component for the resulting FPML model. Accordingly, the proposed branch-and-prune strategy proceeds from coarse structural decisions to finer variable-selection decisions, thereby reducing the number of expensive model evaluations relative to comprehensive enumeration. Additionally, while using the FPML superstructure as building block across multiple stages, these three classes of decisions are re-solved stage-wise over $\mathcal{J}^{(b)}$ and the current candidate input library $\mathcal{U}^{(b)}$ within the same system or envelope. Thus, the branch-and-prune workflow is not based on a fixed global ranking of structures inherited from the initial full-output problem, but instead, on successive reduced optimization problems whose feasible set evolves as outputs are resolved and synthetic variables (i.e., resolved outputs or intermediate variables propagated from previously optimized stages) are propagated forward.

\subsubsection{Structure Selection Problem} \label{sec:structure_selection}

The first discrete subproblem identifies the most appropriate FPML structure over the feasible candidate set defined by the admissible routing operators introduced in Sections \ref{sec:superstructure} and \ref{sec:consistency_constraints}. For superstructure block $b$, this problem is posed over the current set of outputs requiring further hybridization $\mathcal{J}^{(b)}$ and the current candidate input library $\mathcal{U}^{(b)}$. The objective is, therefore, to determine how the FP and ML components should interact within the current superstructure block so that the resulting hybrid model achieves the best trade-off between predictive accuracy and model complexity for the outputs that remain unsatisfactorily modeled in that block.

Let $\mathcal{R}=\{R_1,R_2,\ldots,R_{N_R}\}$ denote the feasible disjunctive configuration set, where $N_R$ denotes the total number of admissible configurations and each $R_r$ corresponds to an admissible assignment of the interaction routing operators $\theta_3$ and $\theta_4$. For the implementation considered in this work, the primary branches are Configs 1-4 defined in Eqs. \eqref{eq:config1}-\eqref{eq:config4}. Within each configuration, additional structural decisions arise from the specific nonzero entries of the corresponding routing operators and from the subsequent input- and output-selection decisions. More general interaction patterns may be incorporated without altering the overall branch-and-prune formulation, provided that they satisfy the structural consistency conditions in Section \ref{sec:consistency_constraints}.

For each candidate configuration $R_r \in \mathcal{R}$, the corresponding hybrid FPML model is trained and evaluated through the information-theoretic objective function defined in Eq. \eqref{eq:overall_minlp_generic}, subject to the governing equations of the FPML superstructure block under the routing combination associated with $R_r$. The resulting local $\mathrm{AIC}_c$ value is then used to rank the candidate structures. This within-block selection criterion is distinct from the global $\mathrm{AIC}_c$ criterion in Eq. \eqref{eq:aic_termination}, which is used only to determine whether the newly selected block should be retained as part of the overall composite FPML architecture. Since the parameter count $k$ appearing in $\mathrm{AIC}_c$ depends implicitly on the selected configuration, Eq. \eqref{eq:overall_minlp_generic} simultaneously accounts for both predictive accuracy and structural complexity. To be specific, for block $b+1$, the candidate structure is not selected simply as the next-ranked structure from the original full-output problem. Instead, a new reduced structure-selection problem is solved over $\left(\mathcal{J}^{(b+1)}, \mathcal{U}^{(b+1)}\right)$, since the most suitable structure for the remaining outputs may change after some outputs have already been assigned to earlier superstructure blocks within the same system.

In contrast, a direct comprehensive enumeration would require evaluating every feasible configuration in $\mathcal{R}$ and solving the corresponding training problem for each case. However, such an approach quickly becomes computationally expensive, particularly when the structure-selection problem is coupled with the subsequent output-selection and input-selection decisions. Therefore, the proposed framework employs a branch-and-prune strategy, in which candidate structures are explored systematically and inferior branches are discarded early whenever they are unlikely to improve the current best value of the objective function.

The 4 disjunctive branches are mutually exclusive with respect to the activation states of $\theta_3$ and $\theta_4$ and collectively span the admissible interaction patterns considered in this work. Model evaluation and subsequent input- and output-selection can therefore be performed independently within each branch and, where computational resources permit, the corresponding branch evaluations may be parallelized. Within each branch, the candidate configuration may be refined further depending on the selected intermediate FP variables associated with $\theta_3$ and the screened parameter-correction candidates associated with $\theta_4$. As discussed previously, the candidate set for $\theta_4$ is not taken to be arbitrarily large; rather, it is pre-screened using physics-based or systems-engineering criteria such as sensitivity analysis so that only the most influential correction targets are retained. This restriction is essential for scalability, since each candidate parameter correction may require an internal optimization step to generate training targets for the ML model. For configurations involving ML-assisted correction of FP model parameters (Configs 3 and 4), additional computational savings is achieved by reusing previously estimated parameter trajectories. Specifically, for a given superstructure block, the parameter trajectories associated with a candidate correction target can be generated once and subsequently stored for reuse during the evaluation of candidate ML models requiring the same targets. This strategy can reduce repeated solutions of the corresponding parameter estimation subproblems.

Accordingly, the structure-selection problem may be interpreted as a hierarchical search over a feasible set of interaction patterns, beginning with the disjunctive Config 1-4 interaction logic and then refining the candidate structure only where justified by the objective function. In this way, the branch-and-prune strategy avoids unnecessary model evaluations and provides a systematic alternative to heuristic trial-and-error selection of the hybrid FPML configuration. Once the interaction structure has been selected, the remaining discrete decisions are narrowed to the corresponding output-selection and input-selection subproblems. Thus, the stage-wise repeated-block formulation preserves adaptivity across successive superstructure blocks and avoids reliance on a fixed initial ranking of canonical structures. Repeated occurrence of the same disjunctive configuration across different superstructure blocks is not excluded \textit{a priori}, but is retained only when that configuration remains optimal for the reduced problem defined by the current set of inadequately modeled outputs and propagated candidate inputs.

\subsubsection{Output Selection Problem} \label{sec:output_selection}

The second discrete subproblem determines which outputs should be assigned to direct FP prediction, hybrid correction, or purely ML-based prediction within the selected superstructure block. Let ${\bf y}(t)\in\mathbb{R}^{n_y}$ denote the full set of candidate outputs and let $\mathcal{J}=\{1,2,\dots,n_y\}$ denote the corresponding output index set. The first step is to identify those outputs for which the standalone FP predictions are already sufficiently accurate according to a prescribed admissible error tolerance.

Let $\varepsilon_{p,j}$ denote the admissible prediction threshold for output $j$. Then the set of adequately predicted outputs is defined as
\begin{equation}
\mathcal{H}=\left\{j\in\mathcal{F}\;\middle|\;
\left\|y_j^{FP}(t)-y_j^{data}(t)\right\|\leq \varepsilon_{p,j}\right\}
\label{eq:hset}
\end{equation}
where the norm may represent an appropriate error metric over the training horizon, such as a root-mean-square error (RMSE) \cite{mukherjee2025nnvsbml}. Outputs belonging to $\mathcal{H}$ are considered to be satisfactorily predicted by the FP model and therefore do not require hybridization within the current superstructure block. The complementary set: 
\begin{equation}
\mathcal{J}_{rem}=\mathcal{J}\setminus\mathcal{H}
\label{eq:remaining_outputs}
\end{equation}
contains the outputs that require either hybrid correction or purely ML-based prediction.

Accordingly, the output-selection problem may be represented through binary indicators $\delta_j\in\{0,1\}$ such that
\begin{equation}
\delta_j=
\begin{cases}
0, & j\in\mathcal{H} \\
1, & j\in\mathcal{J}_{rem}
\end{cases}
\label{eq:delta_output}
\end{equation}
where $\delta_j=1$ indicates that output $j$ participates in the hybridization problem for the current branch. These indicators directly influence the effective output-routing decisions and determine the output dimension of the corresponding ML component.

For a selected structure, the output-selection step is implemented sequentially within the branch-and-prune framework. In particular, once a candidate structure has been identified for the current set of outputs $\mathcal{J}^{(b)}$ that is insufficiently represented by the model, only those outputs that are satisfactorily represented by the optimal hybrid FPML configuration are assigned to the current superstructure block. The corresponding resolved subset is denoted by $\mathcal{H}^{(b)}$, and the remaining outputs are propagated forward through the update in Eq. \eqref{eq:J_update}. In this way, the proposed framework ensures that the overall hybridization scheme remains parsimonious within the same system or envelope of an interconnected process, with each ML component addressing only the residual uncertainty associated with its assigned outputs rather than forming a single hybrid FPML model to address the full output set simultaneously.

In the present work, this output-selection step is implemented by screening FP model outputs through an accuracy threshold, which provides a simple and physically interpretable mechanism for separating satisfactorily predicted outputs from those requiring hybridization. However, more general output-selection strategies are also possible. For example, when outputs are strongly correlated or exhibit significant redundancy, clustering, classification, or other feature-grouping techniques may be used to identify representative outputs or output groups prior to hybridization. Such extensions can further improve scalability in multivariate problems by reducing redundant model evaluations while preserving the essential  information required for hybrid FPML synthesis.

\subsubsection{Input Selection Problem} \label{sec:input_selection}

The third discrete subproblem determines the subset of candidate inputs supplied to the ML model for the selected structure and output set. In the repeated-block setting, this problem is posed over the current candidate input library $\mathcal{U}^{(b)}$, which may include both the original measured inputs and a screened set of synthetic variables propagated from previously accepted blocks. This step is particularly important because the number of selected inputs directly affects the dimension of the ML input layer and, consequently, the total number of trainable parameters in the resulting hybrid FPML model. Since the objective function, as defined in Eq. \eqref{eq:overall_minlp_generic}, penalizes unnecessary model complexity, the input-selection problem must identify a subset of variables that is sufficiently informative for prediction while avoiding excessive redundancy.

Let $\mathcal{U}^{(b)}$ denote the full candidate input library available for block $b$, and let $\mathcal{K}^{(b)} \subseteq \mathcal{U}^{(b)}$ denote the selected input set for the current branch, with cardinality $|\mathcal{K}|=n_{\mathrm{in}}$, where $n_{\mathrm{in}}$ denotes the number of selected ML inputs. The resulting ML model is then constructed using the reduced input vector associated with $\mathcal{K}^{(b)}$. In principle, a comprehensive enumeration of all non-empty candidate subsets would require
\begin{equation}
\sum_{m=1}^{|\mathcal{U}^{(b)}|} \binom{|\mathcal{U}^{(b)}|}{m} = 2^{|\mathcal{U}^{(b)}|}-1
\label{eq:input_enum_count}
\end{equation}
possible input combinations for each candidate structure and output subset. Such a search rapidly becomes computationally intractable, since each candidate subset requires retraining and re-evaluation of the corresponding hybrid FPML model.

To avoid this combinatorial growth, the present framework employs an approach in which candidate inputs are screened and ranked before sequential inclusion into the ML model. In the simplest implementation, this ranking may be based on a relevance metric computed from the training data, such as the average absolute correlation between each candidate input and the outputs retained for hybridization. Let $\mathcal{J}^{(b)}$ denote the current set of outputs at stage $b$ that requires further corrections. Then a scalar screening score for candidate input $u_i$ may be defined as:
\begin{equation}
\bar{\rho}_i^{(b)} =\frac{1}{|\mathcal{J}^{(b)}|}\sum_{j\in\mathcal{J}^{(b)}}
|\rho_{ij}|
\label{eq:avg_corr_stagewise}
\end{equation}
where $\rho_{ij}$ denotes the correlation between candidate input $u_i$ and output $y_j$ over the training data. The candidate inputs are then ranked in descending order of $\bar{\rho}_i^{(b)}$ and incorporated sequentially into the ML model.

At each input-addition step within a branch, the hybrid FPML model is retrained using the currently selected input subset and evaluated through the objective function in Eq. \eqref{eq:overall_minlp_generic}. If inclusion of a new input does not improve the corresponding $\mathrm{AIC}_c$ value, that branch may be pruned and no further refinement is required along that direction. Under the greedy input-addition rule adopted here, a branch that fails to improve the incumbent objective at the current expansion step is therefore assumed to be insufficiently promising to justify further refinement, and its descendants are therefore pruned. In this way, the proposed approach replaces exhaustive subset enumeration with a structured search that reduces the number of model evaluations while retaining the most informative input variables for the hybrid model. Although correlation-based screening is adopted here as an efficient and physically interpretable criterion, the proposed framework is not restricted to raw measured inputs or linear relevance measures. More generally, the candidate input set may be expanded to include transformed or constructed basis functions (e.g., polynomial, bilinear, logarithmic, etc. \cite{adeyemo2024bidsam}). This is particularly useful when the dominant process relationships are not captured adequately by the original measured inputs alone. Such feature-generation strategies are conceptually aligned with systematic candidate-library construction approaches used in data-driven model discovery reported in the literature \cite{wilson2017alamo, brunton2016sindy}. 

When the candidate input library is augmented by propagated synthetic variables from previous block(s), an additional screening layer may be used to control input-library growth. In particular, if the number of admissible propagated synthetic variables exceeds the admissible cap $N_I$ in Eq. \eqref{eq:NI_cap}, the candidate synthetic set may be compressed using clustering over relevance profiles with respect to the inadequately modeled outputs. A representative variable is then selected from each cluster based on maximum relevance and minimum redundancy relative to the current input library. Thus, multiple representative variables may be retained across clusters, subject to the cardinality cap $N_I$. In this way, the dimensionality of the propagated synthetic input set is bounded while preserving diversity in the information supplied to downstream blocks.

The same branch-and-prune approach can, therefore, be extended to a richer candidate feature library, given the use of an appropriate screening metric to control combinatorial growth. Some of the most commonly used screening metrics include mutual information \cite{vergara2014mutualinfo}, sensitivity indices \cite{saltelli2019sensitivity, iooss2015globalsensitivity}, and variance-based relevance measures \cite{guyon2003variableselection, zhang2014variancesensitivity}, all of which quantify the influence of candidate inputs on the selected outputs. Irrespective of the specific screening metric employed, however, the primary role of the input-selection problem remains unchanged, i.e., to identify a compact and informative set of candidate variables, including screened propagated synthetic inputs when admissible, that yields the best trade-off between predictive performance and model sparsity for the selected hybrid FPML structure. The proposed branch-and-prune superstructure optimization algorithm is summarized in Table \ref{tab:fpml_algorithm}. The proposed ADOPT-FPML algorithm summarizes the stage-wise synthesis of a composite hybrid FPML configuration for a single system or envelope within an interconnected system. The same algorithm can be applied independently to each subsystem envelope in an interconnected system, as shown in Figure \ref{fig:composite_structures}.

\begin{table}[htpb]
\centering
\caption{Branch-and-prune superstructure optimization algorithm used in ADOPT-FPML}
\label{tab:fpml_algorithm}
\normalsize
\setlength{\tabcolsep}{4pt}
\renewcommand{\arraystretch}{1.06}
\begin{tabularx}{1\textwidth}{@{}X@{}}
\toprule
\textbf{Algorithm: ADOPT-FPML stage-wise branch-and-prune superstructure optimization for a single system or envelope using FPML building blocks} \\
\midrule
\textbf{Given:} Training dataset $\mathcal{D}=\{ {\bf u}(t_j),{\bf y}^{data}(t_j)\}_{j=1}^{n_s}$, output-specific FP prediction thresholds $\varepsilon_{p,j}$, admissible routing logic, and candidate input library $\mathcal{U}^{(1)}$ \\

Simulate the standalone FP model \\
Initialize $\mathcal{J}^{(1)}$ using all outputs that do not satisfy the FP accuracy threshold \\
Set the block index $b=1$ \\

\textbf{while} $\mathcal{J}^{(b)}\neq\emptyset$ \textbf{do} \\
\hspace*{1em} Generate feasible disjunctive configurations $\mathcal{R}^{(b)}=\left\{ R_1^{(b)},\ldots,R_{N_R^{(b)}}^{(b)}\right\}$ according to Eqs. \eqref{eq:config1}-\eqref{eq:config4} \\
\hspace*{1em} Set $r=1$ \\
\hspace*{1em} \textbf{while} $r \leq N_R^{(b)}$ \textbf{do} \\
\hspace*{2em} Select candidate configuration $R_r^{(b)}$ and impose the corresponding routing disjunction \\
\hspace*{2em} Update routing assignment ${\boldsymbol \theta}^{(b,r)}$ \\
\hspace*{2em} Assign $\mathcal{J}^{(b)}$ as the target output set for the current branch \\
\hspace*{2em} Rank candidate inputs from $\mathcal{U}^{(b)}$ using the chosen screening metric \\
\hspace*{2em} $\ell=1$, $\mathcal{K}^{(b,r)}=\emptyset$ \\
\hspace*{2em} \textbf{while} $\ell \leq |\mathcal{U}^{(b)}|$ \textbf{do} \\
\hspace*{3em} Append the next ranked input to $\mathcal{K}^{(b,r)}$ \\
\hspace*{3em} Train the hybrid FPML model under the candidate configuration $\big(R_r^{(b)},\mathcal{J}^{(b)},\mathcal{K}^{(b,r)}\big)$ \\
\hspace*{3em} Compute $\mathrm{AIC}_c$ \\
\hspace*{3em} Prune the branch if the updated model does not improve the objective \\
\hspace*{3em} $\ell=\ell+1$ \\
\hspace*{2em} \textbf{end while} \\
\hspace*{2em} $r=r+1$ \\
\hspace*{1em} \textbf{end while} \\
\hspace*{1em} Retain the admissible configuration with minimum local $\mathrm{AIC}_c$ for block $b$ and define the resolved output set $\mathcal{H}^{(b)}$ \\
\hspace*{1em} Screen the candidate propagated synthetic variables and construct $\mathcal{S}^{(b)}$ \\
\hspace*{1em} Update $\mathcal{J}^{(b+1)}=\mathcal{J}^{(b)}\setminus\mathcal{H}^{(b)}$ \\
\hspace*{1em} Update $\mathcal{U}^{(b+1)}=\mathcal{U}^{(b)}\cup\mathcal{S}^{(b)}$ \\
\hspace*{1em} Evaluate the updated global $\mathrm{AIC}_c$; if $\mathcal{H}^{(b)}=\emptyset$ or the retained block does not satisfy Eq. \eqref{eq:aic_termination}, terminate \\
\hspace*{1em} $b=b+1$ \\
\textbf{end while} \\

\textbf{Return:} The optimal sequence of routing assignments $\{{\boldsymbol \theta}^{(1)\ast},{\boldsymbol \theta}^{(2)\ast},\ldots\}$, the corresponding selected outputs and input sets for each block, and the resulting composite hybrid configuration for the system under consideration \\
\bottomrule
\end{tabularx}
\normalsize
\end{table}

A subsequent superstructure block is retained only when it resolves at least one additional output in the current set $\mathcal{J}^{(b)}$ according to the prescribed prediction-accuracy criterion and the resulting composite architecture satisfies the global $\mathrm{AIC}_c$ progression criterion. To ensure finite {\textit{termination}} of the repeated-block approach, the stage-wise optimization is continued only while these conditions are satisfied. Let $\mathrm{AIC}_{c,\mathrm{global}}^{(b)}$ denote the global penalized objective associated with the composite FPML architecture retained up to block $b$. Then block $b+1$ is accepted only if:

\begin{equation}
\Delta_{\mathrm{rel}}\mathrm{AIC}_c^{(b+1)}
=\frac{\mathrm{AIC}_{c,\mathrm{global}}^{(b+1)}-
\mathrm{AIC}_{c,\mathrm{global}}^{(b)}}{
\left|\mathrm{AIC}_{c,\mathrm{global}}^{(b)}\right|}
<\tau_{\mathrm{AIC}}
\label{eq:aic_termination}
\end{equation}
where $\tau_{\mathrm{AIC}}\geq 0$ denotes the maximum admissible relative increase in the global penalized objective between successive retained blocks. Thus, $\tau_{\mathrm{AIC}}=0$ requires a strict reduction in the global $\mathrm{AIC}_c$, whereas $\tau_{\mathrm{AIC}}>0$ permits a subsequent block to be retained when the increase in the global $\mathrm{AIC}_c$ remains below the admissible relative tolerance. If the best admissible candidate structure for block $b+1$ fails to resolve at least one additional output or does not satisfy Eq. \eqref{eq:aic_termination}, the algorithm terminates and the retained composite hybrid configuration for the system up to block $b$ is declared to be the {\textit{optimal parsimonious solution}}, even if some outputs remain outside their corresponding admissible accuracy thresholds $\varepsilon_{p,j}$. If the algorithm terminates with one or more outputs remaining outside their admissible accuracy thresholds, further improvements may require enhancing the FP model, expanding the candidate ML model structures or input library, or reconsidering the prescribed modeling requirements.

\section{Case Studies} 
\label{sec:case_studies}

This section presents three case studies used to demonstrate the implementation of the proposed ADOPT-FPML framework. The first two case studies are defined over different subsystem envelopes within an interconnected process for acetic anhydride production, while the third case study considers an industrial superheater system. 

\subsection{Acetic Anhydride Production}\label{sec:acetic_anhydride}

\begin{figure}[htbp]
    \centering
    \includegraphics[width=0.98\textwidth]{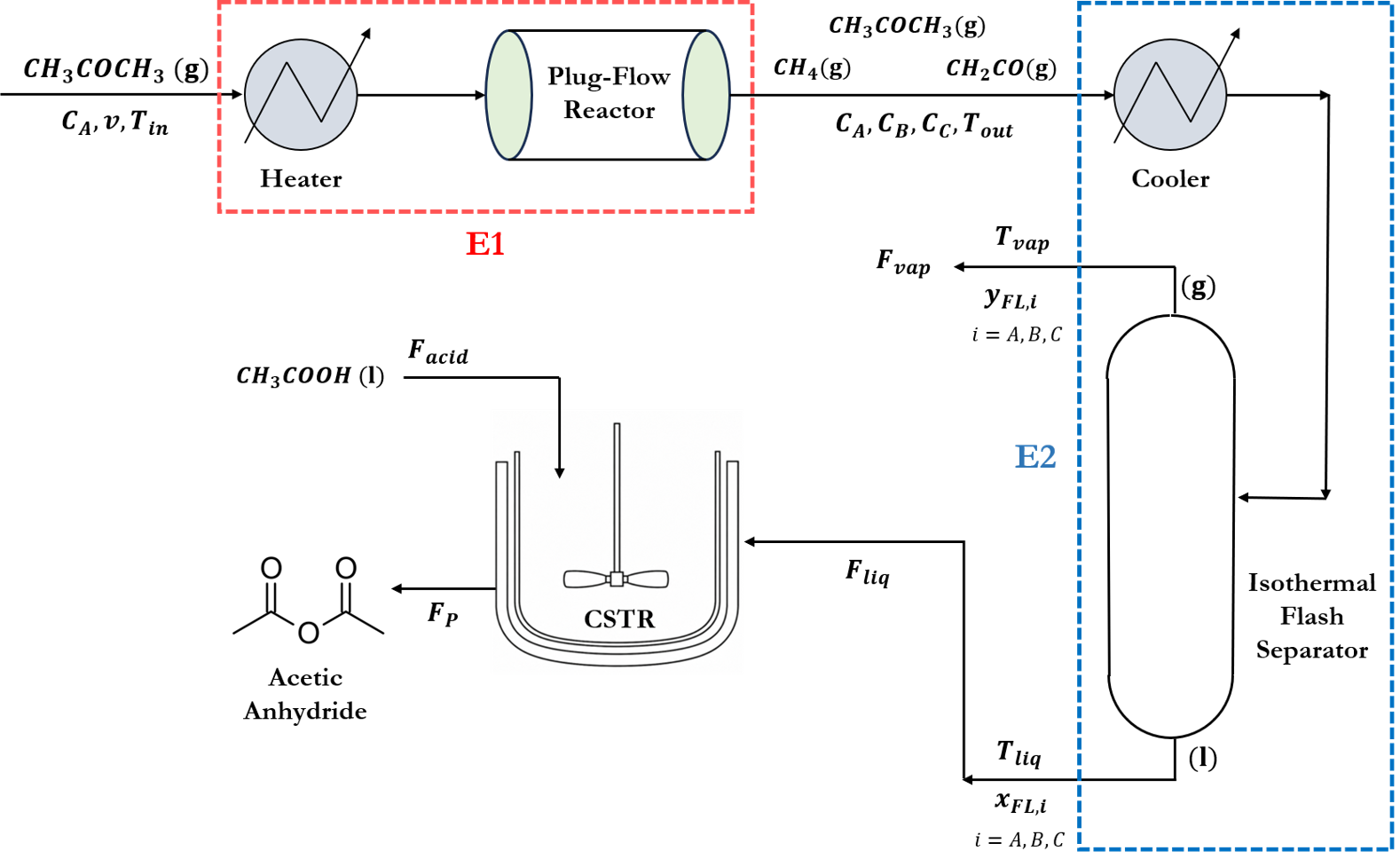}
    \caption{Process flowsheet for the production of acetic anhydride from acetone, showing subsystem envelopes E1 and E2 defined based on the availability of measurements and corresponding FP models.}
    \label{fig:case_study_1}
\end{figure}

The proposed ADOPT-FPML framework is first applied to an interconnected process for the production of acetic anhydride from vapor-phase acetone. This process provides a representative nonlinear system involving reaction, heat-transfer, and phase-separation operations \cite{Jeffreys1964Book,Fogler2017Book,mukherjee2025metcnn}. The overall process considered in this work is shown in Figure \ref{fig:case_study_1}.

Vapor-phase acetone is cracked in an electrically heated tubular plug-flow reactor (PFR) to produce methane and ketene. Because the acetone cracking reaction is endothermic, external heat is supplied along the reactor to sustain the reaction \cite{mukherjee2024mecnn}. The reactor effluent is subsequently cooled to produce a partially condensed stream, which is separated into vapor and liquid phases using a flash separator. The liquid stream (bottom product from the flash separator) is then supplied to a continuous stirred tank reactor (CSTR), where the ketene reacts with acetic acid to produce acetic anhydride \cite{Jeffreys1964Book}. The principal reactions associated with the overall process are:

\begin{align}
\mathrm{CH_3COCH_3(g)}
&\longrightarrow
\mathrm{CH_4(g)}+\mathrm{CH_2CO(g)}
\label{eq:acetone_cracking}
\\
\mathrm{CH_2CO}+\mathrm{CH_3COOH}
&\longrightarrow
\mathrm{(CH_3CO)_2O}
\label{eq:acetic_anhydride_reaction}
\end{align}

As discussed in Section \ref{sec:compositesubstructure}, the application of ADOPT-FPML to an interconnected process does not require every individual unit operation to be represented by a separate hybrid model. Instead, a subsystem envelope may contain one or more connected unit operations, provided that an appropriate FP model and the required measurements are available at the boundaries of that envelope. The envelope boundaries can therefore be selected based on process connectivity, measurement availability, and the scope of the corresponding mechanistic representation.

Accordingly, the case study under consideration is divided into two subsystem envelopes, denoted by E1 and E2. Envelope E1 contains the electrically heated PFR system, whereas Envelope E2 collectively contains the cooler and flash separator. Intermediate quantities generated within a given envelope need not be directly measured unless they are required as boundary variables for another modeling block. This formulation enables multiple unit operations to be incorporated within a single FPML block when intermediate measurements are unavailable, while preserving a modular representation of the overall interconnected system. The output variables predicted by the optimal model for E1 define the measured inlet conditions supplied to E2. Thus, the two envelopes are treated as separate applications of the ADOPT-FPML framework, but they remain connected through the physical flow of information (mass, heat/energy, etc.) across the process. It is worth noting that the downstream CSTR, in which ketene from the liquid product of the flash separator reacts with acetic acid to form acetic anhydride, may also be modeled using the proposed ADOPT-FPML framework. A separate subsystem envelope could be defined around the CSTR based on the availability of an associated FP model and local input/output measurements. However, this additional case is not included in the paper for brevity.  

\subsubsection{Case Study 1: Envelope E1}\label{sec:case_study_e1}

The first case study considers Envelope E1, which contains the electrically heated PFR used for vapor-phase acetone cracking. In this work, the tubular reactor is approximated as three CSTRs connected in series, where the outlet states from one CSTR serve as the inlet conditions for the next CSTR, thereby providing a finite-dimensional dynamic approximation of the PFR \cite{Fogler2017Book}. The measured inputs considered at the inlet boundary of E1 are collected in the vector:
\begin{equation}
{\bf u}_{E1}(t)=\left[C_{A,0}(t),\,T_{\mathrm{in}}(t),\,v(t),\,
\dot{q}_{1}(t),\,\dot{q}_{2}(t),\,\dot{q}_{3}(t)\right]^{\mathsf{T}}
\label{eq:e1_input_vector}
\end{equation}
where $C_{A,0}$ is the inlet acetone concentration, $T_{\mathrm{in}}$ is the inlet temperature, $v$ is the volumetric flow rate, and $\dot{q}_{s}$ denotes the heat supplied to the $s$-th CSTR, for $s\in\{1,2,3\}$. The desired outputs are the outlet concentrations of acetone, methane, and ketene, together with the reactor outlet temperature:
\begin{equation}
{\bf y}_{E1}(t)=\left[C_A(t),\,C_B(t),\,C_C(t),\,T_{\mathrm{out}}(t)\right]^{\mathsf{T}}
\label{eq:e1_output_vector}
\end{equation}
where the subscripts $A$, $B$, and $C$, denote acetone, methane, and ketene, respectively.

Additionally, for the $s$-th CSTR $(s\in\{1,2,3\})$, the FP model within the ADOPT-FPML framework is defined as:
\begin{align}
\frac{dC_{A,s}}{dt}&=\frac{v}{V_s}\left(C_{A,\mathrm{in},s}-C_{A,s}\right)-r_s
\label{eq:e1_fp_ca}
\\
\frac{dC_{B,s}}{dt}&=\frac{v}{V_s}\left(C_{B,\mathrm{in},s}-C_{B,s}\right)+r_s
\label{eq:e1_fp_cb}
\\
\frac{dC_{C,s}}{dt}&=\frac{v}{V_s}\left(C_{C,\mathrm{in},s}-C_{C,s}\right)+r_s
\label{eq:e1_fp_cc}
\\
\frac{dT_s}{dt}&=\frac{v}{V_s}\left(T_{\mathrm{in},s}-T_s\right)+
\frac{\dot{q}_s}{C_{p,A}C_{A,s}}-\frac{\Delta H_r}{C_{p,A}C_{A,s}}r_s
\label{eq:e1_fp_temperature}
\end{align}
where $V_s$ denotes the volume of CSTR $s$, $C_{p,A}$ is the heat capacity of acetone, and $\Delta H_r$ is the heat of reaction. The inlet conditions for CSTRs 2 and 3 are represented by the corresponding outlet states of the immediately preceding CSTR.

The acetone-cracking rate in the FP model used within ADOPT-FPML is represented using a lumped constant kinetic parameter, given by:
\begin{equation}
r_s = k ~C_{A,s}
\label{eq:e1_lumped_rate}
\end{equation}

In this work, the FP model is formulated using a constant nominal kinetic parameter $k$ because sufficient information is assumed to be unavailable for characterizing its temperature dependence. Since E1 is non-isothermal, the use of a constant nominal kinetic parameter under the assumed level of available mechanistic knowledge may lead to discrepancies between the standalone FP predictions and the measured process outputs. Such discrepancies may subsequently be addressed through the hybridization structure identified by ADOPT-FPML.

The transient synthetic measurement data for E1 are generated using the detailed process representation provided in Section \ref{app:e1_data_generation} of Appendix \ref{app:e1}, together with external measurement bias and noise to emulate realistic process observations. This data-generating model is used solely for constructing the synthetic measurement dataset and is not assumed to be available during structure discovery; ADOPT-FPML uses only the FP representation in Eqs. \eqref{eq:e1_fp_ca}-\eqref{eq:e1_lumped_rate}, formulated based on the available mechanistic knowledge, together with the corresponding input-output measurements.

\subsubsection{Case Study 2: Envelope E2}\label{sec:case_study_e2}

The second case study considers Envelope E2, which collectively contains the cooler and downstream flash separator. The stream leaving E1 provides the measured inlet conditions for E2. The cooling of the reactor effluent produces a partially condensed two-phase stream before it enters the flash separator. Since direct measurement of the individual liquid- and vapor-phase flow rates and compositions between the cooler and flash separator may not be readily available, the two unit operations are represented within a single subsystem envelope (E2). Measurements are instead assumed to be available at the inlet boundary of the cooler and at the vapor and liquid outlet boundaries of the flash separator.

The measured inputs considered at the inlet boundary of E2 are collectively defined as:
\begin{equation}
{\bf u}_{E2}(t)=\left[C_A(t),\,C_B(t),\,C_C(t),\,T_{\mathrm{out}}(t),\, v(t)\right]^{\mathsf{T}}
\label{eq:e2_input_vector}
\end{equation}
where $C_A$, $C_B$, and $C_C$ are the concentrations of acetone, methane, and ketene, respectively, in the stream leaving E1; $T_{\mathrm{out}}$ is the corresponding stream temperature; and $v$ is the volumetric flow rate. The desired outputs are the vapor- and liquid-phase outlet flow rates, temperatures, and compositions from the flash separator:
\begin{equation}
{\bf y}_{E2}(t)=\left[F_{\mathrm{vap}}(t),\,F_{\mathrm{liq}}(t),\,
T_{\mathrm{vap}}(t),\,T_{\mathrm{liq}}(t),\,y_A(t),\,y_B(t),\,y_C(t),\,
x_A(t),\,x_B(t),\,x_C(t)\right]^{\mathsf{T}}
\label{eq:e2_output_vector}
\end{equation}
where $F_{\mathrm{vap}}$ and $F_{\mathrm{liq}}$ denote the vapor- and liquid-phase outlet flow rates, $T_{\mathrm{vap}}$ and $T_{\mathrm{liq}}$ denote the corresponding outlet temperatures, and $y_i$ and $x_i$ denote the vapor- and liquid-phase mole fractions of component $i$ for $i\in\{A,B,C\}$, respectively.

For the FP model considered within ADOPT-FPML, the component molar flow rates and overall feed composition entering the cooler are calculated as:
\begin{align}
F_{i,\mathrm{in}} &= v C_i,\qquad i\in\{A,B,C\}
\label{eq:e2_component_feed}
\\
F_{\mathrm{in}} &= \sum_{i\in\{A,B,C\}} F_{i,\mathrm{in}}
\label{eq:e2_total_feed}
\\
z_i &= \frac{F_{i,\mathrm{in}}}{F_{\mathrm{in}}}
\label{eq:e2_feed_composition}
\end{align}
where $z_i$ denotes the overall inlet mole fraction of component $i$.

Additionally, the FP model for the cooler is defined using the following lumped heat transfer relation:
\begin{equation}
T_{\mathrm{c}}=T_{\mathrm{coolant}}+\left(T_{\mathrm{out}}-T_{\mathrm{coolant}}\right)
\exp\left(-\frac{UA}{\dot{C}_{p,\mathrm{in}}}\right)
\label{eq:e2_cooler_temperature}
\end{equation}
where $T_{\mathrm{c}}$ is the cooler outlet temperature, $T_{\mathrm{coolant}}$ is the coolant temperature, $UA$ is the overall heat-transfer parameter, and $\dot{C}_{p,\mathrm{in}}$ is the heat-capacity flow rate of the inlet stream. The partially condensed stream is subsequently separated in the flash separator. The corresponding phase-equilibrium relations can be represented as:
\begin{equation}
y_i = K_i x_i,\qquad i\in\{A,B,C\}
\label{eq:e2_phase_equilibrium}
\end{equation}
together with the Rachford--Rice equation:
\begin{equation}
\sum_{i\in\{A,B,C\}}\frac{z_i(K_i-1)}{1+\beta(K_i-1)}=0
\label{eq:e2_rachford_rice}
\end{equation}
where $\beta$ is the vapor fraction and $K_i$ denotes the vapor-liquid equilibrium ratio of the $i$-th component. Furthermore, the liquid- and vapor-phase compositions are defined as:

\begin{align}
x_i &= \frac{z_i}{1+\beta(K_i-1)}
\label{eq:e2_liquid_composition}
\end{align}
while the corresponding vapor compositions $(y_i)$ can be obtained using Eq. \eqref{eq:e2_phase_equilibrium}. The respective phase flow rates are:
\begin{align}
F_{\mathrm{vap}} &= \beta F_{\mathrm{in}}
\label{eq:e2_vapor_flow}
\\
F_{\mathrm{liq}} &= (1-\beta)F_{\mathrm{in}}
\label{eq:e2_liquid_flow}
\end{align}

In this study, the flash separator is assumed to operate isothermally, such that the vapor and liquid outlet streams leave at a common equilibrium temperature:
\begin{equation}
T_{\mathrm{vap}}=T_{\mathrm{liq}}=T_{\mathrm{flash}}
\label{eq:e2_flash_temperature}
\end{equation}

The FP model considered in this work is formulated using constant nominal values of $UA$ and the equilibrium ratios $K_A$, $K_B$, and $K_C$, because sufficient information is assumed to be unavailable for accurately characterizing their variation with operating conditions. The use of these constant nominal parameters under the assumed level of available mechanistic knowledge may lead to discrepancies between the standalone FP predictions and the measured process outputs. Therefore, similar to E1, such discrepancies may subsequently be addressed through the optimal hybrid configuration identified by ADOPT-FPML.

The transient synthetic measurement data for E2 are generated using the detailed process representation provided in Appendix \ref{app:e2}, including the operating-condition dependent heat-transfer representation, together with external measurement bias and noise to emulate realistic process observations. This data-generating model is used solely for constructing the synthetic measurement dataset and is not assumed to be available during structure discovery; ADOPT-FPML uses only the FP representation in Eqs. \eqref{eq:e2_component_feed}-\eqref{eq:e2_flash_temperature}, formulated based on the available mechanistic knowledge, together with the corresponding input-output measurements.

\subsection{Industrial Superheater System}\label{sec:industrial_superheater}

The proposed ADOPT-FPML algorithm is also demonstrated on an industrial steam superheater system of a natural gas combined cycle (NGCC) power plant previously analyzed by some of the authors of this paper using heuristically selected hybrid FPML configurations \cite{mukherjee2025hybridfpai}. This case study enables a direct comparison between the automatically discovered configuration with optimal input-output selection and the previously developed baseline models that use all available data. Additional details on the industrial datasets, system description, and the corresponding FP model formulation are available in numerous references \cite{mukherjee2025hybridfpai, saini2024nonlinear}.

\subsubsection{Case Study 3: NGCC Plant Superheater}
\label{sec:case_study_3}

The third case study considers the final steam superheater in the heat-recovery steam generator (HRSG) of an operating NGCC power-plant with a rated capacity of 600 MW \cite{saini2024nonlinear}. The industrial system under consideration consists of 74 parallel flow circuits, each comprising a six-pass tube assembly, with the combustion-turbine exhaust gas flowing across the tube banks. Steam leaving the superheater is collected in an outlet header and subsequently supplied to the high-pressure steam turbine, with nominal main-steam outlet conditions of approximately $538 ~^{\circ} \mathrm{C}$ and 10 MPa. Detailed descriptions of the system geometry, instrumentation, and heuristic-based hybrid modeling configurations are provided in our previous publication \cite{mukherjee2025hybridfpai}, while the development and validation of the corresponding distributed FP model are discussed in Saini et al. \cite{saini2024nonlinear}.

The measured inlet boundary conditions obtained from the plant historian and considered as inputs to the ADOPT-FPML framework are collectively referred to as:
\begin{equation}
{\bf u}_{SH}(t)=\left[T_{\mathrm{st,in}}(t),\,\dot{m}_{\mathrm{st}}(t),\,
T_{\mathrm{fg,in}}(t),\,\dot{m}_{\mathrm{fg}}(t)\right]^{\mathsf{T}}
\label{eq:sh_input_vector}
\end{equation}
where $T_{\mathrm{st,in}}$ and $\dot{m}_{\mathrm{st}}$ denote the inlet temperature and mass flow rate of steam, respectively, and $T_{\mathrm{fg,in}}$ and $\dot{m}_{\mathrm{fg}}$ denote the inlet temperature and mass flow rate of the flue-gas stream. Industrial temperature measurements are available at five outer tube-wall locations, denoted by the tags E$6'$, E$49'$, E$75'$, RB1, and RB2; at two flue-gas locations, denoted by G$47'$ and G$71'$; and at the main-steam outlet. Accordingly, the desired output vector is defined as:
\begin{equation}
{\bf y}_{SH}(t)=\left[T_{\mathrm{E6}}(t),\,T_{\mathrm{E49}}(t),\,
T_{\mathrm{E75}}(t),\,T_{\mathrm{RB1}}(t),\,
T_{\mathrm{RB2}}(t),\,T_{\mathrm{G47}}(t),\,
T_{\mathrm{G71}}(t),\,T_{\mathrm{st,out}}(t)\right]^{\mathsf{T}}
\label{eq:e3_output_vector}
\end{equation}

The FP model considered within ADOPT-FPML is represented by a dynamic distributed two-dimensional differential-algebraic equation (DAE) model formulated from mass and energy conservation for the steam and flue-gas streams, together with the tube-wall temperature dynamics. Since mass is conserved trivially through the corresponding flow continuity equations, the FP model is expressed only in terms of the relevant energy balance equations given by:
\begin{align}
\frac{\partial T_{\mathrm{st}}}{\partial t}+
v_{\mathrm{st}}\frac{\partial T_{\mathrm{st}}}{\partial z}&=
\frac{q_{\mathrm{wall,st}}}
{\rho_{\mathrm{st}}C_{p,\mathrm{st}}}
\label{eq:e3_steam_energy}
\\[0.5cm]
\frac{\partial T_{\mathrm{fg}}}{\partial t}+
v_{\mathrm{fg}}\frac{\partial T_{\mathrm{fg}}}{\partial y}&=
-\frac{q_{\mathrm{wall,fg}}}
{\rho_{\mathrm{fg}}C_{p,\mathrm{fg}}}
\label{eq:e3_flue_gas_energy}
\\[0.5cm]
m_{\mathrm{tube}}C_{p,\mathrm{tube}}
\frac{dT_{\mathrm{tube}}}{dt}&=h_{\mathrm{conv,fg}}A_o
\left(T_{\mathrm{fg}}-T_{\mathrm{tube}}\right)-
h_{\mathrm{conv,st}}A_i\left(T_{\mathrm{tube}}-T_{\mathrm{st}}\right)
\label{eq:e3_tube_energy}
\end{align}
where $T$, $v$, $\rho$, and $C_p$ denote temperature, velocity, density, and specific heat capacity, respectively; the subscripts $\mathrm{st}$, $\mathrm{fg}$, and $\mathrm{tube}$ denote steam, flue gas, and tube metal; $q_{\mathrm{wall}}$ denotes the corresponding volumetric heat flux; $h_{\mathrm{conv}}$ denotes the convective heat-transfer coefficient; and $A_i$ and $A_o$ are the inner and outer tube heat-transfer areas. The complete set of model equations, heat-transfer correlations, spatial discretization, and system-specific geometric parameters is available in \cite{saini2024nonlinear, mukherjee2025hybridfpai}. For this case study, the parameters associated with the available FP model, for example, the flue gas and steam side heat transfer coefficients, are {\textit{assumed}} to be sufficiently well characterized and are therefore not treated as uncertain quantities requiring data-driven correction. Consequently, the admissible structural search for Case Study 3 is restricted to Config 1 and Config 2, whereas Config 3 and Config 4, which involve ML-assisted updates to FP-model parameters through the $\theta_4$ pathway, are not considered.

As mentioned previously, this case study is included in the study to evaluate the capability of ADOPT-FPML to automatically identify a parsimonious hybrid configuration using industrial operating data and to compare the resulting structure with the hybrid models developed previously for the same system \cite{mukherjee2025hybridfpai}. In the previous work, the series, parallel, and integrated configurations were constructed and selected based primarily on physical interpretation, observed plant-model discrepancies, and heuristic assessment using all available measured variables. In contrast, the present framework systematically optimizes the FPML interaction configuration together with the corresponding input and output selections. The previously reported hybrid architectures, therefore, provide suitable baseline models for assessing the configuration and predictive performance identified through automated structure discovery.

\section{Results and Discussions} \label{sec:results}

The proposed branch-and-prune-based ADOPT-FPML framework is applied to the three case studies introduced in Section \ref{sec:case_studies}. For each system or subsystem envelope, the optimization proceeds stage-wise over the set of outputs that remain outside their admissible prediction-accuracy thresholds $(\varepsilon_{p,j})$. At every stage, the 4 mutually exclusive disjunctive configurations defined in Eqs. \eqref{eq:config1}-\eqref{eq:config4} are evaluated independently based on the corresponding activation states of the interaction routing operators $\theta_3$ and $\theta_4$. The evaluations associated with Configs 1-4 are therefore independent and may be parallelized. Within each configuration, however, the candidate input set is expanded sequentially according to the input-selection strategy described in Section \ref{sec:input_selection}. Candidate inputs are introduced in their prescribed relevance order, and the resulting FPML model is retrained and evaluated through $\mathrm{AIC}_c$ after each addition. Consistent with the greedy input-addition strategy, refinement of a branch is terminated once inclusion of the next candidate input fails to improve the objective function.

For comparability across successive candidate evaluations, the ML model architecture and training settings are held consistent throughout a given case study. The structural discovery itself is automated once the user specifies the prescribed output-wise prediction thresholds $\varepsilon_{p,j}$, the maximum number of propagated synthetic inputs $N_I$, and the minimum required improvement in the global $\mathrm{AIC}_{c}$ represented as $\tau_{\mathrm{AIC}}$. The first criterion determines whether an output requires further hybridization, $N_I$ limits the dimensionality of the synthetic information propagated between successive superstructure blocks, and $\tau_{\mathrm{AIC}}$ determines whether an additional superstructure block (stage) is retained. Unless otherwise stated, all remaining structural decisions, including the selected disjunctive configuration, output allocation, input selection, and number of retained stages, are determined automatically by ADOPT-FPML.

\subsection{Case Study 1: Envelope E1} \label{sec:results_e1}

For Envelope E1, the admissible output-specific RMSE thresholds are selected as $\varepsilon_{p,C_A} = 5 \ \mathrm{mol/m^3}$, $\varepsilon_{p,C_B} = 2 \ \mathrm{mol/m^3}$, $\varepsilon_{p,C_C} = 3 \ \mathrm{mol/m^3}$, and $\varepsilon_{p,T_{\mathrm{out}}} = 10 \ \mathrm{K}$. The maximum number of propagated synthetic inputs is set to $N_I=10$, while $\tau_{\mathrm{AIC}}=0$ is used such that an additional superstructure block is retained only when it results in a strict reduction of the global $\mathrm{AIC}_c$. As defined in Eq. \eqref{eq:e1_output_vector}, the desired outputs for this envelope are the outlet concentrations $C_A$, $C_B$, and $C_C$, together with the reactor outlet temperature $T_{\mathrm{out}}$.

\begin{figure}[htbp]
    \centering
    \includegraphics[width=0.8\textwidth]{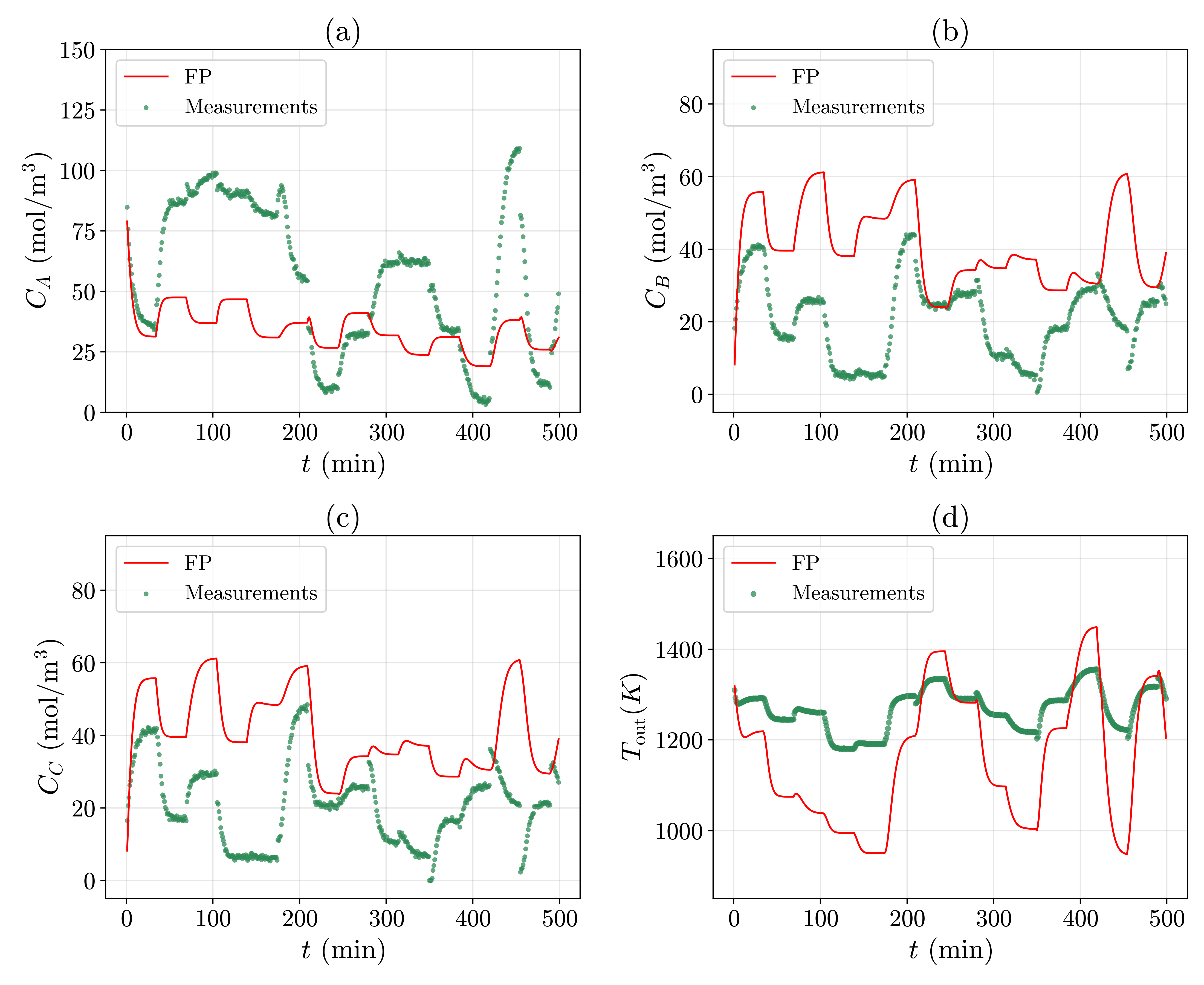}
    \caption{Standalone FP model predictions and corresponding measurements for Envelope E1: (a) $C_A$, (b) $C_B$, (c) $C_C$, and (d) $T_{\mathrm{out}}$.}
    \label{fig:e1_fp_predictions}
\end{figure}

Figure \ref{fig:e1_fp_predictions} compares the measurements with the predictions obtained from the standalone FP model over the training horizon. The standalone FP model exhibits substantial mismatch for all 4 outputs, reflecting the limitations of the available mechanistic knowledge and the resulting uncertainty in the representation of accurate reaction kinetic relative to the observed process behavior. The resulting RMSE values are $33.6 ~\mathrm{mol/m^3}$, $24.3 ~\mathrm{mol/m^3}$, $23.6 ~\mathrm{mol/m^3}$, and $149.8 ~\mathrm{K}$ for $C_A$, $C_B$, $C_C$, and $T_{\mathrm{out}}$, respectively. Since all values exceed their corresponding admissible thresholds, none of the desired outputs is retained directly from the standalone FP model and all 4 outputs are supplied to the first stage of the ADOPT-FPML framework.

Table \ref{tab:e1_stage1} summarizes the candidate evaluations performed during Stage 1. The first column identifies the disjunctive branch using the generic Config 1-4 notation introduced in Section \ref{sec:consistency_constraints}. Within each configuration, the input-selection algorithm sequentially augments the ML input set according to the ranking criterion in Section \ref{sec:input_selection}. Among the Stage 1 candidates, Config 2 provides the minimum local $\mathrm{AIC}_c$ of around 7373 using the selected ML input set $\{T_{\mathrm{out,FP}}, T_{\mathrm{in}}, C_{A,0}, C_{B,\mathrm{FP}}, C_{C,\mathrm{FP}}\}$. The corresponding RMSE values are $6.0 ~\mathrm{mol/m^3}$, $1.5 ~\mathrm{mol/m^3}$, $1.4 ~\mathrm{mol/m^3}$, and $10.3 ~\mathrm{K}$ for $C_A$, $C_B$, $C_C$, and $T_{\mathrm{out}}$, respectively. Therefore, $C_B$ and $C_C$ satisfy their admissible accuracy thresholds and are assigned as the final outputs of the first-stage optimal superstructure block, whereas $C_A$ and $T_{\mathrm{out}}$ remain outside their respective thresholds. However, these intermediate predictions, i.e., $C_{A,\mathrm{stg1}}$ and $T_{\mathrm{out,stg1}}$ from Stage 1, constitute synthetic variables generated by the retained block and are therefore available for selection as candidate inputs (refer Section \ref{sec:consistency_constraints}). The reduced output set propagated to Stage 2 is therefore $\mathcal{J}^{(2)}=\{C_A,T_{\mathrm{out}}\}$.   

Following the Stage 1 reduction, the second-stage search is performed only for $C_A$ and $T_{\mathrm{out}}$. Table \ref{tab:e1_stage2} summarizes the resulting Stage 2 branch-and-prune search. 

\begin{table}[!htbp]
\centering
\caption{Stage 1 branch-and-prune search for Envelope E1.}
\label{tab:e1_stage1}
\footnotesize
\setlength{\tabcolsep}{3pt}
\renewcommand{\arraystretch}{1.08}

\begin{tabularx}{\textwidth}{
@{} c |>{\centering\arraybackslash}X | c | c | c | c | c @{}}
\hline

\multirow{3}{*}{Configuration} &
\multirow{3}{*}{Selected ML inputs} &
\multirow{3}{*}{Local $\mathrm{AIC}_c$} &
\multicolumn{4}{c}{RMSE} \\
\cline{4-7}

& & &
$C_A$ &
$C_B$ &
$C_C$ &
$T_{\mathrm{out}}$ \\

& & &
($\mathrm{mol/m^3}$) &
($\mathrm{mol/m^3}$) &
($\mathrm{mol/m^3}$) &
(K) \\

\hline

Config 1 &
$T_{\mathrm{in}}$ &
14688 &
17.6 & 10.4 & 9.9 & 73.1 \\

Config 1 &
$T_{\mathrm{in}},\,C_{A,0}$ &
13990 $\downarrow$ &
12.0 & 8.9 & 9.4 & 61.6 \\

Config 1 &
$T_{\mathrm{in}},\,C_{A,0},\,\dot{q}_2$ &
13927 $\downarrow$ &
12.4 & 9.3 & 9.4 & 60.1 \\

Config 1 &
$T_{\mathrm{in}},\,C_{A,0},\,\dot{q}_2,\,v$ &
13586 $\downarrow$ &
11.4 & 8.9 & 9.4 & 54.7 \\

Config 1 &
$T_{\mathrm{in}},\,C_{A,0},\,\dot{q}_2,\,v,\,\dot{q}_3$ &
13211 $\downarrow$ &
10.5 & 7.8 & 7.8 & 49.7 \\

Config 1 &
$T_{\mathrm{in}},\,C_{A,0},\,\dot{q}_2,\,v,\,\dot{q}_3,\,\dot{q}_1$ &
13244 $\uparrow$ &
10.4 & 8.0 & 7.9 & 49.9 \\

\hline

Config 2 &
$T_{\mathrm{out,FP}}$ &
9600 &
12.1 & 7.1 & 8.5 & 13.8 \\

Config 2 &
$T_{\mathrm{out,FP}},\,T_{\mathrm{in}}$ &
9262 $\downarrow$ &
11.3 & 6.2 & 7.0 & 13.1 \\

Config 2 &
$T_{\mathrm{out,FP}},\,T_{\mathrm{in}},\,C_{A,0}$ &
7922 $\downarrow$ &
7.6 & 4.1 & 4.1 & 10.2 \\

Config 2 &
$T_{\mathrm{out,FP}},\,T_{\mathrm{in}},\,C_{A,0},\,C_{B,\mathrm{FP}}$ &
7605 $\downarrow$ &
7.4 & 1.6 & 2.0 & 10.2 \\

Config 2 &
$T_{\mathrm{out,FP}},\,T_{\mathrm{in}},\,C_{A,0},\,C_{B,\mathrm{FP}},\,C_{C,\mathrm{FP}}$ &
7373 $\downarrow$ &
6.0 & 1.5 & 1.4 & 10.3 \\

Config 2 &
$T_{\mathrm{out,FP}},\,T_{\mathrm{in}},\,C_{A,0},\,C_{B,\mathrm{FP}},\,C_{C,\mathrm{FP}},\,\dot{q}_2$ &
7930 $\uparrow$ &
7.1 & 1.7 & 2.0 & 11.6 \\

\hline

Config 3 &
$T_{\mathrm{in}}$ &
17277 &
20.9 & 19.2 & 19.8 & 144.3 \\

Config 3 &
$T_{\mathrm{in}},\,C_{A,0}$ &
20272 $\uparrow$ &
17.0 & 19.0 & 20.8 & 310.7 \\

\hline

Config 4 &
$T_{\mathrm{out,FP}}$ &
21762 &
11.4 & 14.9 & 16.7 & 454.6 \\

Config 4 &
$T_{\mathrm{out,FP}},\,T_{\mathrm{in}}$ &
25488 $\uparrow$ &
18.8 & 15.9 & 17.2 & 1150.5 \\

\hline
\end{tabularx}

\normalsize
\end{table}

\noindent For both Tables \ref{tab:e1_stage1} and \ref{tab:e1_stage2}, the reported arrows indicate whether addition of the corresponding input decreases or increases the local $\mathrm{AIC}_c$ relative to the preceding candidate within the same branch. Because the search follows a greedy input-addition strategy, a branch is pruned when the next ranked input fails to improve $\mathrm{AIC}_c$. Subsequent supersets along that branch are not evaluated.

\begin{table}[!htbp]
\centering
\caption{Stage 2 branch-and-prune search for Envelope E1 over the reduced output set $\{C_A,T_{\mathrm{out}}\}$.}
\label{tab:e1_stage2}
\footnotesize
\setlength{\tabcolsep}{4pt}
\renewcommand{\arraystretch}{1.08}

\begin{tabularx}{\textwidth}{
@{} c | >{\centering\arraybackslash}X | c | c | c @{}}
\hline

Configuration &
Selected ML inputs &
Local $\mathrm{AIC}_c$ &
RMSE in $C_A$ ($\mathrm{mol/m^3}$) &
RMSE in $T_{\mathrm{out}}$ (K) \\

\hline

Config 1 &
$T_{\mathrm{out,FP}}$ &
4989 &
7.9 & 14.5 \\

Config 1 &
$T_{\mathrm{out,FP}},\,T_{\mathrm{out,stg1}}$ &
3958 $\downarrow$ &
5.7 & 7.9 \\

Config 1 &
$T_{\mathrm{out,FP}},\,T_{\mathrm{out,stg1}},\,C_{A,\mathrm{stg1}}$ &
3351 $\downarrow$ &
4.3 & 5.7 \\

Config 1 &
$T_{\mathrm{out,FP}},\,T_{\mathrm{out,stg1}},\,C_{A,\mathrm{stg1}},\,T_{\mathrm{in}}$ &
3331 $\downarrow$ &
2.9 & 6.4 \\

Config 1 &
$T_{\mathrm{out,FP}},\,T_{\mathrm{out,stg1}},\,C_{A,\mathrm{stg1}},\,T_{\mathrm{in}},\,C_{B,\mathrm{stg1}}$ &
3672 $\uparrow$ &
3.2 & 7.6 \\

\hline

Config 2 &
$T_{\mathrm{out,FP}}$ &
5191 &
12.1 & 13.7 \\

Config 2 &
$T_{\mathrm{out,FP}},\,T_{\mathrm{out,stg1}}$ &
4679 $\downarrow$ &
11.5 & 8.1 \\

Config 2 &
$T_{\mathrm{out,FP}},\,T_{\mathrm{out,stg1}},\,C_{A,\mathrm{stg1}}$ &
4003 $\downarrow$ &
6.3 & 7.7 \\

Config 2 &
$T_{\mathrm{out,FP}},\,T_{\mathrm{out,stg1}},\,C_{A,\mathrm{stg1}},\,T_{\mathrm{in}}$ &
4220 $\uparrow$ &
5.1 & 9.7 \\

\hline

Config 3 &
$T_{\mathrm{out,FP}}$ &
8331 &
19.5 & 86.6 \\

Config 3 &
$T_{\mathrm{out,FP}},\,T_{\mathrm{out,stg1}}$ &
8507 $\uparrow$ &
19.2 & 94.2 \\

\hline

Config 4 &
$T_{\mathrm{out,FP}}$ &
8331 &
19.5 & 86.6 \\

Config 4 &
$T_{\mathrm{out,FP}},\,T_{\mathrm{out,stg1}}$ &
8507 $\uparrow$ &
19.2 & 94.2 \\

\hline
\end{tabularx}

\normalsize
\end{table} 

\begin{figure}[htbp]
    \centering
    \includegraphics[width=1\textwidth]{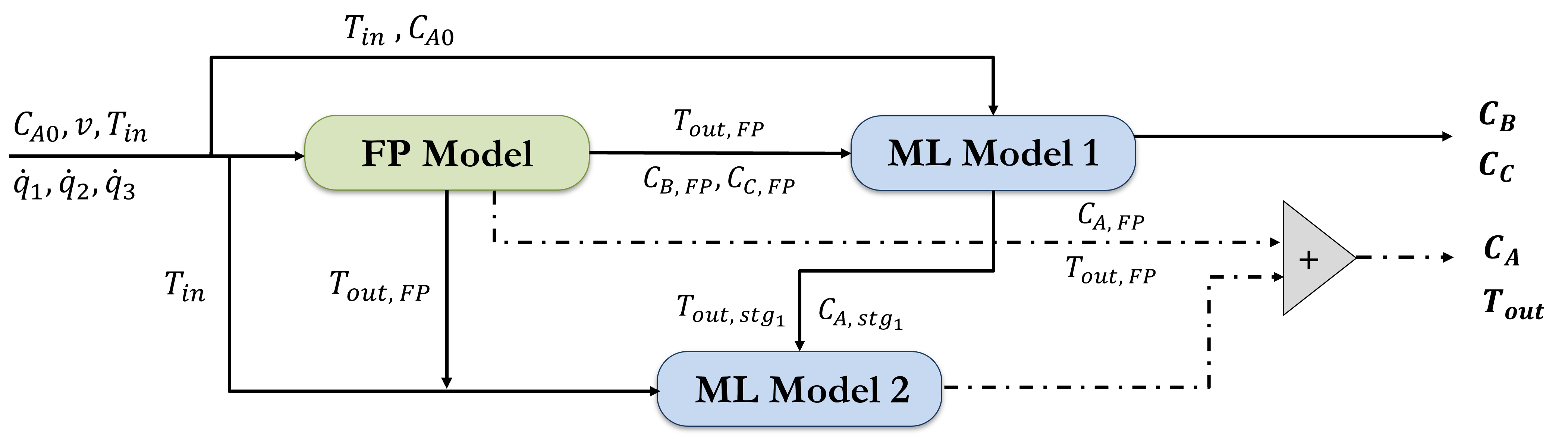}
    \caption{Optimal composite hybrid FPML architecture discovered by ADOPT-FPML for Envelope E1.}
    \label{fig:e1_optimal_structure}
\end{figure}

The minimum Stage 2 objective is obtained for Config 1 using 4 selected ML inputs, namely, $T_{\mathrm{out,FP}}$, $T_{\mathrm{out,stg1}}$, $C_{A,\mathrm{stg1}}$, and $T_{\mathrm{in}}$, yielding an approximate local $\mathrm{AIC}_c$ of 3331. As defined in the formulation of Config 1 (refer Eq. \eqref{eq:config1}), the second-stage ML model uses these selected inputs to estimate corrective contributions for the remaining outputs $C_A$ and $T_{\mathrm{out}}$. These corrections are combined with the corresponding standalone FP predictions $C_{A,\mathrm{FP}}$ and $T_{\mathrm{out,FP}}$ to obtain the final Stage 2 predictions. The resulting RMSE values for $C_A$ and $T_{\mathrm{out}}$ are $2.9~\mathrm{mol/m^3}$ and $6.4~\mathrm{K}$, respectively, both of which satisfy the corresponding accuracy thresholds. No unresolved outputs therefore remain after Stage 2, and the stage-wise search terminates. The resulting optimal composite FPML architecture is shown in Figure \ref{fig:e1_optimal_structure}. 

The first retained block follows Config 2, in which selected FP-generated variables and measured inputs are supplied to ML Model 1. The predictions of $C_B$ and $C_C$ satisfy their prescribed thresholds and are therefore retained as final outputs from this block. The same ML model also generates $C_{A,\mathrm{stg1}}$ and $T_{\mathrm{out,stg1}}$; although these quantities do not satisfy the Stage 1 output thresholds, they are retained as informative synthetic variables for the subsequent block. The second retained block follows Config 1 and uses the selected inputs $T_{\mathrm{in}}$, $T_{\mathrm{out,FP}}$, $C_{A,\mathrm{stg1}}$, and $T_{\mathrm{out,stg1}}$ to generate data-driven corrections to $C_{A,\mathrm{FP}}$ and $T_{\mathrm{out,FP}}$. The corrected predictions constitute the final outputs $C_A$ and $T_{\mathrm{out}}$. The resulting architecture therefore does not correspond to a single globally prescribed canonical FPML structure, but is assembled automatically from two locally optimal and complementary superstructure blocks.

\begin{figure}[!htbp]
    \centering
    \includegraphics[width=0.80\textwidth]{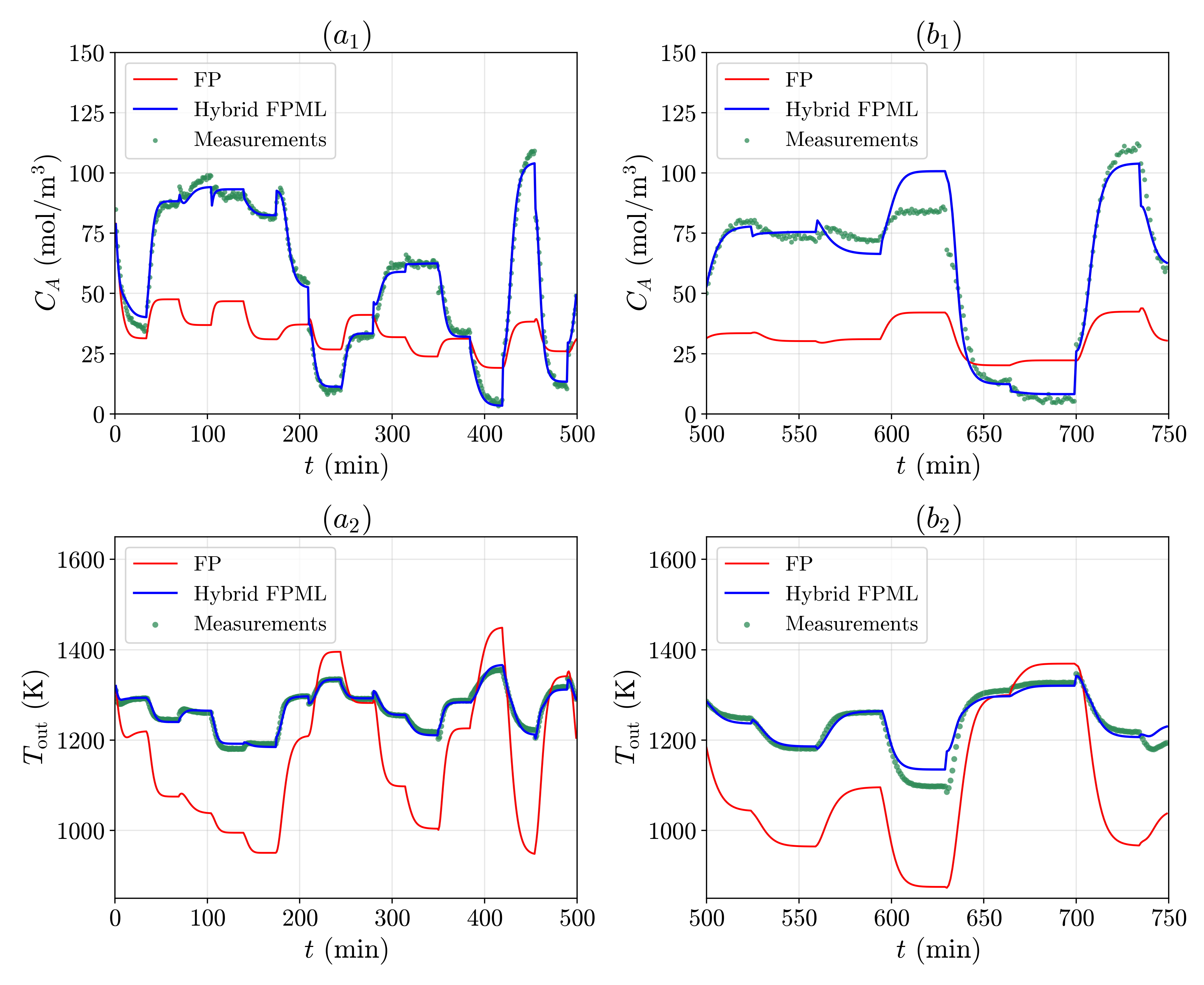}
    \caption{Training and testing performance of the optimal hybrid FPML architecture for Envelope E1: ($a_1$) training and ($b_1$) testing predictions for $C_A$, and ($a_2$) training and ($b_2$) testing predictions for $T_{\mathrm{out}}$.}
    \label{fig:e1_train_test}
\end{figure}

Figure \ref{fig:e1_train_test} compares the predictions of the optimal hybrid FPML architecture for $C_A$ and $T_{\mathrm{out}}$ over both the training and testing horizons. The corresponding standalone FP predictions are also shown for comparison. The hybrid architecture closely follows the transient measurements during training as well as records accurate predictive performance over the unseen testing data. Similar comparisons for $C_B$ and $C_C$, which are resolved in the first superstructure block, are provided in Figure \ref{fig:app_e1_cbcc_train_test} in Appendix \ref{app:e1}. The optimal FPML architecture discovered by ADOPT-FPML substantially reduces the prediction error relative to the standalone FP model for all 4 outputs. For example, the training RMSE decreases from $33.6$ to $2.9~\mathrm{mol/m^3}$ for $C_A$ and from $149.8$ to $6.4~\mathrm{K}$ for $T_{\mathrm{out}}$. Although the corresponding testing errors are larger than the training values, the hybrid model retains a substantial improvement over the FP baseline. The complete training and testing RMSE values, together with the admissible output-specific thresholds, are summarized in Table \ref{tab:app_e1_train_test_rmse} in Appendix A.

The stage-wise branch-and-prune strategy also substantially reduces the number of candidate models that must be evaluated. For E1, a total of only 29 candidate FPML models are evaluated during the complete search, comprising 16 evaluations in Stage 1 and 13 evaluations in Stage 2. By comparison, the largest Stage 1 candidate library contains 10 admissible variables (i.e., 6 input variables and 4 FP-predicted output variables), for which exhaustive enumeration of the non-empty input subsets alone would require $2^{10}-1=1023$ candidate subsets for a single unrestricted configuration. This count does not yet include the additional alternatives arising from the 4 disjunctive configurations, output-selection decisions, or subsequent superstructure blocks. The complete ADOPT-FPML search for E1 requires approximately 22.4 seconds of computational time. Configs 3 and 4 incur a slightly higher computational expense than Configs 1 and 2 because they require an auxiliary parameter estimation step to generate the FP-parameter correction targets used for training the associated ML component. However, as discussed in Section \ref{sec:consistency_constraints}, this parameter estimation problem is solved only once within each stage, and the resulting correction targets are subsequently reused across all candidate ML models associated with the same parameter-correction problem. Thus, the additional computational burden is incurred at the stage level rather than for every candidate evaluation.

Table \ref{tab:e1_full_baselines} further illustrates the trade-off between predictive accuracy and model sparsity that governs the structural selection through the $\mathrm{AIC}_c$ objective. The optimal ADOPT-FPML architecture consists of a Config 2 block in Stage 1 and a Config 1 block in Stage 2, containing 84 and 58 trainable ML parameters, respectively, for a total of 142 parameters. Although the full-size single-stage Config 2 baseline also satisfies the admissible RMSE thresholds for all 4 outputs, its global $\mathrm{AIC}_c$ remains higher than that of the automatically discovered composite architecture. This indicates that restricting the full system to a single interaction configuration does not necessarily provide the most favorable balance between predictive accuracy and model complexity. In contrast, the stage-wise ADOPT-FPML solution assigns different configurations and input/output subsets to different portions of the prediction problem, thereby achieving a lower penalized objective while simultaneously satisfying all admissible accuracy thresholds. It is particularly noteworthy that the Stage 1 Config 2 block retained by ADOPT-FPML contains 84 trainable parameters, compared with 124 parameters in the full-size Config 2 baseline. Although the subsequent Config 1 block adds 58 parameters, increasing the total to 142, the composite architecture still achieves a lower global $\mathrm{AIC}_c$, demonstrating that reusing selected intermediate predictions from the first hybrid block within a complementary downstream correction structure can provide a more favorable accuracy-complexity trade-off than increasing model capacity within a single fixed configuration.

\begin{table}[htbp]
\centering
\caption{Comparison of the optimal ADOPT-FPML architecture with full-size single-stage baseline configurations for Envelope E1.}
\label{tab:e1_full_baselines}
\footnotesize
\setlength{\tabcolsep}{2.5pt}
\renewcommand{\arraystretch}{1.08}

\begin{tabularx}{\textwidth}{
@{}>{\centering\arraybackslash}X |
>{\centering\arraybackslash}X |
c | c | c | c | c | c @{} }
\hline
Model & Structure & ML parameters & Global $\mathrm{AIC}_c$ &
$C_A$ RMSE & $C_B$ RMSE & $C_C$ RMSE & $T_{\mathrm{out}}$ RMSE \\
\hline

ADOPT-FPML & Config 2 $\rightarrow$ Config 1 & $84+58=142$ & 5502 &
2.9 & 1.5 & 1.4 & 6.4 \\

Full Config 1 & Config 1 & 92 & 13369 & 
11.3 & 8.3 & 8.6 & 51.3 \\

Full Config 2 & Config 2 & 124 & 6666 &
4.8 & 1.5 & 1.7 & 8.4 \\

Full Config 3 & Config 3 & 65 & 24417 &
15.0 & 15.7 & 17.3 & 865.4 \\

Full Config 4 & Config 4 & 97 & 27508 &
11.9 & 13.1 & 14.9 & 1842.8 \\
\hline
\end{tabularx}

\normalsize
\end{table}

The remaining full-size baseline configurations exhibit substantially larger global $\mathrm{AIC}_c$ values, reflecting either inadequate predictive performance, excessive model complexity relative to the achieved accuracy, or both. These results demonstrate that the benefit of the proposed framework does not arise solely from minimizing prediction error or minimizing the number of parameters independently, but from optimizing their combined trade-off through the selected information-theoretic objective. Accordingly, the architecture reported in Figure \ref{fig:e1_optimal_structure} should be interpreted as optimal with respect to the objective function adopted in this work. Alternative objective functions could lead to different structural selections; for example, an objective based solely on maximizing likelihood or minimizing prediction error would place no explicit penalty on model complexity and could therefore favor a different hybrid architecture.

\subsection{Case Study 2: Envelope E2}

For Envelope E2, the output-specific RMSE thresholds are selected as $\varepsilon_{p,F_{\mathrm{vap}}}=0.005~\mathrm{mol/min}$, $\varepsilon_{p,F_{\mathrm{liq}}}=0.010~\mathrm{mol/min}$, $\varepsilon_{p,T_{\mathrm{vap}}}=20~\mathrm{K}$, $\varepsilon_{p,T_{\mathrm{liq}}}=20~\mathrm{K}$, $\varepsilon_{p,y_A}=0.010$, $\varepsilon_{p,y_B}=0.050$, $\varepsilon_{p,y_C}=0.010$, $\varepsilon_{p,x_A}=0.050$, $\varepsilon_{p,x_B}=0.010$, and $\varepsilon_{p,x_C}=0.050$. The maximum number of propagated synthetic inputs is retained at $N_I=10$. In contrast to Case Study 1, a relative global objective tolerance of $\tau_{\mathrm{AIC}}=0.05$ is employed, such that an additional superstructure block may be retained when its global $\mathrm{AIC}_c$ is lower than that of the preceding stage or increases by less than 5\%. This tolerance applies only to the acceptance of successive stages; candidate configurations and input sets within each individual stage continue to be ranked strictly according to the minimum local $\mathrm{AIC}_c$.

As defined in Eq. \eqref{eq:e2_output_vector}, Envelope E2 contains 10 desired outputs. Among these, $x_C$ is already predicted by the standalone FP model within its admissible accuracy threshold, with an RMSE value of approximately $3.5\times10^{-2}$ for training data as compared with $\varepsilon_{p,x_C}=0.05$. Therefore, $x_C$ does not require hybridization and is obtained directly from the FP model. The remaining 9 outputs are supplied to the first stage of the ADOPT-FPML approach. Table \ref{tab:e2_stage1} summarizes the Stage 1 branch-and-prune search over the 9 outputs that remain outside their desired accuracy thresholds after standalone FP screening. 

\begin{table}[!htbp]
\centering
\caption{Stage 1 branch-and-prune search for Envelope E2. Downward and upward arrows denote a decrease and an increase, respectively, in local $\mathrm{AIC}_c$ following sequential inclusion of the next ranked candidate input.}
\label{tab:e2_stage1}
\footnotesize
\setlength{\tabcolsep}{3pt}
\renewcommand{\arraystretch}{1.08}

\begin{tabularx}{\textwidth}{
@{} c | >{\centering\arraybackslash}X | c |
>{\centering\arraybackslash}X @{}}
\hline
Configuration & Selected ML inputs & Local $\mathrm{AIC}_c$ & Outputs satisfying thresholds \\
\hline

Config 1 &
$T_{\mathrm{out}}$ &
30532 &
$F_{\mathrm{vap}},F_{\mathrm{liq}},y_A,y_B,x_A,x_B$ \\

Config 1 &
$T_{\mathrm{out}},C_A$ &
28551 $\downarrow$ &
$F_{\mathrm{vap}},F_{\mathrm{liq}},y_A,y_B,x_A,x_B$ \\

Config 1 &
$T_{\mathrm{out}},C_A,C_B$ &
28323 $\downarrow$ &
$F_{\mathrm{vap}},F_{\mathrm{liq}},y_A,y_B,x_A,x_B$ \\

Config 1 &
$T_{\mathrm{out}},C_A,C_B,C_C$ &
28302 $\downarrow$ &
$F_{\mathrm{vap}},F_{\mathrm{liq}},y_A,y_B,x_A,x_B$ \\

Config 1 &
$T_{\mathrm{out}},C_A,C_B,C_C,v$ &
14504 $\downarrow$ &
$F_{\mathrm{vap}},F_{\mathrm{liq}},T_{\mathrm{vap}},T_{\mathrm{liq}},y_A,y_B,x_A,x_B$ \\

\hline

Config 2 &
$x_{A,\mathrm{FP}}$ &
15531 &
$F_{\mathrm{vap}},T_{\mathrm{vap}},T_{\mathrm{liq}},y_B,y_C,x_A$ \\

Config 2 &
$x_{A,\mathrm{FP}},x_{C,\mathrm{FP}}$ &
11781 $\downarrow$ &
$F_{\mathrm{vap}},T_{\mathrm{vap}},T_{\mathrm{liq}},y_B,y_C,x_A,x_B$ \\

Config 2 &
$x_{A,\mathrm{FP}},x_{C,\mathrm{FP}},T_{\mathrm{out}}$ &
12773 $\uparrow$ &
$F_{\mathrm{vap}},T_{\mathrm{vap}},T_{\mathrm{liq}},y_B,y_C,x_A,x_B$ \\

\hline

Config 3 &
$T_{\mathrm{out}}$ &
35497 &
$F_{\mathrm{vap}},F_{\mathrm{liq}},y_A,y_B,y_C,x_A$ \\

Config 3 &
$T_{\mathrm{out}},C_A$ &
35513 $\uparrow$ &
$F_{\mathrm{vap}},F_{\mathrm{liq}},y_A,y_B,y_C,x_A$ \\

\hline

Config 4 &
$x_{A,\mathrm{FP}}$ &
35497 &
$F_{\mathrm{vap}},F_{\mathrm{liq}},y_A,y_B,y_C,x_A$ \\

Config 4 &
$x_{A,\mathrm{FP}},x_{C,\mathrm{FP}}$ &
35513 $\uparrow$ &
$F_{\mathrm{vap}},F_{\mathrm{liq}},y_A,y_B,y_C,x_A$ \\

\hline
\end{tabularx}
\normalsize
\end{table}

For each disjunctive configuration, candidate ML inputs are introduced sequentially according to the input-selection strategy in Section \ref{sec:input_selection}, and the resulting models are ranked using the corresponding local $\mathrm{AIC}_c$. The arrows indicate whether addition of the next candidate input improves or worsens the local objective within that branch, while the final column identifies the outputs satisfying their admissible thresholds for each candidate model. Among the Stage 1 candidates, Config 2 yields the minimum local $\mathrm{AIC}_c$ (around 11,781) using the FP-generated inputs $x_{A,\mathrm{FP}}$ and $x_{C,\mathrm{FP}}$. The resulting model satisfies the admissible thresholds for 7 of the 9 outputs participating in the hybridization problem, namely, $F_{\mathrm{vap}}$, $T_{\mathrm{vap}}$, $T_{\mathrm{liq}}$, $y_B$, $y_C$, $x_A$, and $x_B$. The remaining output set propagated to Stage 2 is therefore $\mathcal{J}^{(2)}=\{F_{\mathrm{liq}},y_A\}$. Following the Stage 1 reduction, only $F_{\mathrm{liq}}$ and $y_A$ remain outside their admissible prediction-accuracy thresholds and are therefore propagated to the second-stage optimization problem. Table \ref{tab:e2_stage2} summarizes the corresponding Stage 2 branch-and-prune search over this reduced output set. 

\begin{table}[htbp]
\centering
\caption{Stage 2 branch-and-prune search for Envelope E2 over the reduced output set $\{F_{\mathrm{liq}},y_A\}$.}
\label{tab:e2_stage2}
\footnotesize
\setlength{\tabcolsep}{4pt}
\renewcommand{\arraystretch}{1.08}

\begin{tabularx}{\textwidth}{
@{} c |
>{\centering\arraybackslash}X |
c |
c |
c @{}
}
\hline
Configuration &
Selected ML inputs &
Local $\mathrm{AIC}_c$ &
RMSE in $F_{\mathrm{liq}}$ &
RMSE in $y_A$ \\
\hline

Config 1 &
$T_{\mathrm{out}}$ &
$-10112$ &
$3.3\times10^{-3}$ & $8.0\times10^{-3}$ \\

Config 1 &
$T_{\mathrm{out}},x_{A,\mathrm{FP}}$ &
$-10936~\downarrow$ &
$2.7\times10^{-3}$ & $5.1\times10^{-3}$ \\

Config 1 &
$T_{\mathrm{out}},x_{A,\mathrm{FP}},x_{A,\mathrm{stg1}}$ &
$-10599~\uparrow$ &
$2.6\times10^{-3}$ & $6.2\times10^{-3}$ \\

\hline

Config 2 &
$F_{\mathrm{liq,FP}}$ &
$-6682$ &
$6.7\times10^{-3}$ & $4.8\times10^{-2}$ \\

Config 2 &
$F_{\mathrm{liq,FP}},T_{\mathrm{out}}$ &
$-7858~\downarrow$ &
$5.0\times10^{-3}$ & $2.6\times10^{-2}$ \\

Config 2 &
$F_{\mathrm{liq,FP}},T_{\mathrm{out}},x_{A,\mathrm{FP}}$ &
$-8466~\downarrow$ &
$3.1\times10^{-3}$ & $1.9\times10^{-2}$ \\

Config 2 &
$F_{\mathrm{liq,FP}},T_{\mathrm{out}},x_{A,\mathrm{FP}},x_{A,\mathrm{stg1}}$ &
$-8828~\downarrow$ &
$3.8\times10^{-3}$ & $1.6\times10^{-2}$ \\

Config 2 &
$F_{\mathrm{liq,FP}},T_{\mathrm{out}},x_{A,\mathrm{FP}},x_{A,\mathrm{stg1}},x_{C,\mathrm{FP}}$ &
$-8243~\uparrow$ &
$5.1\times10^{-3}$ & $2.1\times10^{-2}$ \\

\hline

Config 3 &
$T_{\mathrm{out}}$ &
$-11795$ &
$7.8\times10^{-4}$ & $3.6\times10^{-3}$ \\

Config 3 &
$T_{\mathrm{out}},x_{A,\mathrm{FP}}$ &
$-13385~\downarrow$ &
$5.6\times10^{-4}$ & $1.5\times10^{-3}$ \\

Config 3 &
$T_{\mathrm{out}},x_{A,\mathrm{FP}},x_{A,\mathrm{stg1}}$ &
$-12846~\uparrow$ &
$5.0\times10^{-4}$ & $2.1\times10^{-3}$ \\

\hline

Config 4 &
$F_{\mathrm{liq,FP}}$ &
$-12294$ &
$6.2\times10^{-4}$ & $2.8\times10^{-3}$ \\

Config 4 &
$F_{\mathrm{liq,FP}},T_{\mathrm{out}}$ &
$-13684~\downarrow$ &
$5.0\times10^{-4}$ & $1.3\times10^{-3}$ \\

Config 4 &
$F_{\mathrm{liq,FP}},T_{\mathrm{out}},x_{A,\mathrm{FP}}$ &
$-13159~\uparrow$ &
$4.9\times10^{-4}$ & $1.8\times10^{-3}$ \\

\hline
\end{tabularx}
\normalsize
\end{table}

The minimum Stage 2 local $\mathrm{AIC}_c$ is obtained for Config 4 using $F_{\mathrm{liq,FP}}$ and $T_{\mathrm{out}}$ as the selected ML inputs. In this configuration, the ML-assisted parameter-correction pathway predicts time-varying values of the effective FP-model parameters $UA$, $K_A$, $K_B$, and $K_C$, after which the FP model is re-evaluated to obtain the final predictions of $F_{\mathrm{liq}}$ and $y_A$. The corresponding cooler and flash-separation relations are given in Eqs. \eqref{eq:e2_cooler_temperature}-\eqref{eq:e2_rachford_rice}. The resulting RMSE values for both outputs satisfy their admissible prediction thresholds.

The Stage 2 block is retained even though the resulting global $\mathrm{AIC}_c$ increases from approximately 12539 after Stage 1 to 12665 after Stage 2. The relative increase is approximately 1.0\%, which remains below the admissible stage-retention tolerance of $\tau_{\mathrm{AIC}}=0.05$. As a result, the second stage is accepted. If $\tau_{\mathrm{AIC}}=0$ was instead imposed, corresponding to the strictly decreasing global $\mathrm{AIC}_c$ criterion used for E1, the algorithm would terminate after Stage 1 and $F_{\mathrm{liq}}$ and $y_A$ would remain outside their desired prediction-accuracy thresholds. This result illustrates the role of $\tau_{\mathrm{AIC}}$ in balancing strict improvement of the global penalized objective against further improvement of outputs that remain inadequately predicted after an earlier stage.

\begin{figure}[htbp]
    \centering
    \includegraphics[width=0.9\textwidth]{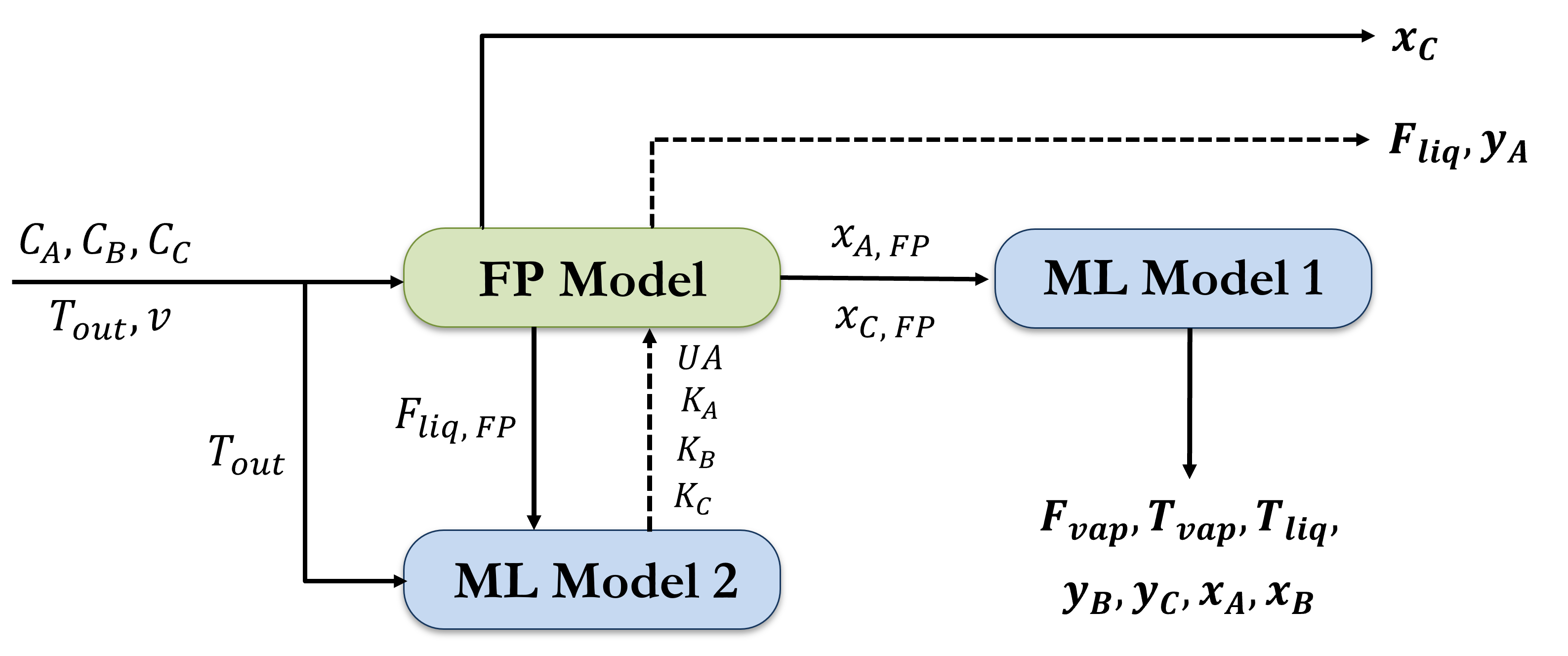}
    \caption{Optimal composite hybrid FPML architecture discovered by ADOPT-FPML for Envelope E2.}
    \label{fig:e2_optimal_structure}
\end{figure}

The resulting optimal composite architecture, consisting of a Stage 1 Config 2 block followed by a Stage 2 Config 4 block, is shown in Figure \ref{fig:e2_optimal_structure}. The output $x_C$ is retained directly from the standalone FP model and therefore does not participate in either hybridization stage.

The optimal ADOPT-FPML architecture improves all 9 outputs participating in hybridization to within their respective admissible training thresholds, while $x_C$ remains directly predicted by the standalone FP model. Significant improvements are also obtained for the unseen testing data. For example, the training RMSE for $T_{\mathrm{vap}}$ decreases from $63.7$ to $7.1~\mathrm{K}$, while the corresponding testing RMSE decreases from $29.4$ to $6.2~\mathrm{K}$. The complete output-wise training and testing RMSE values, together with the admissible thresholds, are reported in Table \ref{tab:app_e2_train_test_rmse} in Appendix \ref{app:e2}. Figure \ref{fig:e2_train_test_main} illustrates representative training and testing predictions for $F_{\mathrm{liq}}$, $T_{\mathrm{vap}}$, and $x_A$. These variables provide examples of flow-rate, thermal, and composition predictions obtained from the discovered hybrid architecture.

Table \ref{tab:e2_full_baselines} further demonstrates the importance of considering the prescribed output-specific prediction requirements in addition to the structural objective. Among the full-size baseline configurations, Full Config 1 violates the prescribed threshold for $y_C$, Full Config 2 violates the threshold for $y_A$, and Full Config 3 and Full Config 4 exhibit substantial errors in both temperature outputs. In contrast, the automatically discovered ADOPT-FPML architecture is the only model among those considered that satisfies the prescribed thresholds for all 10 desired outputs. 

\begin{figure}[!htbp]
    \centering
    \includegraphics[width=0.98\textwidth]{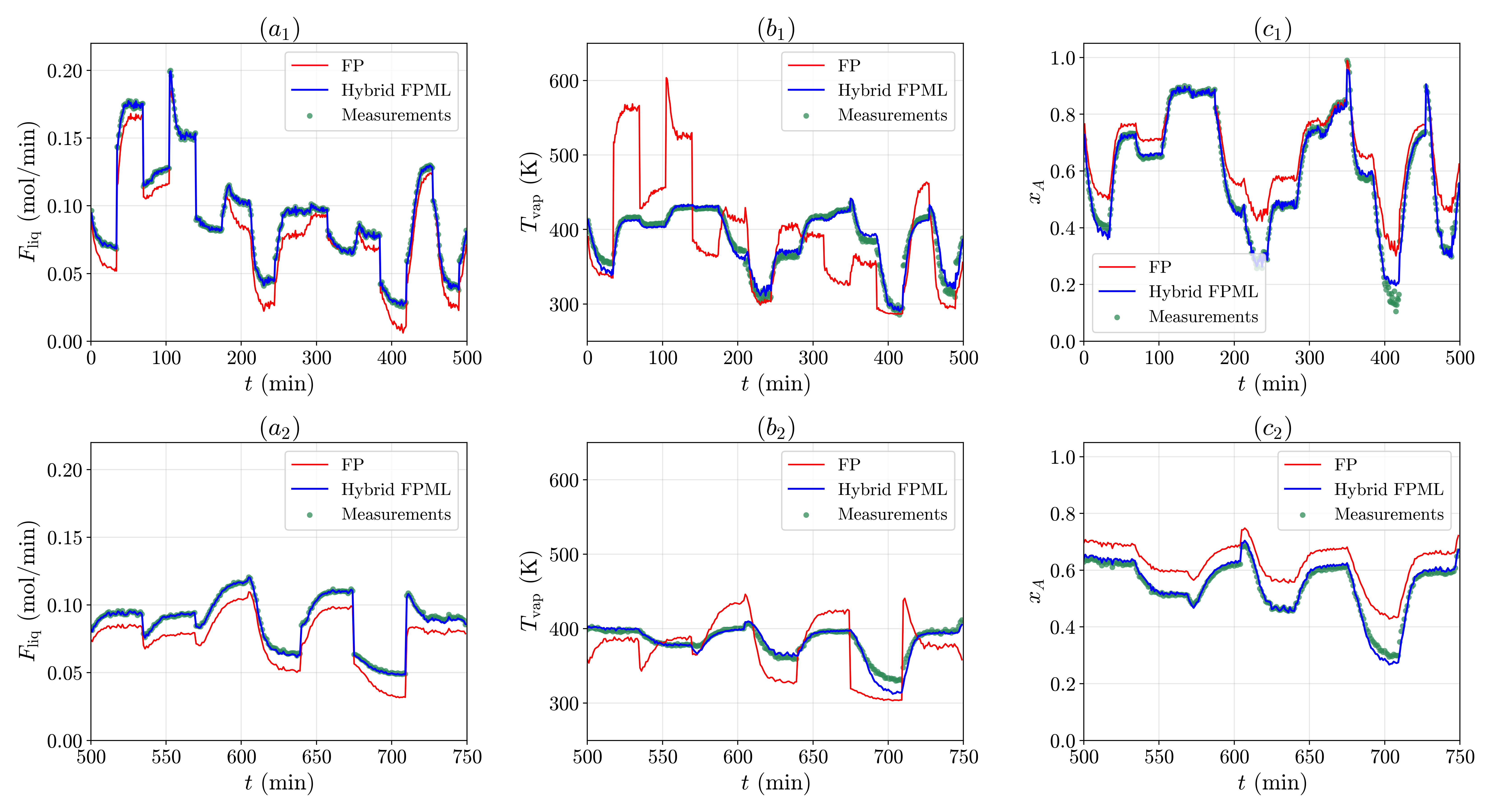}
    \caption{Training and testing performance of the optimal hybrid FPML architecture for Envelope E2: ($a_1$) training and ($a_2$) testing predictions for $F_{\mathrm{liq}}$, ($b_1$) training and ($b_2$) testing predictions for $T_{\mathrm{vap}}$, and ($c_1$) training and ($c_2$) testing predictions for $x_A$.}
    \label{fig:e2_train_test_main}
\end{figure}

For the corresponding structural comparison, the optimal ADOPT-FPML architecture contains 165 trainable ML parameters and yields a global $\mathrm{AIC}_c$ of 12665. The full-size single-stage implementations of Config 1, Config 2, Config 3, and Config 4 contain 138, 218, 84, and 164 trainable ML parameters, respectively, with corresponding global $\mathrm{AIC}_c$ values of 19255, 12307, 38968, and 39769. Although Full Config 2 yields a slightly lower global $\mathrm{AIC}_c$ than the selected ADOPT-FPML architecture, its prediction of $y_A$ remains outside the admissible threshold and therefore it does not satisfy the complete set of output-specific requirements considered in this case study. Moreover, Full Config 2 contains 218 trainable parameters compared with 165 parameters in the selected composite architecture. As an auxiliary comparison, increasing the ML model size within Full Config 2 until the $y_A$ threshold is also satisfied increases the number of trainable parameters to 296 and yields a global $\mathrm{AIC}_c$ of approximately 12892, which is higher than the value of 12665 obtained for ADOPT-FPML. Thus, under the consistent modeling basis adopted in this work, the stage-wise architecture provides a more favorable combination of output-wise admissibility, predictive fidelity, and model complexity.

\begin{table}[htbp]
\centering
\caption{Comparison of output-specific training RMSE values for the optimal ADOPT-FPML architecture and full-size single-stage baseline configurations for Envelope E2. Check and cross symbols indicate whether the corresponding prediction satisfies the prescribed RMSE threshold.}
\label{tab:e2_full_baselines}
\scriptsize
\setlength{\tabcolsep}{3pt}
\renewcommand{\arraystretch}{1.08}

\begin{tabularx}{\textwidth}{
@{}
>{\centering\arraybackslash}X |
c |
c |
c |
c |
c |
c
@{}
}
\hline
Output &
$\varepsilon_{p,j}$ &
ADOPT-FPML &
Full Config 1 &
Full Config 2 &
Full Config 3 &
Full Config 4 \\
\hline

$F_{\mathrm{vap}}~(\mathrm{mol/min})$ &
0.005 &
$4.3\times10^{-3}\,\checkmark$ &
$1.3\times10^{-3}\,\checkmark$ &
$2.9\times10^{-3}\,\checkmark$ &
$8.0\times10^{-4}\,\checkmark$ &
$8.9\times10^{-4}\,\checkmark$ \\

$F_{\mathrm{liq}}~(\mathrm{mol/min})$ &
0.010 &
$5.1\times10^{-4}\,\checkmark$ &
$1.4\times10^{-3}\,\checkmark$ &
$6.3\times10^{-3}\,\checkmark$ &
$8.5\times10^{-4}\,\checkmark$ &
$9.4\times10^{-4}\,\checkmark$ \\

$T_{\mathrm{vap}}~(\mathrm{K})$ &
20.0 &
$7.1\,\checkmark$ &
$16.0\,\checkmark$ &
$5.0\,\checkmark$ &
$108.3\,\times$ &
$115.3\,\times$ \\

$T_{\mathrm{liq}}~(\mathrm{K})$ &
20.0 &
$8.2\,\checkmark$ &
$13.7\,\checkmark$ &
$5.8\,\checkmark$ &
$108.2\,\times$ &
$115.3\,\times$ \\

$y_A$ &
0.010 &
$1.3\times10^{-3}\,\checkmark$ &
$6.2\times10^{-3}\,\checkmark$ &
$1.5\times10^{-2}\,\times$ &
$2.3\times10^{-3}\,\checkmark$ &
$2.8\times10^{-3}\,\checkmark$ \\

$y_B$ &
0.050 &
$2.1\times10^{-2}\,\checkmark$ &
$1.3\times10^{-2}\,\checkmark$ &
$1.3\times10^{-2}\,\checkmark$ &
$5.5\times10^{-3}\,\checkmark$ &
$6.4\times10^{-3}\,\checkmark$ \\

$y_C$ &
0.010 &
$5.4\times10^{-3}\,\checkmark$ &
$1.5\times10^{-2}\,\times$ &
$5.0\times10^{-3}\,\checkmark$ &
$3.7\times10^{-3}\,\checkmark$ &
$4.0\times10^{-3}\,\checkmark$ \\

$x_A$ &
0.050 &
$1.6\times10^{-2}\,\checkmark$ &
$8.5\times10^{-3}\,\checkmark$ &
$3.4\times10^{-2}\,\checkmark$ &
$4.2\times10^{-3}\,\checkmark$ &
$5.0\times10^{-3}\,\checkmark$ \\

$x_B$ &
0.010 &
$8.4\times10^{-3}\,\checkmark$ &
$7.3\times10^{-3}\,\checkmark$ &
$7.8\times10^{-3}\,\checkmark$ &
$6.3\times10^{-3}\,\checkmark$ &
$7.2\times10^{-3}\,\checkmark$ \\

$x_C$ &
0.050 &
$3.5\times10^{-2}\,\checkmark$ &
$9.9\times10^{-3}\,\checkmark$ &
$2.2\times10^{-2}\,\checkmark$ &
$2.4\times10^{-3}\,\checkmark$ &
$2.5\times10^{-3}\,\checkmark$ \\

\hline
\end{tabularx}

\normalsize
\end{table}

For this case study, the complete ADOPT-FPML search for Envelope E2 requires approximately 35.4 s and evaluates 26 candidate hybrid structures across the two stages. For comparison, the corresponding E1 search requires approximately 22.4 s while evaluating 29 candidates. Thus, despite requiring fewer candidate evaluations, E2 incurs a moderately larger computational cost, which can be attributed in part to the larger output dimensionality of the E2 hybridization problem and the resulting increase in the number of trainable parameters in the corresponding ML models. 

Nevertheless, the branch-and-prune strategy substantially reduces the structural search space. At the beginning of Stage 1, the candidate library can contain the 5 measured inputs in Eq. \eqref{eq:e2_input_vector} together with the 10 variables predicted by the FP model, yielding up to 15 candidate variables before structural and output-specific restrictions are imposed. Exhaustive enumeration of all non-empty subsets of this library would therefore require $2^{15}-1=32767$ candidate input subsets for a single unrestricted configuration alone. This count does not include the additional alternatives associated with the 4 disjunctive configurations, output-selection decisions, or subsequent superstructure blocks, highlighting the substantial reduction obtained by the proposed stage-wise branch-and-prune strategy. Additional training and testing comparisons for $T_{\mathrm{liq}}$, $y_C$, and $x_C$ are provided in Figure \ref{fig:app_e2_additional_train_test} in Appendix \ref{app:e2}; notably, $x_C$ is retained directly from the standalone FP model because its prediction already satisfies the prescribed accuracy threshold and therefore does not require hybridization.

The identified hybrid FPML structure should, however, be interpreted as optimal with respect to the modeling basis and objective function adopted in this work. The FP model definition, ML model class and architecture, candidate feature library, training procedure, and information-theoretic objective collectively define the structural optimization problem. Improvements or modifications to any of these components, such as the use of a higher-fidelity FP representation, a more expressive ML architecture, or an alternative objective function, may alter the relative performance of the candidate configurations and consequently lead to a different optimal hybrid structure.

\subsection{Case Study 3: Industrial Superheater System}

The final case study evaluates ADOPT-FPML using industrial operating data from the NGCC superheater system \cite{saini2024nonlinear} described in Section \ref{sec:case_study_3}. In addition to assessing automatic structure discovery, this case study also enables direct comparison with the hybrid FPML models previously developed for the same system using heuristic structure selection and all available input variables \cite{mukherjee2025hybridfpai}. As discussed in Section \ref{sec:case_study_3}, the present structural search is restricted to Config 1 and Config 2 because the parameters of the available FP model are assumed to be sufficiently well characterized and are not considered candidates for ML-assisted correction.

For this study, 3 prediction scenarios are considered to examine how the admissibility criterion influences the automatically discovered FPML architecture through ADOPT-FPML. {\textit{Scenario 1}} adopts the same temperature-specific admissibility requirements considered in our previous work \cite{mukherjee2025hybridfpai}: an RMSE not exceeding $5~^{\circ}\mathrm{C}$ together with a maximum absolute instantaneous error \cite{mukherjee2025nnvsbml, das2026bcd} (MAIE) not exceeding $10~^{\circ}\mathrm{C}$ for each desired temperature output. Here, MAIE is defined as the maximum absolute deviation between the predicted and measured temperature at any individual time point over the considered horizon. In contrast to RMSE, which characterizes the overall prediction error across the complete trajectory, MAIE explicitly captures the worst-case instantaneous error, and is therefore critically relevant when evaluating dynamic models for process monitoring applications. {\textit{Scenario 2}} retains only the RMSE criterion of $5~^{\circ}\mathrm{C}$, whereas {\textit{Scenario 3}} imposes a more restrictive RMSE threshold of $3~^{\circ}\mathrm{C}$. Thus, the 3 modeling scenarios provide a systematic assessment of how the desired prediction criterion influences FP screening, output selection, input selection, and the resulting optimal hybrid structure.

For all 3 scenarios, the maximum dimensionality of the propagated synthetic input space is set to $N_I=10$. For Scenario 1, $\tau_{\mathrm{AIC}}=0.05$ is adopted, similar to the tolerance used for Envelope E2, such that a subsequent block may be retained when the relative increase in the global $\mathrm{AIC}_c$ remains below $5\%$ and at least one additional output satisfies the prescribed accuracy criterion. For Scenarios 2 and 3, $\tau_{\mathrm{AIC}}=0$ is used, corresponding to a strictly decreasing global $\mathrm{AIC}_c$ requirement for retaining any additional stage.

\subsubsection{Scenario 1: Consistent RMSE and MAIE Thresholds}

Scenario 1 provides the closest comparison with the modeling requirements adopted previously for this industrial system \cite{mukherjee2025hybridfpai}. The admissible prediction criterion requires both $\mathrm{RMSE}\leq 5~^{\circ}\mathrm{C}$ and $\mathrm{MAIE}\leq 10~^{\circ}\mathrm{C}$ for a temperature output to be considered satisfactorily (accurately) predicted. The screening of the standalone FP model shows that only $T_{\mathrm{st,out}}$ satisfies both conditions, with an RMSE of approximately $0.9~^{\circ}\mathrm{C}$ and an MAIE of approximately $3.5~^{\circ}\mathrm{C}$. The remaining 7 outputs are therefore supplied to the first ADOPT-FPML superstructure block.

\begin{figure}[htbp]
    \centering
    \includegraphics[width=0.98\textwidth]{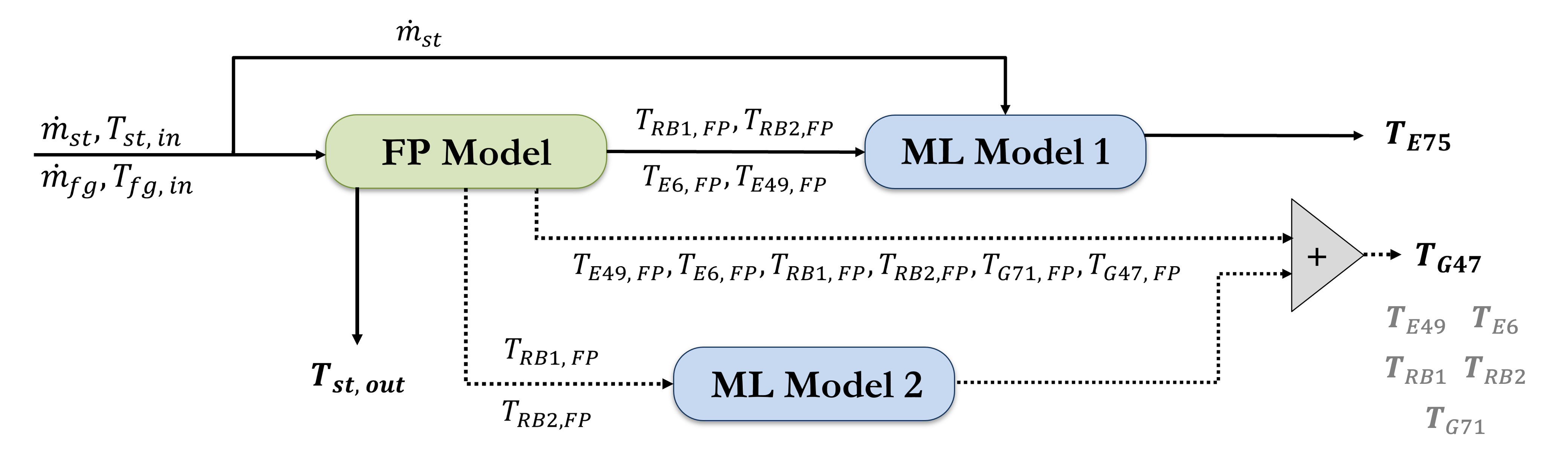}
    \caption{Optimal composite hybrid FPML architecture discovered by ADOPT-FPML under Scenario 1 of Case Study 3.}
    \label{fig:sh_optimal_scenario1}
\end{figure}

At Stage 1, the minimum local $\mathrm{AIC}_c$ among the candidate structures is obtained for Config 2, corresponding to the Series I $(\mathrm{FP}\rightarrow\mathrm{ML})$ interaction structure. The selected ML inputs are $T_{\mathrm{RB2,FP}}$, $T_{\mathrm{RB1,FP}}$, $T_{\mathrm{E6,FP}}$, $\dot{m}_{\mathrm{st}}$, and $T_{\mathrm{E49,FP}}$, as shown in Figure \ref{fig:sh_optimal_scenario1}. Among the 7 outputs participating in this block, only $T_{\mathrm{E75}}$ satisfies both the desired RMSE and MAIE thresholds at Stage 1. The reduced synthetic intermediate variable set propagated to Stage 2 consequently contains the remaining 6 outputs.  

The Stage 2 optimization is then performed over this reduced output set using the updated candidate input library. The minimum local $\mathrm{AIC}_c$ is obtained for Config 1 using $T_{\mathrm{RB2,FP}}$ and $T_{\mathrm{RB1,FP}}$ as the selected ML inputs. This block additionally resolves $T_{\mathrm{G47}}$. The Stage 2 block satisfies the prescribed stage-retention criterion for $\tau_{\mathrm{AIC}}=0.05$ and is therefore incorporated into the composite architecture. No admissible candidate at the subsequent stage resolves any additional output under the combined RMSE and MAIE requirement, and the stage-wise search therefore terminates. In Figure \ref{fig:sh_optimal_scenario1}, the grayed-out final output variables indicate temperatures that are not satisfactorily predicted under the prescribed admissible threshold criteria.

For comparison, the full-input Config 1 considers all 4 measured process inputs and is trained against all 8 desired outputs using a single Config 1 block. Similarly, the full Config 2 baseline uses all 4 measured inputs together with the 8 corresponding FP-predicted temperatures, resulting in 12 ML inputs. These full baseline models respectively refer to the parallel and series model structures defined in our previous paper \cite{mukherjee2025hybridfpai}. As summarized in Table \ref{tab:e3_s1_structural}, for this case study, none of the 3 candidate models satisfies the combined Scenario 1 criterion for all 8 outputs. The ADOPT-FPML architecture satisfies the criterion for $T_{\mathrm{st,out}}$, $T_{\mathrm{E75}}$, and $T_{\mathrm{G47}}$, while the full Config 1 satisfies it only for $T_{\mathrm{st,out}}$ and the full Config 2 satisfies it for $T_{\mathrm{st,out}}$, $T_{\mathrm{E6}}$, $T_{\mathrm{E75}}$, and $T_{\mathrm{RB1}}$.

\begin{table}[htbp]
\centering
\caption{Structural comparison of the candidate hybrid FPML models for Case Study 3 under Scenario 1. An output is considered satisfactory only when both $\mathrm{RMSE}\leq 5~^{\circ}\mathrm{C}$ and $\mathrm{MAIE}\leq 10~^{\circ}\mathrm{C}$ are satisfied.}
\label{tab:e3_s1_structural}
\small
\setlength{\tabcolsep}{5pt}
\renewcommand{\arraystretch}{1.08}
\begin{tabularx}{\textwidth}{
@{}
>{\centering\arraybackslash}X |
>{\centering\arraybackslash}X |
c |
c |
>{\centering\arraybackslash}X
@{}
}
\hline
Model &
Hybrid structure &
No. of parameters &
Global $\mathrm{AIC}_c$ &
Outputs satisfying threshold criteria \\
\hline

ADOPT-FPML &
Config 2 $\rightarrow$ Config 1 &
189 &
11,616 &
$T_{\mathrm{st,out}},\,T_{\mathrm{E75}},\,T_{\mathrm{G47}}$ \\

Full Config 1 &
Config 1 &
112 &
13,330 &
$T_{\mathrm{st,out}}$ \\

Full Config 2 &
Config 2 &
176 &
11,866 &
$T_{\mathrm{st,out}},\,T_{\mathrm{E6}},\,T_{\mathrm{E75}},\,T_{\mathrm{RB1}}$ \\

\hline
\end{tabularx}
\normalsize
\end{table}

It is worth mentioning that the purpose of the structural comparison is not solely to maximize the number of outputs satisfying the desired thresholds, but to identify the FPML structure that provides the most favorable information-theoretic trade-off between predictive accuracy and model complexity under the adopted modeling algorithm. Using the same underlying ML model class and training basis for all candidate structures, the automatically discovered ADOPT-FPML architecture yields the lowest global $\mathrm{AIC}_c$ of 11,616, compared with 11,866 for the full Config 2 and 13,330 for the full Config 1 cases. The ADOPT-FPML architecture contains 189 trainable parameters, which is comparable in magnitude to the 176 parameters required by the full Config 2 model. Therefore, the lower global $\mathrm{AIC}_c$ indicates that the improvement in overall predictive likelihood obtained through the selected output allocation, input selection, and composite Config 2 $\rightarrow$ Config 1 interaction structure is sufficient to offset the associated model complexity. Thus, even under a common ML modeling basis for which the stringent MAIE requirement cannot be satisfied for every output, automated structural selection provides a more favorable overall accuracy-complexity trade-off than the corresponding full-size single-configuration baselines.

The remaining violations of the MAIE criterion across the candidate models should also be interpreted in the context of the type and efficiency of the ML models adopted in the present study. In our previous work \cite{mukherjee2025hybridfpai}, the analysis of the same industrial superheater system employed more specialized ML modeling approaches, such as hybrid series and parallel all-nonlinear static-dynamic NNs \cite{mukherjee2023hybserpar} and a Bayesian ML framework \cite{adeyemo2024bidsam}, that were developed specifically for accurate prediction of highly complex nonlinear transient process data. In contrast, the present study employs a significantly simpler single hidden-layered feedforward NN so that the effects of FPML structure, input selection, and output selection can be investigated on a consistent basis across different candidates without introducing additional complexity associated with the underlying ML architecture. While the NN model considered within ADOPT-FPML produces satisfactory RMSE values, it is less effective at reproducing some of the sharp transient overshoots and undershoots in the temporal temperature profiles of the industrial system. As a result, the corresponding worst-case errors obtained from these models remain comparatively large, thus making MAIE the limiting admissibility criterion for several output variables across all 3 candidate structures. One objective of this scenario analysis is also to convey that it cannot be expected that ADOPT-FPML would be able to satisfy the desired accuracy thresholds of all outputs. For such cases, users need to consider additional candidate ML models and/or enhancement of the FP model and/or lack of sufficient measured inputs for the ML model and/or uncertainty in measurements. 

\subsubsection{Scenario 2: Only RMSE Threshold of $5~^{\circ}\mathrm{C}$}

The results obtained for Scenario 1 illustrate that, when none of the candidate hybrid configurations can satisfy all prescribed admissibility requirements, two general alternatives are generally available. In the first approach, the underlying ML model may be enhanced, for example, by increasing the model size through additional hidden layers, modifying the training settings, tuning the associated hyperparameters, or adopting a more effective class of the ML model. Alternatively, if such additional model complexity is undesirable or unnecessary for the intended application, the prescribed admissibility requirements may be slightly relaxed. As an illustrative example of the latter case, Scenario 2 removes the MAIE requirement and considers an output satisfactorily predicted when its RMSE does not exceed $5~^{\circ}\mathrm{C}$. Under this criterion, the standalone FP model already provides admissible predictions for 6 of the 8 desired outputs: $T_{\mathrm{st,out}}$, $T_{\mathrm{E75}}$, $T_{\mathrm{E49}}$, $T_{\mathrm{E6}}$, $T_{\mathrm{RB2}}$, and $T_{\mathrm{G47}}$. Only $T_{\mathrm{RB1}}$ and $T_{\mathrm{G71}}$, with standalone FP RMSE values of approximately $12.7~^{\circ}\mathrm{C}$ and $8.2~^{\circ}\mathrm{C}$, respectively, therefore require hybridization.

\begin{figure}[!htbp]
    \centering
    \includegraphics[width=0.98\textwidth]{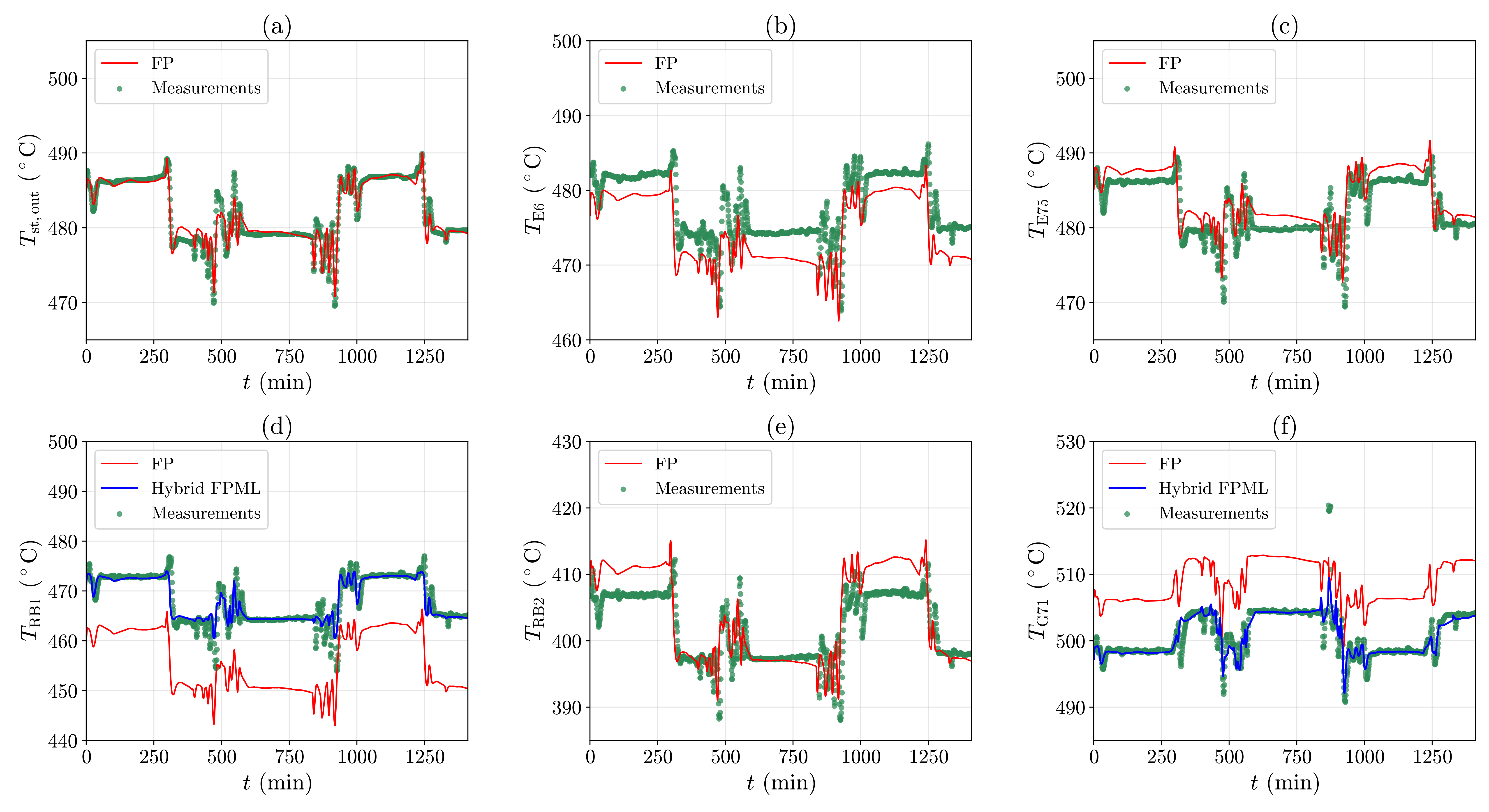}
    \caption{Results from the optimal ADOPT-FPML architecture under Scenario 2 ($\mathrm{RMSE}\leq 5~^{\circ}\mathrm{C}$): (a) $T_{\mathrm{st,out}}$, (b) $T_{\mathrm{E6}}$, (c) $T_{\mathrm{E75}}$, (d) $T_{\mathrm{RB1}}$, (e) $T_{\mathrm{RB2}}$, and (f) $T_{\mathrm{G71}}$.}
    \label{fig:sh_scenario2}
\end{figure}

ADOPT-FPML identifies a single-stage Config 2 architecture for these two outputs, consistent with the series interaction pattern selected heuristically for this system in our previous work \cite{mukherjee2025hybridfpai}. However, it is worth noting that the automated model discovery framework does not retain the complete set of available variables. The selected ML input set is reduced to $T_{\mathrm{RB2,FP}}$, $\dot{m}_{\mathrm{st}}$, $T_{\mathrm{st,in}}$, and $T_{\mathrm{RB1,FP}}$. The resulting block contains only 58 trainable parameters and yields a local $\mathrm{AIC}_c$ of approximately 3,517. The corresponding predicted RMSE values of $T_{\mathrm{RB1}}$ and $T_{\mathrm{G71}}$ decrease to approximately $2.1~^{\circ}\mathrm{C}$ and $1.5~^{\circ}\mathrm{C}$, respectively, such that both satisfy the prescribed threshold for this scenario. Since all remaining outputs are already retained directly from the standalone FP model, no additional hybrid stage is required. Figure \ref{fig:sh_scenario2} illustrates temporal temperature profiles for the optimal Scenario 2 architecture. The hybrid FPML predictions are shown only for $T_{\mathrm{RB1}}$ and $T_{\mathrm{G71}}$, which require hybridization under the prescribed RMSE criterion; the remaining displayed outputs are retained directly from the standalone FP model.

The key takeaway message from this result is that the automatically selected interaction structure is consistent with the physically motivated series configuration reported in the previous publication \cite{mukherjee2025hybridfpai}, while ADOPT-FPML additionally determines that only 2 of the 8 outputs require data-driven correction and only 4 candidate inputs need to be supplied to the ML component. Thus, the principal contribution in this case is not the discovery of a fundamentally different canonical interaction pattern, but the automated identification of a substantially more parsimonious output and input mapping within that structure.

\subsubsection{Scenario 3: Only RMSE Threshold of $3~^{\circ}\mathrm{C}$}

Scenario 3 further tightens the admissible RMSE threshold from $5~^{\circ}\mathrm{C}$ to $3~^{\circ}\mathrm{C}$. As expected, this stricter prediction requirement changes the FP screening result. Only $T_{\mathrm{st,out}}$, $T_{\mathrm{E75}}$, $T_{\mathrm{E49}}$, and $T_{\mathrm{G47}}$ remain within the admissible thresholds, whereas $T_{\mathrm{E6}}$, $T_{\mathrm{RB1}}$, $T_{\mathrm{RB2}}$, and $T_{\mathrm{G71}}$ require hybridization.

\begin{figure}[!htbp]
    \centering
    \includegraphics[width=0.98\textwidth]{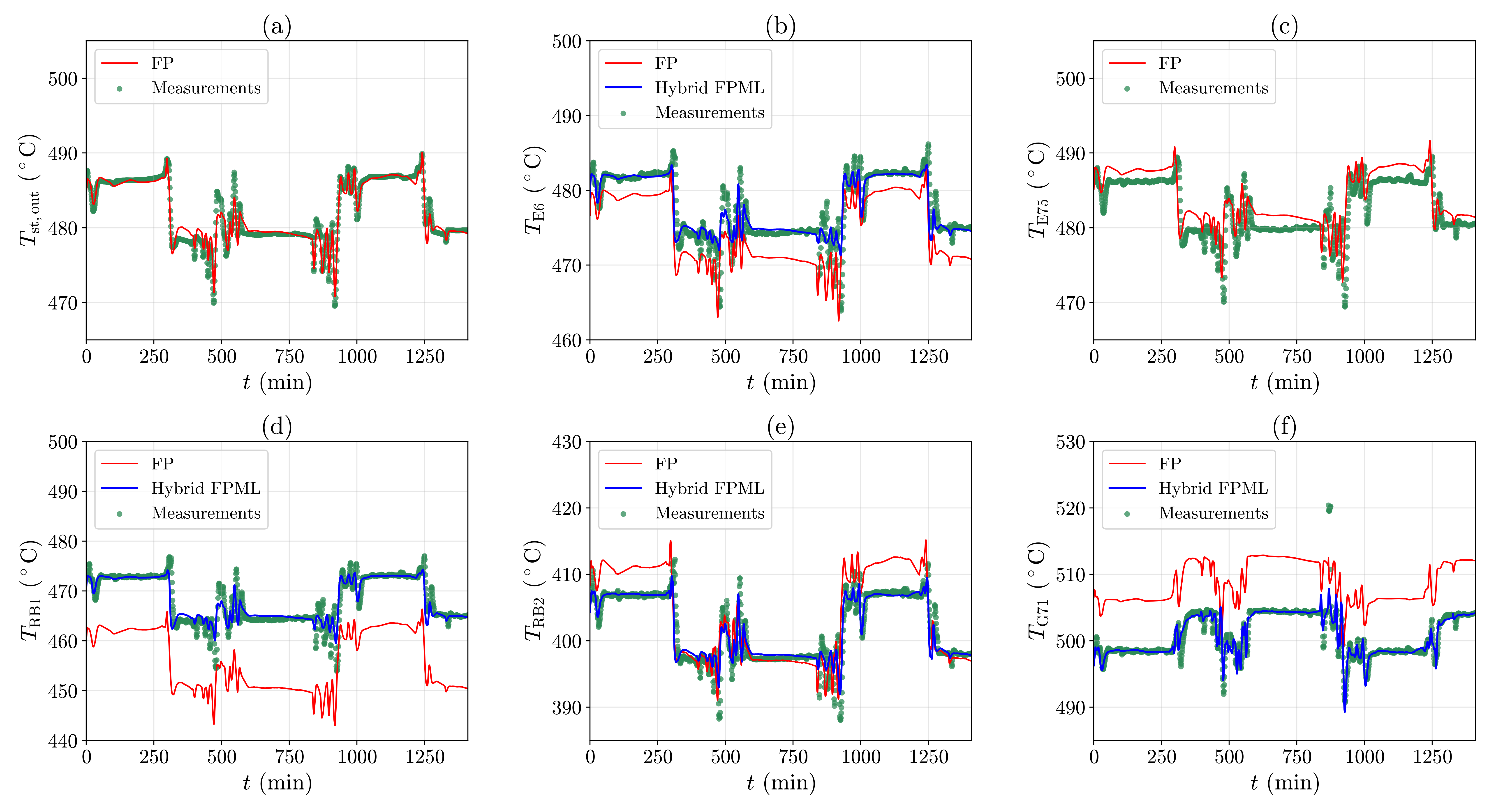}
    \caption{Results from the optimal ADOPT-FPML architecture under Scenario 3 ($\mathrm{RMSE}\leq 3~^{\circ}\mathrm{C}$): (a) $T_{\mathrm{st,out}}$, (b) $T_{\mathrm{E6}}$, (c) $T_{\mathrm{E75}}$, (d) $T_{\mathrm{RB1}}$, (e) $T_{\mathrm{RB2}}$, and (f) $T_{\mathrm{G71}}$.}
    \label{fig:sh_scenario3}
\end{figure}

The optimal model discovered by the ADOPT-FPML framework again corresponds to a single-stage Config 2 structure. However, because the stricter accuracy requirement increases the number of outputs requiring correction from 2 to 4, the automatically selected ML input set also changes. The updated optimal set of inputs are represented by $T_{\mathrm{RB1,FP}}$, $T_{\mathrm{RB2,FP}}$, $T_{\mathrm{E6,FP}}$, $T_{\mathrm{E49,FP}}$, and $T_{\mathrm{st,out,FP}}$. The resulting ML block contains 84 trainable parameters and yields a local $\mathrm{AIC}_c$ of approximately 7,483. As a result, the corresponding RMSE values for $T_{\mathrm{E6}}$, $T_{\mathrm{RB1}}$, $T_{\mathrm{RB2}}$, and $T_{\mathrm{G71}}$ are approximately $2.1$, $2.1$, $1.9$, and $1.5~^{\circ}\mathrm{C}$, respectively, and therefore satisfy the stricter $3~^{\circ}\mathrm{C}$ criterion. The effect of the stricter admissibility criterion under Scenario 3 is illustrated in Figure \ref{fig:sh_scenario3}. 

As discussed before, under the stricter RMSE criterion, $T_{\mathrm{E6}}$, $T_{\mathrm{RB1}}$, $T_{\mathrm{RB2}}$, and $T_{\mathrm{G71}}$ are supplied by the retained hybrid FPML block, while the remaining displayed outputs are retained directly from the standalone FP model. The comparison between Scenarios 2 and 3 demonstrates that the automatically discovered hybrid FPML architecture is sensitive to the desired admissibility requirement, even when the optimal canonical interaction pattern remains unchanged. In both cases, ADOPT-FPML selects Config 2, analogous to the series hybrid $(\mathrm{FP}\rightarrow\mathrm{ML})$ architecture previously discussed for this industrial superheater system through integration of mechanistic and data-driven models \cite{mukherjee2025hybridfpai}. However, changing the desired accuracy threshold modifies both the target output set and the optimal ML input library. Under Scenario 2, the $5~^{\circ}\mathrm{C}$ RMSE criterion requires hybridization of only $T_{\mathrm{RB1}}$ and $T_{\mathrm{G71}}$, and the resulting ML model contains 58 trainable parameters. In contrast, when the RMSE requirement is tightened to $3~^{\circ}\mathrm{C}$ in Scenario 3, 4 outputs require hybridization, namely, $T_{\mathrm{E6}}$, $T_{\mathrm{RB1}}$, $T_{\mathrm{RB2}}$, and $T_{\mathrm{G71}}$. The selected input set consequently changes and the resulting ML component contains 84 trainable parameters. Thus, the proposed framework adapts not only the FPML interaction structure, but also the dimensionality of the associated data-driven model according to the desired prediction fidelity. Table \ref{tab:sh_scenario_summary} summarizes the structural changes obtained across the 3 admissibility scenarios. 

\begin{table}[htbp]
\centering
\caption{Summary of the ADOPT-FPML architectures obtained for the three
admissibility scenarios considered for this case study.}
\label{tab:sh_scenario_summary}
\scriptsize
\setlength{\tabcolsep}{3pt}
\renewcommand{\arraystretch}{1.15}

\renewcommand{\tabularxcolumn}[1]{m{#1}}

\begin{tabularx}{\textwidth}{|
>{\centering\arraybackslash}m{1.2cm}|
>{\centering\arraybackslash}m{2.2cm}|
>{\centering\arraybackslash}m{3.4cm}|
>{\centering\arraybackslash}m{2.0cm}|
>{\centering\arraybackslash}X|
>{\centering\arraybackslash}m{1.4cm}|
}
\hline
\textbf{Scenario} &
\makecell[c]{\textbf{Admissibility}\\\textbf{criterion}} &
\makecell[c]{\textbf{Outputs requiring}\\\textbf{hybridization}} &
\makecell[c]{\textbf{Optimal}\\\textbf{ADOPT-FPML}\\\textbf{structure}} &
\makecell[c]{\textbf{Selected ML inputs}} &
\makecell[c]{\textbf{No. of}\\\textbf{parameters}} \\
\hline

1 &
\begin{tabular}[c]{@{}c@{}}
RMSE $\leq 5\,^{\circ}\mathrm{C}$ \\
and \\
MAIE $\leq 10\,^{\circ}\mathrm{C}$
\end{tabular}
&
\begin{tabular}[c]{@{}c@{}}
$T_{\mathrm{E75}},\,T_{\mathrm{E49}},\,T_{\mathrm{E6}},$ \\
$T_{\mathrm{RB1}},\,T_{\mathrm{RB2}},\,T_{\mathrm{G71}},\,
T_{\mathrm{G47}}$
\end{tabular}
&
\begin{tabular}[c]{@{}c@{}}
Config 2 \\
$\rightarrow$ \\
Config 1
\end{tabular}
&
\begin{tabular}[c]{@{}c@{}}
Stage 1: $T_{\mathrm{RB2,FP}},\,T_{\mathrm{RB1,FP}},\,
T_{\mathrm{E6,FP}},$ \\
$\dot{m}_{\mathrm{st}},\,T_{\mathrm{E49,FP}}$ \\
Stage 2: $T_{\mathrm{RB2,FP}},\,T_{\mathrm{RB1,FP}}$
\end{tabular}
&
189 \\
\hline

2 &
RMSE $\leq 5\,^{\circ}\mathrm{C}$
&
\begin{tabular}[c]{@{}l@{}}
$T_{\mathrm{RB1}},\,T_{\mathrm{G71}}$
\end{tabular}
&
Config 2
&
\begin{tabular}[c]{@{}c@{}}
$T_{\mathrm{RB2,FP}},\,\dot{m}_{\mathrm{st}},\,
T_{\mathrm{st,in}},\,T_{\mathrm{RB1,FP}}$
\end{tabular}
&
58 \\
\hline

3 &
RMSE $\leq 3\,^{\circ}\mathrm{C}$
&
\begin{tabular}[c]{@{}l@{}}
$T_{\mathrm{E6}},\,T_{\mathrm{RB1}},\,T_{\mathrm{RB2}},\,
T_{\mathrm{G71}}$
\end{tabular}
&
Config 2
&
\begin{tabular}[c]{@{}c@{}}
$T_{\mathrm{RB1,FP}},\,T_{\mathrm{RB2,FP}},\,
T_{\mathrm{E6,FP}},$ \\
$T_{\mathrm{E49,FP}},\,T_{\mathrm{st,out,FP}}$
\end{tabular}
&
84 \\
\hline

\end{tabularx}
\normalsize
\end{table}

Scenario 1 requires a two-stage composite Config 2 $\rightarrow$ Config 1 architecture because the combined RMSE-MAIE threshold criterion is sufficiently restrictive that only a subset of the desired outputs can be satisfactorily predicted at each stage. In contrast, Scenarios 2 and 3 both terminate after a single Config 2 block because all outputs requiring hybridization are satisfactorily resolved within the first stage. Table \ref{tab:sh_scenario_summary} provides the complete RMSE comparison for Scenarios 2 and 3. For the ADOPT-FPML models, the reported value corresponds to the final composite prediction, i.e., the standalone FP prediction is retained whenever that output already satisfies the prescribed admissibility criterion, while the hybrid FPML prediction is used only for outputs selected for correction. The full Config 1 and full Config 2 baselines, in contrast, employ ML predictions for all 8 desired outputs.

\begin{table}[htbp]
\centering
\caption{Comparison of RMSE, $\mathrm{AIC_c}$, and number of parameters for the optimal ADOPT-FPML architecture and the full Configs 1 and 2 baselines under Scenarios 2 and 3.}
\label{tab:sh_scenario23_rmse}
\scriptsize
\setlength{\tabcolsep}{4pt}
\renewcommand{\arraystretch}{1.08}

\begin{tabular}{c|ccc|ccc}
\hline
\multirow{2}{*}{\textbf {Output $(^{\circ}\mathrm{C})$}} &
\multicolumn{3}{c|}{\textbf {Scenario 2: RMSE $\leq5~^{\circ}\mathrm{C}$}} &
\multicolumn{3}{c}{\textbf {Scenario 3: RMSE $\leq3~^{\circ}\mathrm{C}$}} \\
\cline{2-7}
&
ADOPT-FPML &
Full Config 1 &
Full Config 2 &
ADOPT-FPML &
Full Config 1 &
Full Config 2 \\
\hline

$T_{\mathrm{st,out}}$
& 0.9 & 0.8 & 1.0
& 0.9 & 0.8 & 1.0 \\

$T_{\mathrm{E75}}$
& 2.7 & 1.8 & 1.6
& 2.7 & 1.8 & 1.6 \\

$T_{\mathrm{E49}}$
& 2.4 & 2.0 & 2.0
& 2.4 & 2.0 & 2.0 \\

$T_{\mathrm{E6}}$
& 4.0 & 2.0 & 1.8
& 2.1 & 2.0 & 1.8 \\

$T_{\mathrm{RB1}}$
& 2.1 & 2.0 & 1.6
& 2.1 & 2.0 & 1.6 \\

$T_{\mathrm{RB2}}$
& 3.9 & 2.1 & 1.9
& 1.9 & 2.1 & 1.9 \\

$T_{\mathrm{G71}}$
& 1.5 & 1.6 & 1.7
& 1.5 & 1.6 & 1.7 \\

$T_{\mathrm{G47}}$
& 2.3 & 1.6 & 1.6
& 2.3 & 1.6 & 1.6 \\

\hline

$\boldsymbol{\mathrm{AIC}_c}$
& 3,517 & 13,330 & 11,866
& 7,483 & 13,330 & 11,866 \\
\hline

\textbf{No. of parameters}
& 58 & 112 & 176
& 84 & 112 & 176 \\

\hline

\end{tabular}
\normalsize
\end{table}

As shown in Table \ref{tab:sh_scenario23_rmse}, all 3 candidate structures satisfy the prescribed RMSE requirements under both Scenarios 2 and 3. The distinguishing feature of the ADOPT-FPML solutions is therefore not a reduction in prediction error relative to every full-size baseline, but the automatic elimination of unnecessary data-driven modeling components. Under Scenario 2, the standalone FP model already provides satisfactory predictions for 6 of the 8 outputs, and ADOPT-FPML therefore introduces an ML component only for $T_{\mathrm{RB1}}$ and $T_{\mathrm{G71}}$. The resulting model contains 58 trainable parameters, compared with 112 and 176 parameters for the corresponding full Config 1 and full Config 2 baselines, respectively. Under the stricter Scenario 3 criterion, 4 outputs require hybridization and the ADOPT-FPML model consequently increases to 84 trainable parameters, while still remaining substantially smaller than the full baseline models.

These results further emphasize that the results obtained from ADOPT-FPML is conditional on the prescribed output-specific admissibility criteria and the stage-wise structural search defined in Section \ref{sec:optimization}. Outputs that are already predicted with sufficient accuracy by the FP model are deliberately excluded from unnecessary hybridization, even when a larger full-size ML model could further reduce their prediction errors. The framework therefore prioritizes a parsimonious formulation of data-driven models to those portions of the prediction problem for which the mechanistic (first-principles) representation is insufficient. 

\section{Conclusions} \label{sec:conclusions}

This work presents ADOPT-FPML, a first-of-a-kind systematic framework for the automatic discovery of optimal hybrid FPML model architectures. We seek to address, within a unified optimization framework, the fundamental structural question of how information should be best exchanged or transferred between an available FP model and an ML model to obtain the most suitable hybrid representation. ADOPT-FPML formulates this problem through a disjunctive superstructure optimization problem in which the interaction between FP and ML components, the outputs requiring hybridization, and the variables supplied to the ML model are treated as structural decisions. A key contribution of this work is the decomposition of the resulting MINLP problem into coupled structure selection, output selection, and input selection subproblems, which are solved using a stage-wise branch-and-prune algorithm defined through the interaction routing operators. The use of the corrected Akaike Information Criterion ($\mathrm{AIC}_c$) as the objective function further provides an information-theoretic basis for balancing predictive accuracy against the complexity introduced by additional ML inputs, outputs, parameters, and hybridization stages. Importantly, the repeated superstructure formulation (building blocks) also allows different subsets of outputs within the same system to be represented through different locally optimal FPML interaction patterns through a greedy search strategy, rather than requiring a single canonical hybrid configuration to be prescribed {\textit{a priori}}.

The case studies demonstrate the ability of the proposed approach to discover parsimonious and, in several cases, non-trivial composite hybrid structures. For Envelope E1 of the acetic anhydride production system, ADOPT-FPML identifies a Config 2 $\rightarrow$ Config 1 architecture that satisfies all prescribed training accuracy thresholds and yields a global $\mathrm{AIC}_c$ of 5,502, compared with 6,666 for the full-size Config 2 baseline. The resulting model reduces the training RMSE of $C_A$ from $33.6$ to $2.9~\mathrm{mol/m^3}$ and that of $T_{\mathrm{out}}$ from $149.8$ to $6.4~\mathrm{K}$. Moreover, only 29 candidate models are evaluated during the complete E1 search, whereas exhaustive enumeration of the 10-variable Stage 1 input library alone would require 1023 non-empty input subsets for a single unrestricted configuration. For Envelope E2, the automatically discovered Config 2 $\rightarrow$ Config 4 architecture is the only candidate among the considered baseline structures that satisfies the prescribed accuracy requirements for all 10 outputs. It contains 165 ML parameters, whereas extending the full Config 2 baseline until the remaining $y_A$ accuracy requirement is also satisfied requires 296 parameters and results in a higher global $\mathrm{AIC}_c$ of approximately 12,892 compared with 12,665 for ADOPT-FPML. Finally, the industrial superheater study demonstrates that the optimal architecture also depends directly on the desired admissibility criterion. For RMSE thresholds of $5~^{\circ}\mathrm{C}$ and $3~^{\circ}\mathrm{C}$, ADOPT-FPML automatically recovers the series (Config 2) interaction pattern previously developed for this system through physical reasoning and heuristic model construction, while reducing the ML model to only 58 and 84 parameters, respectively, compared with 176 parameters for the corresponding full Config 2 baseline. The associated architecture-specific $\mathrm{AIC}_c$ values are approximately 3,517 and 7,483, compared with 11,866 for the full-input series model. These results show that the benefit of automated structure discovery is not restricted to identifying a canonical interaction pattern; it also arises from determining where hybridization is actually required and allocating data-driven model capacity only to those outputs and information pathways for which it provides sufficient value. It should be noted that the optimal FPML model may still fail to satisfactorily represent some outputs under the prescribed accuracy criteria. In such cases, users should consider alternative ML models, improvements to the FP model, additional measured or estimated inputs, and uncertainties in the measurements.  

The individual algorithmic components introduced in ADOPT-FPML provide several opportunities for further development. First, although the correlation-based input-ranking strategy adopted in this work provides a computationally efficient screening mechanism, more general measures of input relevance may be considered to better capture nonlinear dependencies in multivariate systems. Second, the present branch-and-prune implementation follows a greedy optimality principle, which substantially reduces computational expense but does not guarantee global optimality. Future developments may therefore explore alternative search and pruning strategies that better balance computational efficiency and the quality of the resulting hybrid FPML architectures. Additional advances may be required to extend the proposed framework to larger and more complex process systems. In particular, more efficient output-selection strategies and automated partitioning of interconnected systems into suitable subsystem envelopes represent promising directions for improving scalability. Overall, ADOPT-FPML provides a foundational framework for addressing a longstanding question in hybrid modeling: {\textit{Given an available FP model and an ML model, what is the most effective way to fuse their information?}} By treating the information-transfer mechanism as an explicit structural decision, the proposed approach represents a step toward the systematic discovery of hybrid FPML architectures rather than relying on configurations prescribed {\textit{a priori}}.

\section*{Code Availability Statement} \label{sec:code_git}

All codes required to generate the synthetic data for the case studies considered in this work, together with the implementation of the ADOPT-FPML optimization framework and installation instructions for the \texttt{adopt-fpml} Python package, are publicly available at: \url{https://github.com/mukherjeeangan26/ADOPT-FPML}. The package can also be installed directly using ~\texttt{pip install adopt-fpml}.

\section*{Acknowledgements} \label{sec:acknowledgement}

The project is funded by the U.S. DOE through the project titled “Boiler Health Monitoring Using a Hybrid First Principles – Artificial Intelligence Model” (Grant \#: DE – FE0031768). The DOE financial support is gratefully acknowledged.

\section*{Appendices}
\addcontentsline{toc}{section}{Appendices}

\appendix

\counterwithin{figure}{section}
\counterwithin{table}{section}

The Appendices provide additional information on the data-generation models and supplementary results associated with the application of ADOPT-FPML to the three case studies considered in this work.

\section{Additional Information and Results for Case Study 1}
\label{app:e1}

Appendix \ref{app:e1} provides the detailed model used to generate the synthetic measurement data for Case Study 1 (Envelope E1), together with additional results from the automated structure-discovery framework.

\subsection{Data Generation for Envelope E1}
\label{app:e1_data_generation}

As discussed in Section \ref{sec:case_study_e1}, the synthetic measurement data used for Envelope E1 are generated using a dynamic representation of the vapor-phase acetone-cracking reactor consisting of three CSTRs connected in series. For the $s$-th CSTR, where $s\in\{1,2,3\}$, the corresponding mass and energy balances are given by:

\begin{align}
\frac{dC_{A,s}}{dt}&=\frac{v}{V_s}\left(C_{A,\mathrm{in},s}-C_{A,s}\right)-r_s
\label{eq:app_e1_ca}
\\
\frac{dC_{B,s}}{dt}&=\frac{v}{V_s}\left(C_{B,\mathrm{in},s}-C_{B,s}\right)+r_s
\label{eq:app_e1_cb}
\\
\frac{dC_{C,s}}{dt}&=\frac{v}{V_s}\left(C_{C,\mathrm{in},s}-C_{C,s}\right)+r_s
\label{eq:app_e1_cc}
\\
\frac{dT_s}{dt}&=\frac{v}{V_s}\left(T_{\mathrm{in},s}-T_s\right)
+\frac{\dot q_s}{C_{p,A}C_{A,s}}-\frac{\Delta H_r}{C_{p,A}C_{A,s}}r_s
\label{eq:app_e1_temperature}
\end{align}

\noindent For the first reactor stage,

\begin{equation}
C_{A,\mathrm{in},1}=C_{A,0}, \qquad
C_{B,\mathrm{in},1}=C_{C,\mathrm{in},1}=0, \qquad
T_{\mathrm{in},1}=T_{\mathrm{in}}
\label{eq:app_e1_stage1_inlet}
\end{equation}
while the outlet conditions of each preceding CSTR provide the inlet conditions to the subsequent stage, such that:

\begin{equation}
\left[C_{A,\mathrm{in},s},C_{B,\mathrm{in},s},C_{C,\mathrm{in},s},
T_{\mathrm{in},s}\right]^{\mathsf T}=
\left[C_{A,s-1},C_{B,s-1},C_{C,s-1},T_{s-1}\right]^{\mathsf T},
\qquad s\in\{2,3\}
\label{eq:app_e1_stage_coupling}
\end{equation}

For generation of the synthetic process dynamic data, the reaction rate is represented using a reaction kinetic expression with order $n=1.2$, as:
\begin{equation}
r_s=k(T_s)C_{A,s}^{n_r},
\qquad n_r=1.2
\label{eq:app_e1_rate}
\end{equation}
where the temperature-dependent rate constant follows the Arrhenius relation:
\begin{equation}
k(T_s)=A\exp\left(-\frac{E_a}{RT_s}\right)
\label{eq:app_e1_arrhenius}
\end{equation}
The kinetic parameters used for data generation are defined as:
\begin{equation}
A=1.0\times10^{10},
\qquad
\frac{E_a}{R}=34222~\mathrm{K}
\label{eq:app_e1_kinetic_parameters}
\end{equation}
while $\Delta H_r=60000$ and $C_{p,A}=163$ are used in the corresponding energy balances. The total reactor volume is $V=1.0\times10^{-2}$ and is divided equally among the three CSTR stages, such that:
\begin{equation}
V_s=\frac{V}{3},
\qquad s\in\{1,2,3\}
\label{eq:app_e1_stage_volume}
\end{equation}

The transient operating trajectory is generated by independently perturbing the 6 measured inputs defined in Eq. \eqref{eq:e1_input_vector}. Each input is represented as a piecewise-constant random signal:
\begin{equation}
u_m(t_k)=\xi_{m,b},
\qquad
\xi_{m,b}\sim\mathcal{U}
\left(u_m^{\mathrm{L}},u_m^{\mathrm{U}}\right),
\qquad
b=\left\lfloor\frac{k}{N_h}\right\rfloor
\label{eq:app_e1_input_excitation}
\end{equation}
where $N_h=35$ denotes the number of samples for which each randomly generated operating condition is held constant, and $u_m^{\mathrm{L}}$ and $u_m^{\mathrm{U}}$ denote the admissible lower and upper bounds for input $m$. Independent perturbations are generated for $C_{A,0}$, $T_{\mathrm{in}}$, $v$, $\dot q_1$, $\dot q_2$, and $\dot q_3$.

To emulate measurement characteristics encountered in practical process systems, the clean outputs generated from Eqs. \eqref{eq:app_e1_ca}-\eqref{eq:app_e1_stage_volume} are subsequently modified using dynamic bias and stochastic noise. Let

\begin{equation}
{\bf y}^{\mathrm{clean}}(t)=
\left[C_A^{\mathrm{clean}}(t),C_B^{\mathrm{clean}}(t),C_C^{\mathrm{clean}}(t),
T_{\mathrm{out}}^{\mathrm{clean}}(t)\right]^{\mathsf T}
\label{eq:app_e1_clean_outputs}
\end{equation}
denote the clean outlet trajectories obtained from the third CSTR. The first measurement-perturbation layer is defined as:
\begin{equation}
{\bf y}^{(1)}(t)={\bf y}^{\mathrm{clean}}(t)+
{\boldsymbol \eta}^{(1)}(t)+{\boldsymbol \epsilon}^{(1)}(t)
\label{eq:app_e1_bias_layer1}
\end{equation}
where
\begin{equation}
{\boldsymbol \eta}^{(1)}(t)={\bf f}^{(1)}
\left(
\widetilde C_{A,0}(t),
\widetilde T_{\mathrm{in}}(t)
\right)
\label{eq:app_e1_bias_function1}
\end{equation}
represents a dynamic input-dependent bias function. Here, $\widetilde{(\cdot)}$ denotes a standardized variable. The function ${\bf f}^{(1)}(\cdot)$ contains linear, quadratic, and bilinear terms of the corresponding normalized process variables, resulting in temporally varying and correlated biases across the measured outputs.

A second perturbation layer is introduced to additionally represent biases that depend on the evolving measured process state:
\begin{equation}
{\bf y}^{\mathrm{data}}(t)=
{\bf y}^{(1)}(t)+{\boldsymbol \eta}^{(2)}(t)+{\boldsymbol \epsilon}^{(2)}(t)
\label{eq:app_e1_bias_layer2}
\end{equation}
where
\begin{equation}
{\boldsymbol \eta}^{(2)}(t)=
{\bf f}^{(2)}\left(\widetilde C_A^{(1)}(t),\widetilde C_B^{(1)}(t),
\widetilde T_{\mathrm{out}}^{(1)}(t)\right)
\label{eq:app_e1_bias_function2}
\end{equation}

The function ${\bf f}^{(2)}(\cdot)$ similarly contains linear, quadratic, and bilinear interactions among the corresponding process variables. Consequently, the overall perturbation is both operating-condition dependent and correlated with the evolving process response, rather than being represented solely by independent additive white noise.

For perturbation layer $r\in\{1,2\}$, the stochastic measurement noise for output $j$ is generated according to:
\begin{equation}
\epsilon_j^{(r)}(t)\sim\mathcal{N}\left(0,\left(\gamma_j^{(r)}s_j\right)^2
\right)
\label{eq:app_e1_noise}
\end{equation}
where $s_j$ denotes the standard deviation of the corresponding clean output trajectory and $\gamma_j^{(r)}$ denotes its admissible relative noise magnitude. The noise-scale vectors used for the first and second perturbation layers are:
\begin{equation}
{\boldsymbol \gamma}^{(1)}=[0.010,\;0.012,\;0.007,\;0.005]^{\mathsf T},
\qquad
{\boldsymbol \gamma}^{(2)}=[0.035,\;0.035,\;0.035,\;0.005]^{\mathsf T}
\label{eq:app_e1_noise_scales}
\end{equation}
corresponding sequentially to $C_A$, $C_B$, $C_C$, and $T_{\mathrm{out}}$.

The dynamic process model is numerically integrated using a fourth-order Runge-Kutta scheme with eight integration sub-steps per reported sample. A continuous trajectory of 750 samples is generated using the above procedure. The first 500 samples are used for training and automated structure discovery, while the subsequent 250 samples are retained as an unseen testing dataset. The process states evolve continuously across the training-testing boundary; no model re-initialization, structure selection, parameter refitting, or threshold screening is performed using the testing data.

\subsection{Additional Results for Envelope E1}
\label{app:e1_additional_results}

\begin{figure}[htbp]
    \centering
    \includegraphics[width=0.80\textwidth]{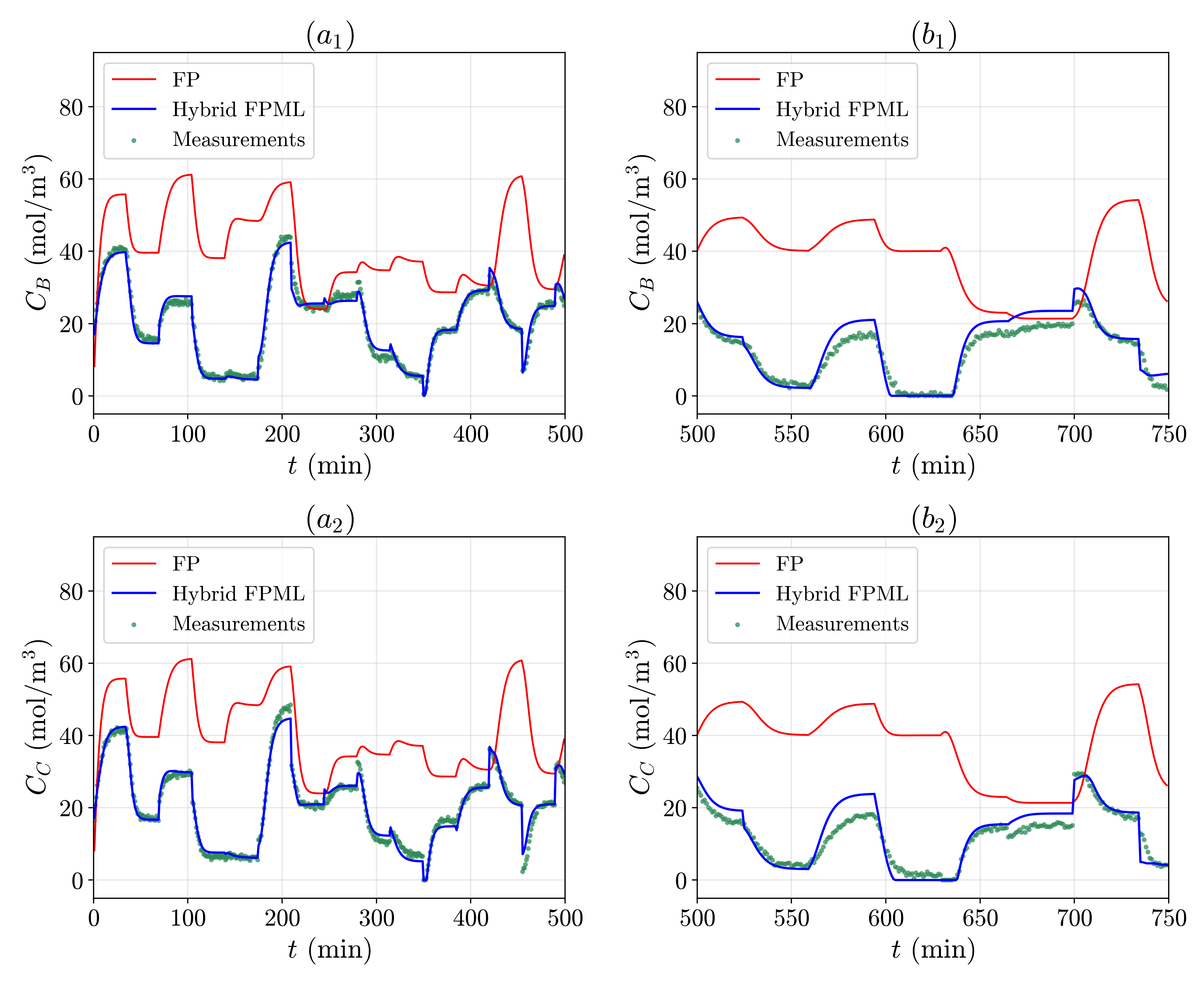}
    \caption{Training and testing performance of the optimal hybrid FPML architecture for Envelope E1: ($a_1$) training and ($b_1$) testing predictions for $C_B$, and ($a_2$) training and ($b_2$) testing predictions for $C_C$.}
    \label{fig:app_e1_cbcc_train_test}
\end{figure}

Figure \ref{fig:app_e1_cbcc_train_test} provides the corresponding training and testing predictions for $C_B$ and $C_C$, which are resolved by the first stage optimal configuration obtained from the ADOPT-FPML. Similar to the results reported for $C_A$ and $T_{\mathrm{out}}$ in Figure \ref{fig:e1_train_test}, the discovered hybrid architecture substantially improves the predictions relative to the standalone FP model over both the training and testing horizons. The complete training and testing RMSE values for all 4 E1 outputs, together with their corresponding admissible prediction thresholds, are summarized in Table \ref{tab:app_e1_train_test_rmse}.

\begin{table}[htbp]
\centering
\caption{Output-specific thresholds and RMSE values evaluated on the training and testing data for the standalone FP model and the optimal ADOPT-FPML architecture for Envelope E1.}
\label{tab:app_e1_train_test_rmse}
\small
\setlength{\tabcolsep}{5pt}
\renewcommand{\arraystretch}{1.08}

\begin{tabular}{c|c|c|c|c|c}
\hline
\multirow{2}{*}{Output} &
\multirow{2}{*}{$\varepsilon_{p,j}$} &
\multicolumn{2}{c|}{RMSE on Training Data} &
\multicolumn{2}{c}{RMSE on Testing Data} \\
\cline{3-6}
& & FP & ADOPT-FPML & FP & ADOPT-FPML \\
\hline

$C_A~(\mathrm{mol/m^3})$
& 5.0 & 33.6 & 2.9 & 38.5 & 7.3 \\

$C_B~(\mathrm{mol/m^3})$
& 2.0 & 24.3 & 1.5 & 29.9 & 2.8 \\

$C_C~(\mathrm{mol/m^3})$
& 3.0 & 23.6 & 1.4 & 29.1 & 3.0 \\

$T_{\mathrm{out}}~(\mathrm{K})$
& 10.0 & 149.8 & 6.4 & 174.6 & 19.0 \\

\hline
\end{tabular}
\normalsize
\end{table}

\section{Additional Information and Results for Case Study 2}
\label{app:e2}

Appendix \ref{app:e2} provides the detailed process representation used to generate the synthetic measurement data for Case Study 2 (Envelope E2), together with supplementary prediction results from the optimal ADOPT-FPML architecture.

\subsection{Data Generation for Envelope E2}
\label{app:e2_data_generation}

As discussed in Section \ref{sec:case_study_e2}, the outlet trajectory generated for Envelope E1 provides the inlet boundary conditions for Envelope E2. Thus, the time-dependent quantities $C_A(t)$, $C_B(t)$, $C_C(t)$, $T_{\mathrm{out}}(t)$, and $v(t)$ obtained from E1 are used directly as the feed trajectory for the cooler-flash system. The component molar flow rates entering E2 are calculated as:

\begin{equation}
F_i(t)=C_i(t)v(t),
\qquad i\in\{A,B,C\}
\label{eq:app_e2_component_flow}
\end{equation}
with total feed flow rate:

\begin{equation}
F(t)=\sum_i F_i(t)
\label{eq:app_e2_total_flow}
\end{equation}
and overall feed mole fractions:

\begin{equation}
z_i(t)=\frac{F_i(t)}{F(t)}
\label{eq:app_e2_feed_composition}
\end{equation}

For the cooler, the equilibrium ratios are represented as temperature- and pressure-dependent quantities:

\begin{equation}
K_i(T,P)=K_{i,\mathrm{ref}}
\left(\frac{P_{\mathrm{ref}}}{P}\right)
\exp\left[-\frac{\Delta H_{K,i}}{R}\left(
\frac{1}{T}-\frac{1}{T_{\mathrm{ref}}}\right)\right]
\label{eq:app_e2_k_tp}
\end{equation}
The cooler outlet temperature $T_c(t)$ is obtained by specifying a target vapor fraction $\beta_c=0.35$ and solving the Rachford-Rice relation, given by:

\begin{equation}
\sum_i\frac{z_i(t)\left[K_i(T_c,P_c)-1\right]}{1+\beta_c\left[K_i(T_c,P_c)-1\right]}=0
\label{eq:app_e2_cooler_rr}
\end{equation}
The corresponding liquid- and vapor-phase compositions are:

\begin{align}
x_{i,c}(t)&=\frac{z_i(t)}
{1+\beta_c\left[K_i(T_c,P_c)-1\right]}
\label{eq:app_e2_cooler_x}
\\[0.2cm]
y_{i,c}(t)&=K_i(T_c,P_c)x_{i,c}(t)
\label{eq:app_e2_cooler_y}
\end{align}
The mixture heat-capacity rate is calculated as:

\begin{equation}
\dot C_p(t)=F(t)\sum_i z_i(t)C_{p,i}^{\mathrm{vap}}
\label{eq:app_e2_heat_capacity_rate}
\end{equation}
and the corresponding time-dependent effective cooler conductance is obtained from:

\begin{equation}
UA(t)=-\dot C_p(t)\ln\left[\frac{T_c(t)-T_{\mathrm{coolant}}}
{T_{\mathrm{out}}(t)-T_{\mathrm{coolant}}}\right]
\label{eq:app_e2_ua}
\end{equation}
The feed and two-phase cooler outlet enthalpy rates are evaluated as:

\begin{equation}
\dot H_{\mathrm{feed}}=\sum_i F_i
\left[C_{p,i}^{\mathrm{vap}}(T_{\mathrm{out}}-T_{\mathrm{ref}})
+
\Delta H_{\mathrm{vap},i}\right]
\label{eq:app_e2_feed_enthalpy}
\end{equation}
and

\begin{equation}
\dot H_c=F\left[\beta_c h_{\mathrm{vap}}+
(1-\beta_c)h_{\mathrm{liq}}\right]
\label{eq:app_e2_cooler_enthalpy}
\end{equation}
where,

\begin{align}
h_{\mathrm{vap}}&=\sum_i y_{i,c}\left[
C_{p,i}^{\mathrm{vap}}
(T_c-T_{\mathrm{ref}})+\Delta H_{\mathrm{vap},i}\right]
\\[0.2cm]
h_{\mathrm{liq}}&=\sum_i x_{i,c}C_{p,i}^{\mathrm{liq}}
(T_c-T_{\mathrm{ref}})
\end{align}
The corresponding cooler duty is:

\begin{equation}
\dot Q_c=\dot H_{\mathrm{feed}}-\dot H_c
\label{eq:app_e2_cooler_duty}
\end{equation}
The downstream flash separator is represented as an isothermal flash, such that:

\begin{equation}
T_{\mathrm{flash}}(t)=T_c(t)
\label{eq:app_e2_flash_temperature}
\end{equation}
while the flash equilibrium ratios are pressure dependent, defined as:

\begin{equation}
K_i^{\mathrm{flash}}=K_{i,\mathrm{ref}}
\left(\frac{P_{\mathrm{ref}}}{P_{\mathrm{flash}}}\right)
\label{eq:app_e2_flash_k}
\end{equation}
The flash vapor fraction $\beta(t)$ is obtained from:

\begin{equation}
\sum_i\frac{z_i(t)\left(K_i^{\mathrm{flash}}-1\right)
}{1+\beta(t)\left(K_i^{\mathrm{flash}}-1\right)}=0
\label{eq:app_e2_flash_rr}
\end{equation}
and the equilibrium compositions are given by:

\begin{align}
x_i(t)&=\frac{z_i(t)}{1+\beta(t)\left(K_i^{\mathrm{flash}}-1\right)}
\\[0.2cm]
y_i(t)&=K_i^{\mathrm{flash}}x_i(t)
\end{align}

Finally,

\begin{align}
F_{\mathrm{vap}}(t)
&=
\beta(t)F(t)
\\[0.2cm]
F_{\mathrm{liq}}(t)&=\left[1-\beta(t)\right]F(t)
\\[0.2cm]
T_{\mathrm{vap}}(t)&=T_{\mathrm{liq}}(t)=T_{\mathrm{flash}}(t)
\end{align}
The numerical constants used for the data-generating model are $T_{\mathrm{ref}}=330$, $P_{\mathrm{ref}}=1$, $P_c=P_{\mathrm{flash}}=2$, $T_{\mathrm{coolant}}=285$, and,

\begin{equation}
{\bf K}_{\mathrm{ref}}
=
[0.15,\;12.0,\;0.35]^{\mathsf T},
\qquad
\Delta{\bf H}_{K}
=
[29000,\;8000,\;22000]^{\mathsf T}
\end{equation}
together with,

\begin{equation}
{\bf C}_p^{\mathrm{vap}}=
[75,\;36,\;44]^{\mathsf T},
\qquad
{\bf C}_p^{\mathrm{liq}}
=
[125,\;55,\;90]^{\mathsf T}
\end{equation}
and $\Delta{\bf H}_{\mathrm{vap}}=[29000,\;8000,\;22000]^{\mathsf T}$, using units consistent with the data-generation implementation.

The clean flash outputs are subsequently perturbed to emulate process measurements. The vapor- and liquid-flow measurements are generated as:

\begin{equation}
F_{\alpha}^{\mathrm{data}}(t)=
F_{\alpha}(t)+\epsilon_{F_\alpha}(t),
\qquad
\epsilon_{F_\alpha}(t)\sim
\mathcal{N}\left(0,[0.003F_{\alpha}(t)]^2\right),
\quad
\alpha\in\{\mathrm{vap},\mathrm{liq}\}
\label{eq:app_e2_flow_noise}
\end{equation}
The phase-composition measurements are generated as:

\begin{equation}
{\bf x}^{\mathrm{data}}(t)=\mathcal{N}_{c}
\left[{\bf x}(t)+{\boldsymbol\epsilon}_x(t)\right],
\qquad
{\bf y}^{\mathrm{data}}(t)=\mathcal{N}_{c}
\left[{\bf y}(t)+{\boldsymbol\epsilon}_y(t)\right]
\label{eq:app_e2_composition_noise}
\end{equation}
where $\mathcal{N}_{c}[\cdot]$ denotes normalization to unit composition sum and the individual perturbations are independently sampled from $\mathcal{N}(0,0.001^2)$. The temperature measurements are generated according to:

\begin{equation}
T_{\alpha}^{\mathrm{data}}(t)=
T_{\mathrm{flash}}(t)+\epsilon_{T_\alpha}(t),
\qquad
\epsilon_{T_\alpha}(t)\sim\mathcal{N}(0,0.1^2),
\quad
\alpha\in\{\mathrm{vap},\mathrm{liq}\}
\label{eq:app_e2_temperature_noise}
\end{equation}
A continuous E2 trajectory corresponding to the E1 operating trajectory is generated using this formulation. The first 500 samples are used for training and automated structure discovery, while the subsequent 250 samples are retained exclusively for testing. As in E1, the testing data are not used for structure selection, input/output selection, parameter fitting, or threshold screening.

\subsection{Additional Results for Envelope E2}
\label{app:e2_additional_results}

\begin{table}[htbp]
\centering
\caption{Output-specific prediction thresholds and RMSE values evaluated on the training and testing data for the standalone FP model and the optimal ADOPT-FPML architecture for Envelope E2.}
\label{tab:app_e2_train_test_rmse}
\footnotesize
\setlength{\tabcolsep}{4pt}
\renewcommand{\arraystretch}{1.08}

\begin{tabular}{c|c|c|c|c|c}
\hline
\multirow{2}{*}{Output} &
\multirow{2}{*}{$\varepsilon_{p,j}$} &
\multicolumn{2}{c|}{RMSE on Training Data} &
\multicolumn{2}{c}{RMSE on Testing Data} \\
\cline{3-6}
& & FP & ADOPT-FPML & FP & ADOPT-FPML \\
\hline

$F_{\mathrm{vap}}~(\mathrm{mol/min})$
& 0.005
& $1.2\times10^{-2}$
& $4.3\times10^{-3}$
& $1.3\times10^{-2}$
& $2.6\times10^{-3}$ \\

$F_{\mathrm{liq}}~(\mathrm{mol/min})$
& 0.010
& $1.2\times10^{-2}$
& $5.1\times10^{-4}$
& $1.3\times10^{-2}$
& $7.9\times10^{-4}$ \\

$T_{\mathrm{vap}}~(\mathrm{K})$
& 20.0
& 63.7
& 7.1
& 29.4
& 6.2 \\

$T_{\mathrm{liq}}~(\mathrm{K})$
& 20.0
& 63.7
& 8.2
& 29.4
& 7.2 \\

$y_A$
& 0.010
& $2.5\times10^{-2}$
& $1.2\times10^{-3}$
& $6.2\times10^{-3}$
& $1.0\times10^{-3}$ \\

$y_B$
& 0.050
& $1.4\times10^{-1}$
& $2.1\times10^{-2}$
& $1.5\times10^{-1}$
& $7.4\times10^{-3}$ \\

$y_C$
& 0.010
& $1.4\times10^{-1}$
& $5.4\times10^{-3}$
& $1.4\times10^{-1}$
& $2.1\times10^{-3}$ \\

$x_A$
& 0.050
& $8.6\times10^{-2}$
& $1.6\times10^{-2}$
& $8.3\times10^{-2}$
& $1.3\times10^{-2}$ \\

$x_B$
& 0.010
& $5.8\times10^{-2}$
& $8.4\times10^{-3}$
& $6.8\times10^{-2}$
& $3.1\times10^{-3}$ \\

$x_C$
& 0.050
& $3.5\times10^{-2}$
& -- 
& $2.0\times10^{-2}$
& -- \\

\hline
\end{tabular}

\normalsize
\end{table}

\begin{figure}[htbp]
    \centering
    \includegraphics[width=0.95\textwidth]{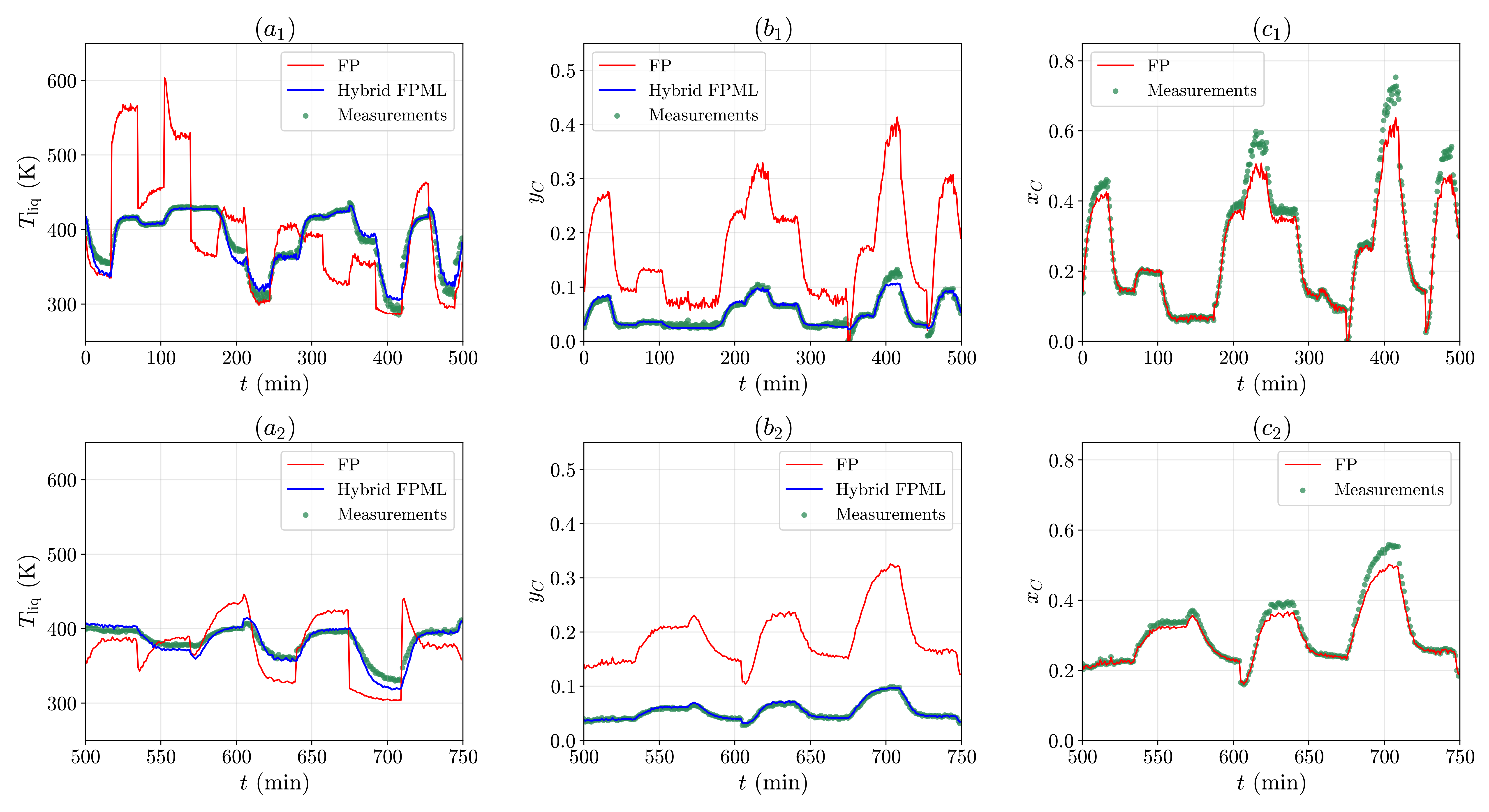}
    \caption{Additional training and testing results for Envelope E2: ($a_1$) training and ($a_2$) testing predictions for $T_{\mathrm{liq}}$, ($b_1$) training and ($b_2$) testing predictions for $y_C$, and ($c_1$) training and ($c_2$) testing predictions for $x_C$. The output $x_C$ satisfies the prescribed standalone FP accuracy threshold and therefore does not require hybridization.}
    \label{fig:app_e2_additional_train_test}
\end{figure}

Figure \ref{fig:app_e2_additional_train_test} provides additional training and testing comparisons for $T_{\mathrm{liq}}$, $y_C$, and $x_C$. While $T_{\mathrm{liq}}$ and $y_C$ are predicted through the discovered hybrid architecture, $x_C$ already satisfies the prescribed accuracy threshold using the standalone FP model and therefore does not require hybridization.

\bibliography{ref}
\bibliographystyle{unsrt}

\end{document}